\documentclass[floats,floatfix,amssymb,prd,twocolumn,superscriptaddress,nofootinbib]{revtex4-2}

\usepackage[english]{babel}
\usepackage[utf8]{inputenc}
\usepackage{amsmath}
\usepackage{mathbbol}
\usepackage{amssymb}
\usepackage{bbold}
\usepackage{graphicx,amsfonts}
\usepackage{epsfig}
\usepackage{hyperref}
\usepackage{bm}
\usepackage{mathrsfs}
\usepackage{mathtools}
\usepackage{enumerate}
\usepackage{amsthm}
\usepackage{bbm}
\usepackage{comment}
\usepackage{physics}
\usepackage{url}
\usepackage{upgreek}
\usepackage[svgnames]{xcolor}
\usepackage[title]{appendix}
\definecolor{teal}{RGB}{0, 128, 128}
\definecolor{myred}{RGB}{179, 27, 27}
\hypersetup{
    colorlinks=true,
    linkcolor=teal, 
    citecolor=teal, 
    urlcolor=teal  
 }

\begin{document}

\title{Environmentally-induced chaos: Extreme-mass-ratio systems of rotating black holes in astrophysical environments}
%
\author{Kyriakos Destounis}
\email{kyriakosdestounis@tecnico.ulisboa.pt}
\affiliation{CENTRA, Departamento de Física, Instituto Superior Técnico – IST, Universidade
de Lisboa – UL, Avenida Rovisco Pais 1, 1049-001 Lisboa, Portugal}
\author{Pedro G. S. Fernandes}
\email{fernandes@thphys.uni-heidelberg.de}
\affiliation{Institut für Theoretische Physik, Universität Heidelberg, Philosophenweg 12, 69120 Heidelberg, Germany}
%
\begin{abstract}
Extreme-mass-ratio inspirals, in which a stellar-mass object orbits a supermassive black hole, are prime sources of millihertz gravitational waves for upcoming space-based detectors. While most studies assume idealized vacuum backgrounds, realistic extreme-mass-ratio binaries are embedded in astrophysical environments such as accretion disks, stellar clusters, or dark matter spikes, disks, and halos, which can significantly alter the orbital dynamics. We explore bound geodesics around general-relativistic solutions describing rotating black holes surrounded by matter halos for the first time, mapping how environmental effects interfere with the spacetime symmetries of vacuum spinning (Kerr) black holes. In particular, we find that the loss of a Carter-like constant leads to geodesic non-integrability and the onset of chaos. This manifests through the formation of resonant islands and chaotic layers around transient orbital resonances in phase space--features that are otherwise completely absent in integrable Kerr geodesics. Resonant islands, which are extended, non-zero volume regions in phase space, encapsulate periodic orbit points. Non-integrability dictates that all geodesics inside the resonant island share the periodicity of the resonance. Thus, the lifespan of resonances around non-Kerr objects can be significantly enhanced beyond the predicted lifetime of Kerr resonances. Consequently, these effects can leave distinct imprints on gravitational-wave signals, with significant implications for gravitational-wave modeling and parameter inference of astrophysical extreme-mass-ratio inspirals.
\end{abstract}

\maketitle

\section{Introduction}

As gravitational-wave (GW) detections from ground-based interferometers are increasing exponentially \cite{KAGRA:2021vkt,LIGO:2021ppb}, our understanding of gravitation becomes more precise \cite{Barack:2018yly}. In the verge of an era where GW detections are systematically becoming far more sensitive and accurate than the initial, though still ground-breaking, detection of 2015 \cite{LIGOScientific:2016aoc}, novel space-borne detectors, like the Laser Interferometer Space Antenna (LISA) \cite{LISA:2017pwj}, will unlock a new window of the Cosmos through the detection of gravitational radiation emitted by enticing theorized sources, thus transforming our understanding of GW astrophysics and astronomy \cite{Schutz:1999xj,Bailes:2021tot}. 

The main categories of LISA targets that regard fundamental physics \cite{Barausse:2020rsu,LISA:2022kgy,Karnesis:2022vdp} and astrophysics \cite{Amaro-Seoane:2022rxf,LISA:2022yao} alone are enthralling; supermassive black-hole (BH) mergers, such as those occurring during galactic collisions that include such cosmic giants in their very core \cite{Ghez:1998ph,Genzel:2010}, compact binary systems in the Milky Way, and infalling stellar-mass compact objects into supermassive BHs, known as extreme-mass-ratio inspirals (EMRIs) \cite{Amaro-Seoane:2007osp,Gair:2017ynp,Babak:2017tow}. LISA will also be sensitive to the measurement of the stochastic GW background which is very important for gravitation, cosmology and particle physics. LISA will contribute to the understanding of the early Universe, the nature of dark matter and dark energy, as well as potential deviations from standard cosmological models and ``beyond the Standard Model'' physics \cite{LISACosmologyWorkingGroup:2022jok}. 

EMRIs are one of the most promising targets of LISA due to their expected formation rates in galactic centers \cite{Raveh:2020jxg,Amaro-Seoane:2012lgq,Pan:2021ksp,Pan:2021oob}, and the fact that the secondary stellar-mass companion is anticipated to perform thousands of revolutions in the strong-field regime before plunging into the supermassive primary. Their extremely large lifetime instantly renders numerical relativity impractical. New techniques are being developed, in order to speed up the process of waveform modeling in EMRIs \cite{Chua:2020stf,Katz:2021yft,LISAConsortiumWaveformWorkingGroup:2023arg}, though there is still an abundance of phenomenology that has to be taken into account. Their unique cosmic design is tailored for precision tests of General Relativity (GR) \cite{Glampedakis:2005cf,Gair:2012nm,Yunes:2011aa,Cardoso:2016ryw,Pani:2011xj,Cardoso:2011xi,Canizares:2012is,Berti:2015itd,Chua:2018yng,Berti:2019xgr,Zhang:2018kib,Berry:2019wgg,Cardenas-Avendano:2024mqp,Berti:2024orb,Datta:2019euh}, probing the existence of new fundamental fields \cite{Hannuksela:2018izj,Maselli:2020zgv,Maselli:2021men,Barsanti:2022vvl,Barsanti:2022ana,Mitra:2023sny,Zhang:2024ogc,Kuntz:2020yow,Lestingi:2023ovn,DellaRocca:2024pnm,Zi:2025lio}, and even testing the existence of astrophysical environments around EMRIs \cite{Gondolo:1999ef,Sigl:2006cg,Preto:2009kd,Yunes:2011ws,Kocsis:2011dr,Macedo:2013qea,Sadeghian:2013laa,Ferrer:2017xwm,Hannuksela:2019vip,Cardoso:2019rou,Polcar:2022bwv,Speri:2022upm,Duque:2023seg,Brito:2023pyl,Traykova:2021dua,Traykova:2023qyv,Chatterjee:2023cry,Santini:2023ukl,Roy:2024rhe,Duque:2024mfw,Lyu:2024gnk,Dyson:2025dlj,Mitra:2025tag,Blas:2024duy,Vicente:2025gsg,Polcar:2025yto,Rahman:2023sof,Rahman:2025mip}. 

Even though, significant work has been done under the assumption of a vacuum spacetime, the majority of binary merger events should, in principle, take place inside astrophysical environments that occupy our Universe \cite{CanevaSantoro:2023aol,Zwick:2025wkt,Zwick:2025ine}. In fact, astrophysical environments pose a key feature in various contexts of gravitational radiation; from the inspiral till the ringdown \cite{Destounis:2023ruj,Jaramillo:2020tuu,Jaramillo:2021tmt,Cheung:2021bol,Destounis:2021lum,Boyanov:2022ark,Sarkar:2023rhp,Destounis:2023nmb,Boyanov:2023qqf,Arean:2023ejh,Cownden:2023dam,Courty:2023rxk,Rosato:2024arw,Boyanov:2024fgc,Cai:2025irl,Spieksma:2024voy,Datta:2023zmd,Datta:2025ruh}. Realistic EMRIs, are, thus, expected to interact with surrounding matter. These environmental components can alter the orbital structure of EMRIs. We now have more than enough evidence that matter environments, luminous or otherwise, are present in the surroundings of compact objects \cite{Pease:1916,Pease:1918,Schwarzschild:1954,Lindblad:1959,deVaucouleurs:1959,deVaucouleurs:1972,Burbidge:1975,vanderKruit:1978,Sofue:2000jx,Canizares:1979,Canizares:1982,Boughn:1983,EventHorizonTelescope:2019dse,EventHorizonTelescope:2022wkp}. Therefore, a huge effort is being spent into employing the addition of environmental effects in GW generation and propagation from EMRIs and systematically streamlining the process.

The bulk of EMRI analyses are treating environments within perturbation theory or with Newtonian and post-Newtonian schemes that approximate the self-gravity of matter, estimate dynamical friction, and simplify gravitational redshift or peculiar motion \cite{Taylor:2008xy,Poisson:2018qqd,Eda:2013gg,Tamanini:2019usx,Kavanagh:2020cfn,Ylla:2025hia}. Even though successful at the time, due to the demands of LISA parameter inference a growing need for first-principle calculations of BHs in astrophysical environments is inevitable. To go beyond crude estimates, a general-relativistic, exact solution of the field equations has been obtained \cite{Cardoso:2021wlq}, that describes a BH immersed in a Hernquist-type dark matter halo \cite{Hernquist:1990be}. This technique helped the community to extend further the construction of general-relativistic spacetimes to other exact and numerical BH solutions with different dark matter distributions around them  \cite{Stuchlik:2021gwg,Konoplya:2022hbl,Jusufi:2022jxu,Pezzella:2024tkf}. The axial and polar GW fluxes of circular equatorial EMRIs including general-relativistic, static and spherically-symmetric BHs surrounded by generic environments have been investigated in a relativistic setup in Refs. \cite{Cardoso:2022whc,Speeney:2024mas,Figueiredo:2023gas,Gliorio:2025cbh}. The results strongly suggest the need for a paradigm-shifting strategy in order to accurately capture environmental effects in EMRIs. 

A milestone regarding BHs in environments was achieved recently; a fully-relativistic numerical configuration of a rotating BH embedded in a Hernquist-type matter halo \cite{Fernandes:2025osu}. These solutions, and the analysis of EMRIs that will be performed with them, will lead us one step closer to the ultimate goal; a generic perturbation scheme for general-relativistic BHs in astrophysical environments that calculates the GW fluxes of generic EMRIs. This technique will provide unlimited access to precise EMRI waveform models, and to date it only exists in spherical symmetry \cite{Cardoso:2022whc,Datta:2025ruh}. EMRIs evolve adiabatically due to the large mass disparity between the primary and secondary. This allows one to perceive the evolution of the secondary, at leading-order, as a geodesic of a massive test-particle in a fixed background. At next-to-leading order, radiation reaction effects take place and drive the secondary through a drift of successively-damped geodesics. 

Formally, a general, stationary and axisymmetric spacetime possesses two Killing vector fields associated with stationarity and axisymmetry. These Killing vector fields are associated with two constants (integrals) of motion; the conserved energy and azimuthal angular momentum. Vacuum Kerr spacetimes have the privilege of possessing yet another constant of motion, namely the Carter constant \cite{Carter:1968rr}, which is associated to a rank-two Killing tensor field. The conservation of the energy, azimuthal angular momentum and Carter constant of the test particle, are enough to deem the geodesics separable, i.e., \emph{integrable}. This results to four, decoupled, first-order ordinary differential equations for the degrees of freedom of the metric \cite{Contopoulos_book}. 

On the other hand, if the, very fragile, Carter symmetry is broken, then even though the temporal and azimuthal geodesics still form decoupled and first-order differential equations, the radial and polar motion of geodesics consist of a coupled second-order dynamical system. In this case, the geodesics of spacetime are \emph{non-integrable}, i.e. they lack a separation constant, and chaotic phenomenology in the orbital phase space emanates \cite{Contopoulos_book}. These phenomena appear in many distinctive ways in various dynamical systems. The most relevant indicator of chaos in EMRIs is the formation of a Birkhoff chain around resonant points in a Poincar\'e surface of section. The chain includes nested island formations around the periodic stable points, also known as resonant islands or islands of stability. 

Resonances in Kerr EMRIs are by themselves challenging for GW modeling and data analysis \cite{Flanagan:2010cd}. This is due to the fact that when the secondary crosses a Kerr resonance, it undergoes a transient phase from a quasi-periodic to periodic orbit and back that can last up to dozens of cycles \cite{Ruangsri:2013hra,Berry:2016bit,Speri:2021psr,Gupta:2022fbe}. If the geodesics are non-integrable, the resonant islands that surround stable periodic points become regions of shared periodicity for all geodesics that occupy the island. This inevitably leads to resonances that last much longer \cite{Apostolatos:2009vu,Lukes-Gerakopoulos:2010ipp,Destounis:2021mqv,Destounis:2021rko} ($\sim$ hundreds of orbits of purely resonant motion) and designates the existence of non-integrability and indirect chaos. 

Astrophysical chaos has proven to be a prominent effect in non-Kerr EMRIs, that can significantly affect the GW frequency evolution when a resonant island is crossed. The studies conducted so far have employed theory-agnostic primaries that are tailored to break integrability \cite{Destounis:2020kss,Eleni:2024fgs,Chen:2022znf}, pathological solutions of GR \cite{Contopoulos:2011dz,Lukes-Gerakopoulos:2021ybx,Destounis:2023gpw}, exotic compact object primaries \cite{Destounis:2023khj,Chen:2023gwm}, and even EMRI analogs \cite{Mukherjee:2022dju,Eleni:2023mjx}. Even though the above classes of EMRIs discussed might lead to chaotic phenomena both at the orbital and GW level, they are either exotic, arise from modifications of gravity or include secondary effects that modify the geodesic equations, such as the secondary spin \cite{Zelenka:2019nyp} and other relativistic effects \cite{DeFalco:2020yys,DeFalco:2021uak}. 

In this work, we embark into a tentative geodesic treatment of general-relativistic solutions of the field equations, describing rotating BHs surrounded by a Hernquist-type dark matter distribution \cite{Fernandes:2025osu}, in an attempt to examine if there are imprints of non-integrability when, relativistically-constructed, rotating BHs in matter environments are present\footnote{Very early attempts for unraveling chaotic phenomena in Schwarzschild BHs \emph{perturbed} with matter halos, that are introduced in the form of dipolar, quadrupolar and octupolar perturbations to background have been successful, though the contribution of the matter halos was added by hand to vacuum GR backgrounds and introduced various pathologies \cite{Vieira:1996zf,deMoura:1999wf}.}. This analysis results to a \emph{novel class of chaos that is environmentally-driven and purely relativistic}. Thus here, we lay the groundwork for future EMRI studies, that will include radiation reaction, and provide powerful insights into how realistic astrophysical environments influence spacetime symmetries.

\section{Black holes in astrophysical environments}

In what follows, we review the first exact, fully-relativistic, solution to the field equations that describes a static and spherically-symmetric BH in the center of a matter halo environment\footnote{Of course, other general-relativistic solutions exist but they do not include a halo of matter but rather a thin disk around the BH, see e.g. \cite{Will:1974,Lemos:1993qp,Kotlarik:2022spo}.} \cite{Cardoso:2021wlq}. Afterwards, we revisit the generalization of this exact solution to a general-relativistic and fully-numerical spacetime that describes a stationary and axially-symmetric rotating BH surrounded by a matter environment \cite{Fernandes:2025osu}.

\subsection{Exact, static and spherically-symmetric BHs in matter halos}

The first, exact solution of Einstein's equations that describes a static and spherically-symmetric BH placed in the core of a dark matter halo was obtained in \cite{Cardoso:2021wlq}. The basic principle of the construction of the environment is the assumption of many gravitating masses, that follow all possible spherical geodesics around a central point where the BH exists, thus building an Einstein cluster \cite{Einstein:1939ms,hoganReconstructionMinkowskianSpacetime1978,1968ApJ...153L.163Z,1974RSPSA.337..529F,Comer:1993rx,kumardattaNonstaticSphericallySymmetric1970,bondiDattasSphericallySymmetric1971,Gair:2001qu,Szybka:2018hoe,Mahajan:2007vw,Magli:1997qf,Acharyya:2023rnq,Boehmer:2007az,Lake:2006pp,Geralico:2012jt,Jusufi:2022jxu,Fernandes:2025lon}. By integrating all geodesics over a finite radius range leads to introducing an anisotropic fluid with zero radial pressure (since the Einstein cluster assumes geodesics) and non-zero tangential pressure $P_t$ that drives the fluid, such that
\begin{equation}
    T^\mu_\nu=\text{diag}(-\varepsilon,0,P_t,P_t),
\end{equation}
where $\varepsilon$ is the energy density of the matter profile. Even though in \cite{Cardoso:2021wlq} a density profile was used that describes dark matter halos around galaxies \cite{King:1962wi,Jaffe,Navarro:1995iw,Zhao:1995cp}, namely the Hernquist density profile \cite{Hernquist:1990be}
\begin{equation}\label{Hernquist}
	\varepsilon=\frac{M_{\rm halo}\, a_0}{2\pi r (r+a_0)^3},
\end{equation}
where $M_{\rm halo}$ is the mass of the halo, $a_0$ its length scale and $M_{\rm halo}/a_0$ defines the halo compactness, there have been other configurations that have been constructed in an equivalent manner though they do not lead to an exact, but rather a numerical solution. The assumption of spherical symmetry, together with a Hernquist-inspired matter distribution
\begin{equation}
	m(r)=M_\text{BH}+\frac{M_{\rm halo}\, r^2}{(a_0+r)^2}\left(1-\frac{2M_\text{BH}}{r}\right)^2,
\end{equation}
with $M_\text{BH}$ the mass of the primary central BH, leads to the geometry\footnote{Recently, it was shown in Ref. \cite{Fernandes:2025lon} that this, and other geometries following from the Einstein cluster construction, are also solutions to a particular vector-tensor theory, where the vector encodes the astrophysical environment.}
\begin{equation}\label{metric}
	ds^2=-f(r)dt^2+\frac{dr^2}{1-2m(r)/r}+r^2 d\Omega^2,
\end{equation}
with 
\begin{align}
	f(r)&=\left(1-\frac{2 M_\textrm{BH}}{r}\right)e^\Upsilon,\\\nonumber
	\Upsilon&=-\pi\sqrt{M_{\rm halo}/\xi}\\&+2\sqrt{M_{\rm halo}/\xi}\arctan\left[\frac{r+a_0-M_{\rm halo}}{\sqrt{M_{\rm halo}\,\xi}}\right],\\
	\xi&=2a_0-M_{\rm halo}+4 M_\textrm{BH}.
\end{align}
At small scales, Eq. \eqref{metric} describes a BH of mass $M_\text{BH}$, while at large scales the Newtonian potential corresponds to that of the Hernquist profile \eqref{Hernquist}, dominated by the halo's mass $M_{\rm halo}$. The spacetime consists of a BH event horizon at $r=r_h=2 M_\text{BH}$, a curvature singularity at $r=0$, while the configuration has an Arnowitt-Deser-Misner (ADM) mass equal to $M_\textrm{ADM}=M_{\rm halo}+M_\text{BH}$. For galactic dark matter halos, the inequality $M_\text{BH}\ll M_{\rm halo}\ll a_0$ must hold while the compactness should satisfy the inequality $M_{\rm halo}/a_0\lesssim 10^{-4}$ \cite{Navarro:1995iw}, in order to describe the dark matter halo of a galaxy. Even so, in the context of BH environments, the compactness, in principle, can be treated as a parameter of spacetime as long as $M_{\rm halo}< 2(a_0+2M_\text{BH})$, in order to avoid spacetime ambiguities and pathologies.

\subsection{Including spin: general-relativistic, rotating BHs in matter halos}

Astrophysical BHs are generally expected to rotate and to reside in galaxies that themselves possess intrinsic angular momentum. Incorporating rotation into the models discussed in the previous section is therefore crucial for a complete understanding of how environments influence astrophysical BHs. However, the absence of spherical symmetry appears to rule out the possibility of obtaining closed-form analytic solutions.

Using numerical methods, Ref. \cite{Fernandes:2025osu} has recently obtained stationary and axisymmetric solutions to the Einstein equations, describing rotating BHs immersed in an astrophysical environment. In the following, we describe briefly the formalism and the numerical method used in Ref. \cite{Fernandes:2025osu}, to which the reader is referred to for a more detailed discussion.

We consider solutions to the Einstein field equations, $G_{\mu \nu} = 8\pi T_{\mu \nu}$, sourced by an anisotropic fluid, with stress-energy tensor
\begin{equation}
    \begin{aligned}
        T_{\mu \nu} =& \left( \varepsilon + p_{1} \right) u_\mu u_\nu + p_{1} g_{\mu \nu} \\&+ (p_r-p_{1}) k_\mu k_\nu + (p_{2} - p_{1})s_\mu s_\nu,
    \end{aligned}
    \label{eq:T}
\end{equation}
where $\varepsilon$, $p_r$, $p_1$ and $p_2$ are the energy density, radial pressure and transverse pressures in the co-moving frame of the fluid, respectively. The vectors $u^\mu$, $k^\mu$ and $s^\mu$ are, respectively, the four-velocity of the fluid and spacelike vectors that define the directions of anisotropy such that $u^\mu u_\mu = -1$, $k^\mu k_\mu = 1$, $s^\mu s_\mu =1$, $u^\mu k_\mu = 0$, $u^\mu s_\mu = 0$, and $k^\mu s_\mu = 0$.

We use a metric that is stationary, axially-symmetric and circular, in quasi-isotropic coordinates, described by four functions, $f$, $g$, $h$, $\omega$, of $r$ and $\theta$, namely
\begin{equation}
    \begin{aligned}
        ds^2 = -f \frac{N_-^2}{N_+^2} dt^2  + & \frac{g}{f}N_+^4 \bigg[ h\left(dr^2 + r^2 d\theta^2 \right)\\&
        +r^2 \sin^2\theta \left(d\varphi - \omega dt \right)^2\bigg],
    \end{aligned}
    \label{eq:metric}
\end{equation}
where $N_\pm = (1\pm r_h/r)$. In these coordinates, the four-velocity of the fluid is
\begin{equation}
    u^\mu = \frac{1}{\sqrt{-(g_{tt} + 2 \Omega g_{t\varphi} + \Omega^2 g_{\varphi \varphi})}} (1,0,0,\Omega),
\end{equation}
where $\Omega\equiv \Omega(r,\theta)$ is the angular velocity of the fluid. Following Ref. \cite{Fernandes:2025osu} we focus on the simplest case, where $\Omega=\omega$, such that the angular velocity of the dark matter halo is simply the frame-dragging angular velocity.

The total mass of the spacetime obeys $M_{\rm ADM} = M_{\rm BH} + M_{\rm halo}$, where $M_{\rm BH} = \frac{\kappa}{4\pi} A_h + 2\Omega_h J_h$, is the mass of the BH\footnote{The quantities $\kappa$, $A_h$, $\Omega_h$ and $J_h$ are, respectively, the surface gravity, horizon area of the BH, angular velocity of the horizon, and the angular momentum of the BH.}, and $M_{\rm halo}$ is the mass of the halo. The total mass $M_{\rm ADM}$ and angular momentum $J_h$\footnote{Since $\Omega=\omega$, the total angular momentum is equal to the angular momentum of the BH.} can also be computed from the asymptotic decay of the metric components $g_{tt} = -1 + 2M_{\rm ADM}/r + \mathcal{O}(r^{-2})$, and $g_{t\varphi} = - 2J_h\sin^2\theta/r  + \mathcal{O}(r^{-2})$, while the surface gravity and the area of the event horizon for the metric \eqref{eq:metric} are computed as
\begin{equation}
    \kappa = \frac{f}{8r_h \sqrt{g h}}, \quad A_h = 32\pi r_h^2 \int_{0}^\pi \frac{g\sqrt{h}}{f} \sin \theta \dd \theta,
    \label{eq:kappaAH}
\end{equation}
at $r=r_h$.

To solve the field equations and obtain stationary, axially-symmetric BH solutions, we use the six independent combinations of equations that follow from the Einstein equations (see Ref. \cite{Fernandes:2025osu}). The system, however, contains eight undetermined functions of $r$ and $\theta$, the four metric functions and the four eigenvalues of the stress-energy tensor. The field equations, can only determine six of these functions, that we choose as the four metric functions and the two tangential pressures $p_1$ and $p_2$. To close the system we choose, from physical considerations, a profile for the energy density $\varepsilon$ of the dark matter halo, and provide an equation of state for $p_r$, namely, $p_r = 0$, in accordance with the Einstein cluster. Motivated by the Hernquist profile, and by Ref. \cite{Cardoso:2021wlq}, we use the following profile for the energy density
\begin{equation}\label{eq:edensity_supp}
    \varepsilon = \frac{\mathcal{M} \left( a_0 + r_h \right)}{2\pi r (r+a_0)^3} \left(1 - \frac{r_h}{r} \right)^2 b^{-5},
\end{equation}
where $\mathcal{M}$ and $a_0$ are parameters with length dimensionality, and $b\equiv b(r)$ is a complicated function of $r$ presented in Ref. \cite{Fernandes:2025osu}. The parameter $a_0$ represents a typical length scale associated with the halo, i.e. the limiting radius that includes $95\%$ of the dark matter content of Eq. \eqref{eq:edensity_supp} beyond which its mass falls of as $r^{-4}$. Finally, the parameter $\mathcal{M}$ represents, to leading order in an expansion of powers of $\mathcal{M}/a_0$, the mass of the halo $M_{\rm halo}$, see Ref. \cite{Fernandes:2025osu}. The function $b(r)$ guarantees that the parameters $\mathcal{M}$ and $a_0$ have suitable physical meaning. Importantly, $b>0$, and therefore $\varepsilon \geq 0$. In the large $r$ limit, this profile for $\varepsilon$ agrees with the Hernquist profile, up to terms proportional to $r_h$.

The functions obey the following boundary conditions. At the horizon ($r=r_h$) we have $\partial_r f = \partial_r g = \partial_r h = \partial_r p_1 = \partial_r p_2 = 0, \quad \omega = \Omega_h$, while asymptotically they obey $f=g=h=1, \quad \omega = p_1 = p_2 =0$. Regularity, axial symmetry and parity considerations imply $\partial_\theta f = \partial_\theta g = \partial_\theta h = \partial_\theta \omega = \partial_\theta p_1 = \partial_\theta p_2 = 0$, at $\theta=0$ and $\theta=\pi/2$.

The input parameters are those describing the halo, i.e., $a_0$ and $\mathcal{M}$, and the ones associated with the spinning BH, namely $r=r_h$, $\omega=\Omega_h$, that altogether fully describe the spacetime. Once a solution is obtained with the numerical method, where we keep 20 digits of precision in every calculation, the ADM mass, BH mass, total angular momentum and mass of the halo can be computed with the expressions provided in the previous discussion. We find that there is a one-to-one correspondence between these quantities and the input parameters; solutions can be fully specified either by $(a_0,\mathcal{M},r_h,\Omega_h)$ or by the quantities $(a_0, M_{\rm halo}, M_{\rm BH}, J_h)$. In the following, we choose to present our results in terms of dimensionless values constructed from the second set of quantities, namely the normalized halo compactness $a_0/M_{\rm halo}$, the ratio between the halo and BH mass $M_{\rm halo}/M_{\rm BH}$, and the dimensionless spin parameter $J_h/M_{\rm BH}^2$. These values are presented with an approximate sign ($\simeq$), in what will follow, due to the numerical nature of the solutions. Nonetheless, we present in Table \ref{tab:initialconditions} the precise values of $(a_0,\mathcal{M},r_h,\Omega_h)$ used as initial conditions for each solution used throughout this work, as to ensure reproducibility of our results.

To solve the system of partial differential equations, we use the same code as in Ref. \cite{Fernandes:2025osu}, which is publicly available at \cite{HighPrecisionSpinningBHs}. This code is a high-precision version of the code presented in Ref. \cite{Fernandes:2022gde}, originally developed by one of the authors. This code combines a pseudospectral method with the Newton-Raphson root-finding algorithm. Details on the code, its validation and accuracy are provided in the appendix of Ref. \cite{Fernandes:2025osu}. We observe exponential convergence of the code as we increased its resolution, and only accepted a solution when the absolute estimated error was below $\mathcal{O}\left( 10^{-8}\right)$.

\begin{table}[]
    \centering
    \begin{tabular}{|c||c|c|c|c|c|c|}
        \hline
        $J/M_{\rm BH}^2$ & $r_h$ & $\Omega_h$ & $\mathcal{M}$ & $a_0$ \\
        \hline
        0.17514520 & 0.49649779 & 0.030211615 & 9.9998521 & 99.998521 \\
        \hline
        0.30021574 & 0.48609324 & 0.052252992 & 9.9992814 & 99.992814 \\
        \hline
        0.44286783 & 0.46705390 & 0.078376538 & 9.9976902 & 99.976902 \\
        \hline
        0.63465392 & 0.42787085 & 0.11637563 & 9.9924688 & 100.00646 \\
        \hline
        0.79463800 & 0.34784164 & 0.19207557 & 9.9936337 & 201.33424 \\
        \hline
        0.80351794 & 0.36333864 & 0.17138511 & 10.005515 & 130.07169 \\
        \hline
        0.80479191 & 0.37707168 & 0.15483751 & 9.9837236 & 100.05776 \\
        \hline
        0.99565384 & 0.29486837 & 0.20766210 & 9.9991003 & 99.991003 \\
        \hline
    \end{tabular}
    \caption{Initial conditions used in the code to obtain all solutions used throughout this work. In all cases, $M_{\rm BH}=1$.}
    \label{tab:initialconditions}
\end{table}

\section{Geodesic motion}

\subsection{Equations of motion and conserved quantities}

We operate on a generic metric, i.e. a stationary and axisymmetric spacetime written as
\begin{equation}\label{line_element}	ds^2=g_{tt}dt^2+2g_{t\varphi}dtd\varphi+g_{rr}dr^2+g_{\theta\theta}d\theta^2+g_{\varphi\varphi} d\varphi^2,
\end{equation}
where the metric tensor components in \eqref{line_element} are functions of $r$ and $\theta$, and the coordinate system $\left(t,r,\theta,\varphi\right)$ can be chosen to be of Boyer-Lindquist \cite{Boyer:1966qh} or quasi-isotropic type \cite{Misner:1973prb}. A zero-order assumption for the evolution of an EMRI is to think of the motion of the secondary as a test-particle following the geodesics of the spacetime geometry of the primary. Under this simplistic assumption, which serves as a good proxy of a short-timescale EMRI, that corresponds to some dozens of revolutions around the primary, can be described by the geodesic equations
\begin{equation}\label{geodesic_equation}
\ddot{x}^\kappa+\Gamma^\kappa_{\lambda\nu}\dot{x}^\lambda\dot{x}^\nu=0,
\end{equation}
where $\Gamma^\kappa_{\lambda\nu}$ are the Christoffel symbols of the primary, $x^\kappa$ is the four-position of the secondary, and an overdot denotes differentiation with respect to proper time $\tau$. The system of second-order, coupled differential equations resulting from Eq. \eqref{geodesic_equation} can be considerably simplified using constants of motion. The spacetime metric \eqref{line_element} bears two Killing vector fields resulting from stationarity and axisymmetry, giving rise to two constants of motion throughout geodesic motion, i.e. the specific energy $E$ and azimuthal angular momentum $L_z$ of the test particle of mass $\mu$
\begin{equation}\label{energy_momentum}
-E/\mu=g_{tt}\dot{t}+g_{t\varphi}\dot{\varphi},\,\,\,\,\,\,\,\,\,
L_z/\mu=g_{t\varphi}\dot{t}+g_{\varphi\varphi}\dot{\varphi}.
\end{equation}
Equations \eqref{energy_momentum} can be written as two, first-order, decoupled, ordinary differential equations for the $t$ and $\varphi$ momenta, so that
\begin{equation}\label{tphi_dot}
\dot{t}=\frac{E g_{\varphi\varphi}+L_zg_{t\varphi}}{\mu\left(g^2_{t\varphi}-g_{tt}g_{\varphi\varphi}\right)},\,\,\,\,\,\,\,\,\,
\dot{\varphi}=\frac{E g_{t\varphi}+L_zg_{tt}}{\mu\left(g_{tt}g_{\varphi\varphi}-g^2_{t\varphi}\right)}.
\end{equation}
There are two remaining geodesics for $r$ and $\theta$ which, in general, form a second-order dynamical system of coupled differential equations. Nevertheless, test particles in geodesic motion provide a third constant of motion, i.e., the conservation of their rest mass, or equivalently their four-velocity $g_{\lambda\nu}\dot{x}^\lambda\dot{x}^\nu=-1$. This leads to a constraint equation
\begin{equation}\label{constraint_equation}
\dot{r}^2+\frac{g_{\theta\theta}}{g_{rr}}\dot{\theta}^2+V_\text{eff}=0,
\end{equation}
with $V_\text{eff}$ a potential of the form
\begin{equation}
V_\text{eff}\equiv\frac{1}{g_{rr}}\left(1+\frac{g_{\varphi\varphi} E^2+g_{tt}L_z^2+2g_{t\varphi}E L_z}{\mu^2\left(g_{tt}g_{\varphi\varphi}-g_{t\varphi}^2\right)}\right).
\end{equation}
Equation \eqref{constraint_equation} characterizes bound geodesic motion, through the potential $V_\text{eff}$. In turn, when $V_\text{eff}=0$ a curve appears (under a choice of $\theta\in[0,\pi]$); the curve of zero velocity (CZV), where the $r$ and $\theta$ momenta are bound to be null there, i.e., $\dot{r}=\dot{\theta}=0$. 

Under the hypothesis of another constant of motion, the geodesics for all degrees of freedom can be separated into decoupled, first-order, differential equations. If though the primary spacetime does not allow for further spacetime symmetries then the $r$ and $\theta$ motion remains coupled, and of second differential order. A Kerr BH, though, possesses a rank-two Killing tensor field, that associates the remaining components of the angular momentum with the Carter constant \cite{Carter:1968rr}. This successfully decouples the motion of $r(\tau)$ and $\theta(\tau)$. Since the primary configurations we will use here do not necessarily have a Carter constant (or any constant related to a higher-rank Killing tensor field), we will work under the assumption of the absence of a fourth constant of motion, and the phase space of orbits will inform us if a separation constant exists in our numerical spacetime. To evolve orbits we will utilize the coupled second-order differential equation system for $r$ and $\theta$, together with Eqs. \eqref{tphi_dot} and \eqref{constraint_equation}, without any further symmetry assumptions.

\subsection{Evolving geodesics on numerical backgrounds}

Adopting quasi-isotropic coordinates, the line element \eqref{line_element} can be described through four independent functions $(f,g,h,\omega)$ of $(r,\theta)$, where the metric tensor components result directly from Eq. \eqref{eq:metric}, i.e.,
\begin{subequations}\label{metric_tensor_components}
\begin{align}
\nonumber \\[-10pt]
g_{tt}(r,\theta)&=-\frac{(1-r_h/r)^2 f(r,\theta)}{(1+r_h/r)^2}\nonumber\\&+\frac{(1+r_h/r)^4 g(r,\theta) \sin^2\theta\, \omega^2(r,\theta)}{r^2f(r,\theta)},  \\[10pt]
g_{rr}(r,\theta)&=\frac{(1+r_h/r)^4 g(r,\theta)\,h(r,\theta)}{f(r,\theta)},\\[10pt]
g_{\theta\theta}(r,\theta)&= r^2 g_{rr}(r,\theta),\\[10pt]
g_{\varphi\varphi}(r,\theta)&=\frac{r^2(1+r_h/r)^4 g(r,\theta)\sin^2\theta}{f(r,\theta)},\\[10pt]
g_{t\varphi}(r,\theta)&=-\frac{(1+r_h/r)^4 g(r,\theta)\sin^2\theta\,\omega(r,\theta)}{f(r,\theta)}.
\end{align}
\end{subequations}
The solutions have been found numerically following the procedure described in the previous section. The details of the numerical implementation are provided also in \cite{Fernandes:2022gde,Fernandes:2025osu}. The geodesic evolution of massive particles around rotating BHs in a halo of Hernquist-type matter, uses the numerically-constructed metric tensor components \eqref{metric_tensor_components}, as in \cite{Destounis:2023khj}, and geodesics are evolved with the use of high-order interpolation functions that are constructed from the data of the metric tensor. For all simulations performed herein, we find that the constraint equation \eqref{constraint_equation} is satisfied to within one part in $10^8$ for the first $\sim20$ thousand revolutions (and $\sim20$ thousand intersections through the equatorial plane). In all cases we fix the mass ratio of the EMRI to $\mu/M_\textrm{BH}=10^{-6}$ (even though at the geodesic level it just leads to a change of scales) and choose the constant energy and azimuthal angular momentum as $E/\mu=0.9$ and $L_z/\mu=3M_\textrm{BH}$, respectively. This choice stems from the fact that the most dominant resonances are located in the strong-field regime, and the aforementioned values for $E$ and $L_z$ gives us access to these resonances. Furthermore, due to the effect of circularization of orbits that takes place during GW emission, the eccentricity should be significantly decreased, with respect to the initial eccentricity at the astrophysical capture of the secondary, that can be very high. The particular choice of $E$ and $L_z$ facilitate eccentricities approximately smaller than $0.4$, thus are good quantities to initialize a somewhat more realistic inspiral. Nevertheless, there is a large choice of parameter space values where small eccentricities can be achieved, thus our choice is neither fine-tuned, nor random.

\subsection{Non-integrability and signatures of chaos}\label{subsec:chaos_theory}
 
The geodesics of massive test-particles around Kerr BHs possess four constants of motion, namely the energy, the azimuthal angular momentum, the Carter constant and the four-velocity of the particle. Thus, the system of equations is (Liouville-)integrable and does not present chaotic features \cite{Contopoulos_book}. Since the Carter symmetry is very fragile it may not exist when the multipolar structure of spacetime has been significantly deformed (see however Ref. \cite{Kocherlakota:2025cwq}). The absence of a Carter symmetry, or in general any other (higher-rank) Killing tensor field, leads to the non-integrability of geodesics; the radial and polar sectors of the geodesics are not separable, but rather form a second-order, coupled dynamical system. Some cases, where the Carter constant ceases to exist, have been discussed in the Introduction. The absence of a fourth constant of motion inevitably leads to non-integrability, which allows for chaotic effects \cite{Contopoulos_book}. Full-blown chaos is extremely rare, if not completely absent, at least in astrophysical EMRIs. The underlying effects of non-integrability, though, can manifest themselves indirectly in the phase space of orbits and in particular close to orbital resonances. Transient orbital resonances already affect EMRI evolution, even in Kerr, where geodesics are integrable \cite{Berry:2016bit,Speri:2021psr,Lynch:2024ohd,Levati:2025ybi}, and introduce significant dephasing effects. Non-integrable EMRIs are very likely to amplify these dephasing effects and can introduce even stronger, clear-cut phenomenology of indirect chaos \cite{Apostolatos:2009vu,Lukes-Gerakopoulos:2010ipp,Destounis:2021mqv}.

In integrable systems, geodesics are orbits that occupy the surface of a two-dimensional torus in the available four-dimensional phase space. The torus of each geodesic is called an \emph{invariant torus} and is described by the minimum and maximum radii of the orbit, associated with the eccentricity, and by the minimum and maximum polar angles, associated with the inclination. Together with the azimuthal orbital period, any trajectory can be described by three fundamental frequencies; the frequency of oscillation from the periapsis (minimum radius) to the apoapsis (maximum radius) and back, $\omega_r$, the oscillation frequency through the equatorial plane, between the maximum and minimum polar angle, $\omega_\theta$, and the revolution frequency $\omega_\varphi$.

If one considers a two-dimensional surface that cuts through a foliage of invariant tori, named \emph{Poincar\'e surface of section}, then each torus defines a closed curve on this surface \cite{lichtenberg2013regular}. This curve is called an \emph{invariant curve}. Each torus corresponds to a characteristic pair of frequencies, as discussed above. Not only the two frequencies, but also the ratio between them varies continuously over successive torii. If the ratio of the frequencies is an irrational number, an orbit continuously winds around its corresponding torus covering its surface densely. This kind of orbit is called \emph{quasi-periodic}, or \emph{generic}. A generic orbit goes repeatedly through a surface of section defining a succession of points which eventually cover densely the corresponding invariant curve on the surface of section. In the case where the ratio of frequencies is a rational number $n/m$, where $n, m \in \mathbb{N}$, the orbit repeats itself after $m$ windings. The orbit is \emph{periodic} and the corresponding torus is called \emph{resonant}. In this case, the resonant invariant curve consists of an infinite number of $m$-multiplets of periodic points. Each $m$-multiplet represents an $m$-multiple periodic orbit.

When non-integrable perturbations are presented in an integrable system, two theorems, namely the Kolmogorov-Arnold-Moser (KAM) \cite{Moser:430015,Arnold_1963} and Poincar\'e-Birkhoff~\cite{Birkhoff:1913} theorems, dictate the modification of orbital dynamics. The KAM theorem ensures that orbits sufficiently away from resonances, are smoothly shifted but, in general, most of the tori are deformed but not destroyed. These tori are called \emph{KAM tori}. Thus, the corresponding surface of section looks very much like the surface of section of the corresponding integrable system. The generic invariant curves in non-integrable systems are called \emph{KAM curves}. Successive KAM curves organize around a common central point, which corresponds to a spherical orbit (zero eccentricity).

Close to resonances, the KAM curves disintegrate into two sets of periodic points, in accord with the Poincar\'e-Birkhoff theorem. The stable periodic points, from which stable resonant orbits emanate, are encapsulated by KAM curves that form a nested island, i.e., a \emph{resonant island}. A phase-orbit of such a resonant case visits all the $m$ islands of the $n/m$-resonance, moving successively to the next $n$-th island along the aforementioned closed curve at every winding, forming eventually the KAM curves inside every island. On the other hand, the unstable periodic points generate chaotic orbits that surround the resonant islands with thin layers. The structure around resonances of non-integrable systems is called a \emph{Birkhoff chain}. The crucial aspect of resonant islands, and their ultimate significance in EMRI dynamics, is the fact that the rational ratio of the central, stable periodic point $\omega_r/\omega_\theta=n/m$ is shared throughout all KAM curves residing inside the island, no matter on which KAM curve inside the chain of islands it belongs. This property is not shared by the non-resonant KAM curves, since the ratio of frequencies on them is irrational and it varies smoothly from one KAM curve to another.

In a sense, integrable EMRIs experience resonances that occupy a ``zero-volume point'' in phase space, while non-integrable EMRIs exhibit prolonged ``non-zero-volume'' resonances where the secondary is locked in perfect resonance for a significant amount of revolutions around the primary. For example, in non-Kerr EMRIs, these prolonged resonances may last up to $\sim200-300$ cycles depending on the primary \cite{Destounis:2021mqv,Destounis:2021rko}, without taking into account pre- and post-resonant effects \cite{Speri:2021psr,Berry:2016bit,Ruangsri:2013hra} or the conservative part of gravitational self-force \cite{Barack:2009ux}.

\begin{figure*}
    \centering
    \includegraphics[width=\textwidth]{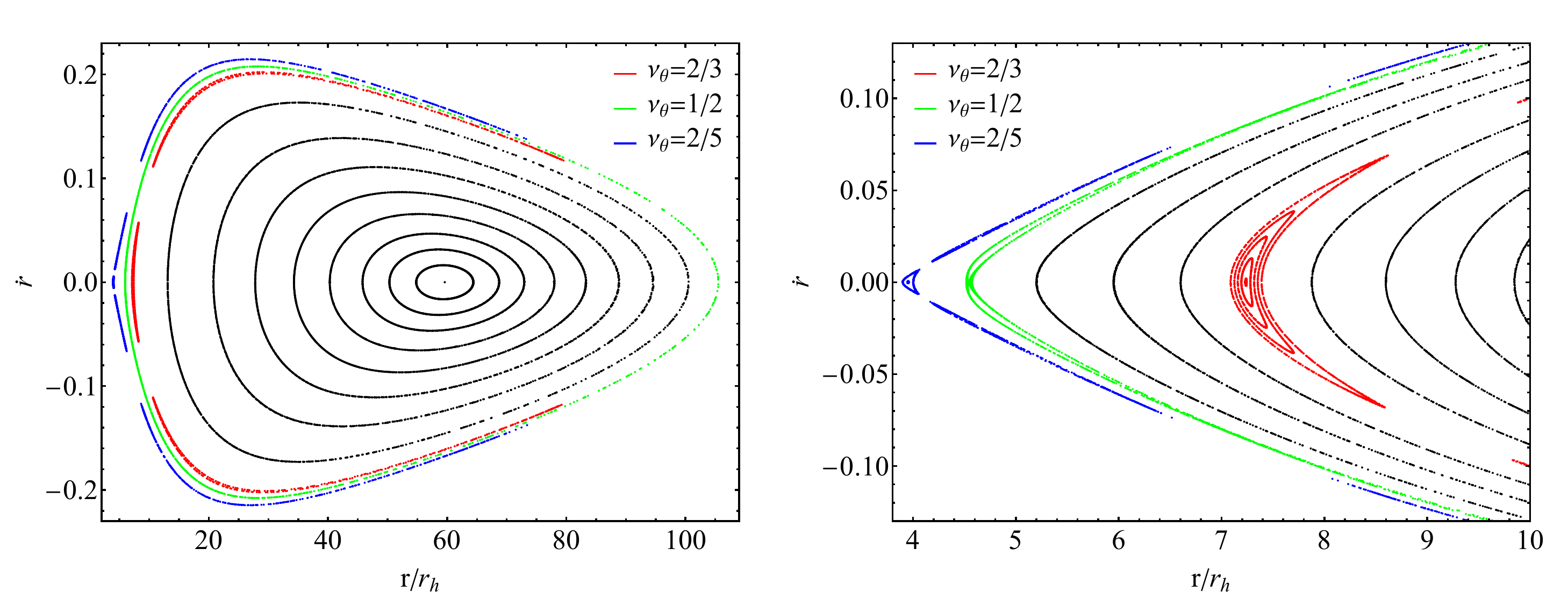}
    \caption{\emph{Left:} Equatorial Poincar\'e map of generic geodesics with fixed $E/\mu=0.9$ and $L_z/\mu=3M_{\rm BH}$ and varying initial position $r(0)/r_h$, where $r=r_h$ is the radius of the event horizon in quasi-isotropic coordinates. The rotating BH embedded in the Hernquist-type environment, where the geodesics are evolved, has spin $J/M_{\textrm{BH}}^2\simeq0.996$, 
    halo compactness $M_\textrm{halo}/a_0\simeq10^{-1}$, and halo mass $M_\textrm{halo}\simeq 10 M_\textrm{BH}$. The black curves formed around the central fixed point of the map are KAM curves that arise from orbits with irrational ratios $\omega_r/\omega_\theta$. On the other hand, the blue and red curves designate resonant islands, with rational frequency ratios $\omega_r/\omega_\theta=2/5, \,2/3$, respectively. These islands encapsulate their corresponding stable periodic points. In turn, the green formation that appears in the map is a chaotic zone, associated with the thin chaotic layer that encircle the resonant islands with rational ratio $\omega_r/\omega_\theta=1/2$ (not shown here). \emph{Right:} Same as left, zoomed-in region at the strong-field regime. The red and blue islands depicted surround the stable periodic points of the resonance $\omega_r/\omega_\theta=2/3$ and $2/5$, respectively, while the green points form a chaotic layer around the $\omega_r/\omega_\theta=1/2$ resonant islands (not shown here).}
    \label{fig:Poincare_map}
\end{figure*}
	
The existence of resonant islands serves as an indirect proof of non-integrability in EMRIs, therefore sketching a Poincar\'e map can inform us regarding the system's symmetries. Furthermore, through the Poincar\'e map, the rotation number can be calculated in order to scan for resonant islands. It practically tracks the angle $\vartheta$ between successive intersections on KAM curves, relative to the fixed central point of the Poincar\'e map. The \emph{rotation number} is defined as the summation of $\mathcal{N}$ angles $\vartheta$ measured between successive intersections of each geodesic in the surface of section, i.e., 
\begin{equation}\label{rotation}
\nu_\vartheta=\frac{1}{2 \pi \mathcal{N}}\sum_{i=1}^{\mathcal{N}}\vartheta_i.
\end{equation}
When $\mathcal{N}\rightarrow\infty$, Eq. \eqref{rotation} asymptotes to the ratio $\nu_\vartheta=\omega_r/\omega_\theta$, thus the rotation number can serve as a resonance browser. Consecutive rotation numbers found from different initial conditions of geodesics, by smoothly varying one of the system's parameters while keeping the rest fixed, forms a \emph{rotation curve}.
	
Non-integrable systems display discontinuities in the monotonicity of a rotation curve through the formation of plateaus with non-zero widths, when geodesics transverse resonant islands. Therefore, the width of the plateau corresponds to the width of the resonant island and is a general measure of the timescales involved in resonant islands of non-integrable EMRIs. Inflection points can also appear in the rotation curve when trajectories pass through unstable periodic points. In what follows, we will examine the characteristic features of rotating BHs in halos of matter with geodesic evolutions, in order to search for signatures of non-integrability driven by astrophysical environments. We note that Poincar\'e maps and rotation curves are not the only diagnostics of chaotic phenomena. For more diagnostics that can be applied in curved spacetimes see \cite{Lukes-Gerakopoulos:2013qva}.

\section{Chaos in relativistic, rotating, non-vacuum black holes}

In what follows, we will use relativistic, rotating BHs in a Hernquist-type halo, with a variety of spin parameters and halo compactness, while fixing the central BH mass to unity, since all our numerical evolutions lie at the geodesic level, where all observables can be appropriately rescaled for any BH mass to simulate an EMRI, thus $M_\textrm{BH}=1$ is merely chosen for simplicity. In turn, the test particle that will be used has the fixed constants of motion $E/\mu=0.9$, $L_z/\mu=3 M_\textrm{BH}$. We assume a test-particle mass so that the mass-ratio is kept to $\mu/M_\textrm{BH}=10^{-6}$, even though this is merely a scale factor in the geodesic evolution scheme. We utilize the numerical configurations in quasi-isotropic coordinates as described in Eq. \eqref{line_element} and normalize the radial quasi-isotropic coordinate $r$ with respect to the event horizon radius $r=r_h$ of each configuration.

As a proof-of-principle, we numerically construct a rapidly-rotating BH with spin $J/M^2_{\rm BH} \simeq 0.996$ and a Hernquist-type halo with compactness $M_\textrm{halo}/a_0\simeq 10^{-1}$, $M_\textrm{halo}\simeq 10 M_{\rm BH}$. The choice of parameters is exaggerated in order to sketch a clear-cut phase-space structure of various bound geodesics that we evolved, which might show chaotic phenomenology, as discussed in Subsection \ref{subsec:chaos_theory}. The initial position of geodesics is chosen such that $r(0)$ is varying inside the CZV, while the rest of the position vector components are fixed to $\dot{r}(0)=0,\,\theta(0)=\pi/2$ and $\dot\theta(0)$ is chosen in accord with the constraint in Eq. \eqref{constraint_equation} so that the resulting orbits remain bounded inside the CZV.

\begin{figure*}
    \centering
    \includegraphics[width=\textwidth]{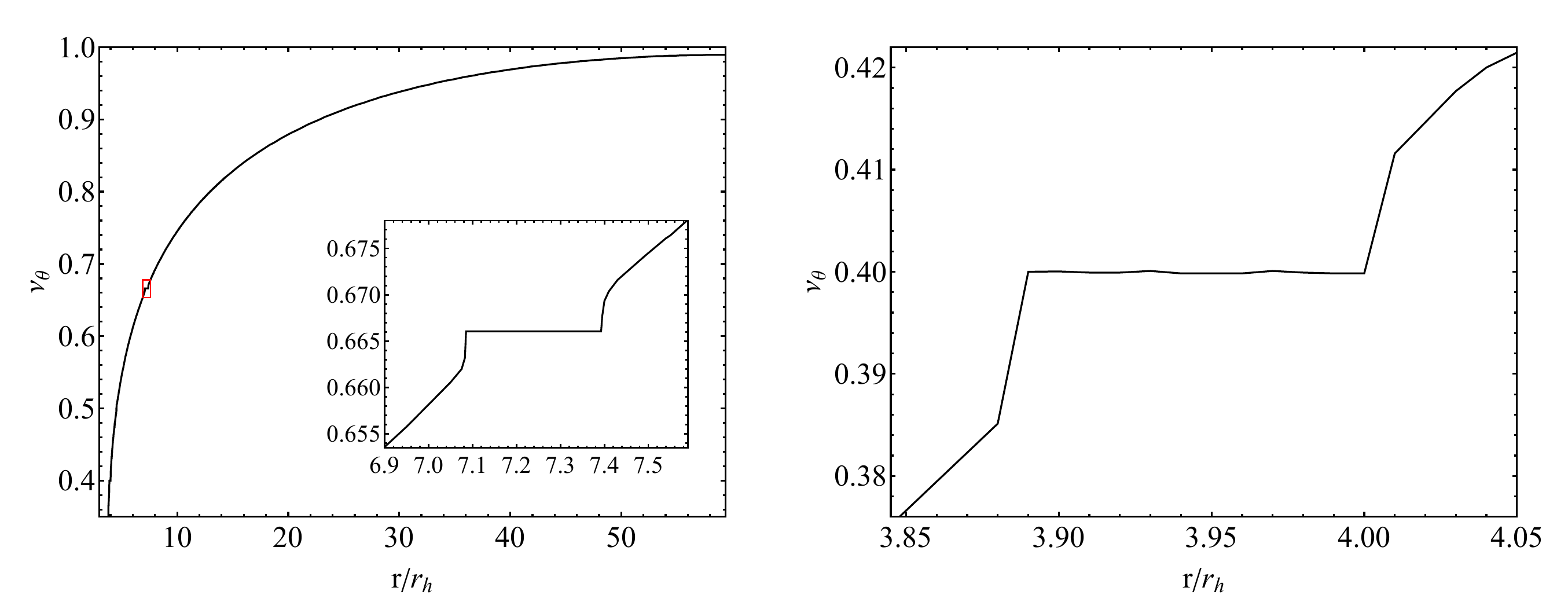}
    \caption{\emph{Left:} Rotation curve of geodesics with fixed $E/\mu=0.9$, $L_z/\mu=3M_{\textrm{BH}}$ and varying initial position $r(0)/r_h$, resulting from an equatorial surface of section of successive geodesics. The rotating BH embedded in the Hernquist-type environment, where the geodesics are evolved, has spin $J/M_{\textrm{BH}}^2\simeq0.996$, 
    halo compactness $M_\textrm{halo}/a_0\simeq10^{-1}$, and halo mass $M_\textrm{halo}\simeq 10 M_\textrm{BH}$. The red box corresponds to the zoomed-in region of a resonant island, where a plateau is formed at $\nu_\vartheta=2/3$. \emph{Right:} Same as left configuration in the phase space vicinity of the $2/5$-resonant island. A plateau is formed exactly at $\nu_\vartheta=2/5$.}
    \label{fig:Rotation_curve}
\end{figure*}

As each geodesic evolves, we dynamically capture the position and velocity $(r(t),\,\dot{r}(t))$ of the orbit, each time it crossed the equatorial plane, with $\dot{r}(t)>0$. This process defines a curve of successive intersections for each geodesic. Putting together the various curves that correspond to different geodesics we sketch the Poincar\'e map in Fig. \ref{fig:Poincare_map}. It is evident that we have managed to capture, at least, three resonances, i.e., two resonant islands namely the $\omega_r/\omega_\theta=2/3$, $2/5$, shown with red and blue, respectively, and a thin chaotic layer, that surrounds the $1/2$-resonant island, shown in green.

Figure \ref{fig:Poincare_map}, left panel, shows the structure of the map. The black curves encircle the central point of the map. These correspond to the KAM curves, away from resonances, that smoothly depart from the invariant curves of integrable systems. In fact, most non-integrable systems associated with EMRIs possess KAM curves that are similar to invariant ones, due to the fact that the perturbation inserted in the Hamiltonian is sufficiently small, in order to not completely disintegrate the whole phase space of orbits. Therefore, the structure of phase space does not entirely break down. Into the strong-field regime, though, numerous islands appear that do not encircle the central point of the Poincar\'e map, but rather encapsulate the stable periodic points of resonances. The island structure around the resonance $2/5$ and $2/3$, shown with blue and red colors, respectively, serve as imprints of chaos and non-integrability of the system under consideration. Figure \ref{fig:Poincare_map}, right panel, shows a clear, zoomed-in, depiction of the islands' nested structure. Interestingly, we do not only observe resonant islands, but we also show a chaotic layer, that surrounds densely the $1/2$-resonant islands. This occurs due to the fact that we have not crossed through the resonant island, but rather traversed within the vicinity of the chaotic zone forming around them. This discussion is very relevant when one sketches the corresponding rotation curve that results from the structure of successive intersections in the map.

In Fig. \ref{fig:Rotation_curve}, left panel, we demonstrate the rotation curve resulting from successive bound geodesics in the BH configuration of our study. Here, we used a rather dense arrangement of successive orbits to achieve a smooth curve. We observe that in most of its parts, the curve increases monotonously as $r(0)$ is increased. A more thorough look in the curve reveals the outcome of the emergence of resonant islands in the Poincar\'e surface. When $\nu_\vartheta$ encounters the resonance $2/3$, we can clearly identify a plateau formation (see red box in Fig. \ref{fig:Rotation_curve}, that corresponds to the inset in the left panel) that breaks the curve's monotonicity and effectively renders the system in study non-integrable. The inset in the left panel of Fig. \ref{fig:Rotation_curve} shows the corresponding plateau encountered when $\nu_\vartheta=2/3=\omega_r/\omega_\theta$, while the right panel shows the plateau formed due to the $\nu_\vartheta=2/5$ resonant island. We note that we do not show how the crossing through the unstable $1/2$-periodic point is imprinted in the rotation curve. This is due the chaoticity of successive intersections in the Poincar\'e map, that leads to rotation numbers that are not well-defined. The result is an erratic oscillation (or overall an inflection point) in the rotation curve (see e.g. Fig. 9, bottom left panel, in \cite{Destounis:2023khj}). The appearance of plateaus, and inflection points, instantly corroborates the Poincar\'e-Birkhoff theorem, which states that geodesics residing inside resonant islands share the same rational ratio $\omega_r/\omega_\theta$, or $\nu_\vartheta$ when the number of intersections measured ($\mathcal{N}$) is sufficiently large. In this analysis, we have recorded approximately 5 to 10 thousand intersections per geodesic, that corresponds to an accuracy of 3 to 4 digits in the resulting rotation numbers. A thorough check of geodesics in further non-vacuum, rotating BH configurations, with smaller BH spin, or smaller compactness, reveals a qualitatively similar structure in the Poincar\'e map and the rotation curve. Hence, adding rotation to the static solution of Ref. \cite{Cardoso:2021wlq}, breaks integrability and leads to chaotic phenomenology, that extends the lifetime of an non-integrable EMRI in resonance \cite{Apostolatos:2009vu,Lukes-Gerakopoulos:2010ipp,Destounis:2021mqv}.

From Fig. \ref{fig:Poincare_map} and \ref{fig:Rotation_curve}, we can conclude that the resonance $2/3$ is the strongest, i.e., having the largest island width (and corresponding plateau in the rotation curve) in non-Kerr EMRIs. This is due to the multiplicity and location of the resonance $2/3$ in the strong-field regime \cite{Contopoulos_book}. This is the underlying reason why the literature is usually studying the effects of the $2/3$ resonance in Kerr \cite{Flanagan:2010cd,Berry:2016bit,Ruangsri:2013hra,Speri:2021psr} and non-Kerr EMRIs \cite{Apostolatos:2009vu,Lukes-Gerakopoulos:2010ipp,Destounis:2021mqv,Destounis:2021rko}. 

In what follows, we present results regarding the behavior of the $2/3$-resonant island, and in particular its width, when the spacetime's BH spin and the halo compactness are varied. The following is an order-of-magnitude analysis and will help us qualitatively understand the role of the BH spin and the halo in EMRIs, which combined break the Carter constant and the eventual integrability of geodesics around these objects.

\begin{figure*}
    \centering
    \includegraphics[width=\textwidth]{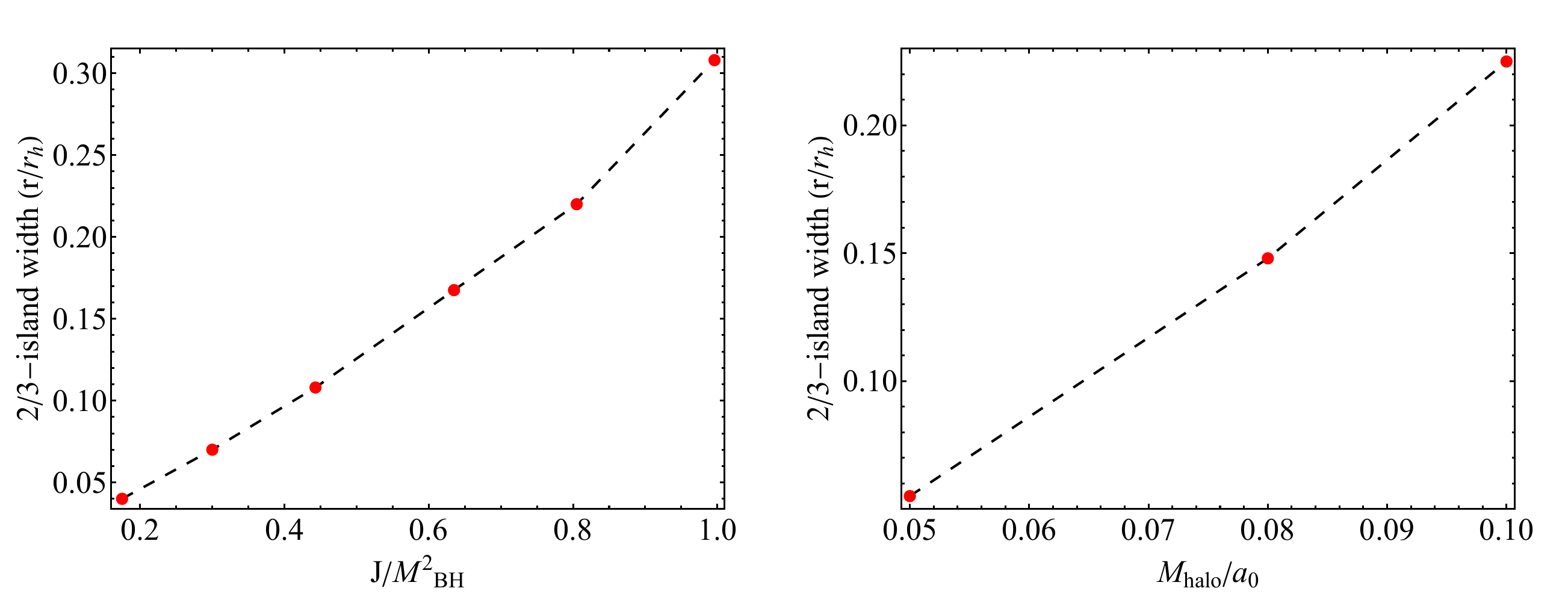}
    \caption{\emph{Left:} Width of the $2/3$-resonant islands formed in the phase space of BHs in matter halos with varying spin parameter $J/M_{\textrm{BH}}^2$ and fixed compactness $M_\textrm{halo}/a_0\simeq10^{-1}$, with $M_\textrm{halo}\simeq 10 M_\textrm{BH}$. The island widths result from geodesic evolutions around the aforementioned configurations, with $E/\mu=0.9$ and $L_z/\mu=3M_{\textrm{BH}}$. The island width has been normalized with respect to the event horizon radius $r=r_h$ of each configuration in quasi-isotropic coordinates. \emph{Right:} Same as left, but with fixed BH spin $J/M_{\textrm{BH}}^2\simeq 0.8$ and varying compactness $M_\textrm{halo}/a_0$, with $M_\textrm{halo}\simeq 10 M_\textrm{BH}$. For both panels, the red dots correspond to the nine different BH configurations analyzed, while the dashed black lines correspond to crude interpolations of these values.}
    \label{fig:island_width_vs_parameters}
\end{figure*}

\subsection{$2/3$ island-width dependence on the BH spin and halo parameters}

Besides providing a proof-of-principle for the non-integrability of general-relativistic BHs surrounded by a matter halo, it is interesting to examine how the different parameters of spacetime affect the chaotic signatures, i.e., the width of the resonant islands. For simplicity, we focus on the $2/3$-resonant island since it provides the widest islands in phase-space, and plateaus in the corresponding rotation curves. Nevertheless, we have checked that the effect observed for the $2/3$-resonance holds for all other resonant islands that we could probe.

Figure \ref{fig:island_width_vs_parameters} shows how the $2/3$-island width changes when we vary the BH spin and the halo compactness. In both cases, increasing the respective parameter leads to an enhancement of the island's width. Therefore, rapidly rotating BHs surrounded by compact matter halos are expected to provide the most pronounced imprints of non-integrability and chaos\footnote{See also Ref. \cite{Gutierrez-Ruiz:2018tre}.}. Thus, although EMRIs of supermassive BHs in galaxies, surrounded by a dark matter halo may not provide detectable chaotic signatures, stellar-mass BHs surrounded by a compact environment, such as BH hair or boson clouds, can produce much more prominent imprints of chaotic phenomenology. Interestingly, the latter configuration can be supported in lower mass-gap EMRIs \cite{Pan:2021lyw}. 

Other studies regarding geodesics and EMRIs around non-Kerr objects \cite{Lukes-Gerakopoulos:2010ipp,Destounis:2023cim} have also demonstrated that the increment of the BH's spin leads to an increase in the resonant-island widths, therefore this result was expected and simultaneously corroborates the validity of our analysis. On the other hand, the island widths are also increasing with the halo compactness. Hence, we may extrapolate that the mass quadrupole moment, and higher multipoles are deformed in a prolate manner due to the increment of the halo's compactness (see e.g., the similarities in the multipole moments of Hartle-Thorne \cite{Destounis:2023cim} and Manko-Novikov objects \cite{Lukes-Gerakopoulos:2010ipp} that produce broader islands for larger prolate deformation parameters; exactly as we observe in Fig. \ref{fig:island_width_vs_parameters}, right panel). 

\section{Conclusions}

Resonances are choreographically well-defined periodic orbits, that return to their initial position after a number of oscillations. When a dynamical system is non-integrable, resonances are extended to resonant islands, which occupy an atypical, non-zero volume in phase space, where the periodicity of the central resonance is shared throughout the geodesics in the island. This leads to orbital and, most importantly, GW observables \cite{Destounis:2021mqv} that may be detected with LISA through abnormally-large cumulative dephasing events, resulting in imprecise parameter inference \cite{Flanagan:2010cd}.

In order to study astrophysical EMRIs, a vacuum Kerr background is typically utilized with manual additions that express the environment's contribution to the dynamics of orbits. These approximates are not solutions to GR, but rather include the environmental effects through, either Newtonian/post-Newtonian schemes, or with fully-relativistic treatments that mimic the dynamical interaction of the secondary with the environment, known as dynamical friction \cite{Vicente:2025gsg}. A reasonable understanding of the main astrophysical features in GW generation and propagation has, thus, been achieved \cite{Macedo:2013qea,Polcar:2022bwv,Cardoso:2019rou,CanevaSantoro:2023aol}. The results so far are quite straightforward; the dynamics of EMRIs in astrophysical environments need to be treated in a fully-relativistic manner, rather than with post-Newtonian methods, in order to capture the correct environmental dephasing of GWs from EMRIs \cite{Polcar:2025yto,Vicente:2025gsg,Dyson:2025dlj}. Nevertheless, at the level of the metric, the spacetime used in contemporary studies still assumes a vacuum (integrable) Kerr BH perturbed by a non-backreacting matter environment.

Here, we surpass crude estimates, at the level of the metric, by constructing and employing general-relativistic solutions of non-vacuum GR, that describe rotating BHs at the center of matter halo environments \cite{Fernandes:2025osu}. We have evolved geodesics in order to capture the leading-order effects in EMRI dynamics. Our analysis demonstrates that when the environment backreacts on the rotating Kerr BH, it produces a non-vacuum solution of GR that significantly deforms the multipolar spacetime structure to the point that the Carter symmetry is broken. The absence of a Carter constant alters the dynamics of geodesics around transient resonances, in a manner that designates the emergence of chaotic phenomena and non-integrability, dubbed \emph{environmental chaos}. We have explicitly found resonant islands in the phase space of orbits, and thin chaotic layers that form around them, thus astrophysical environments can indeed lead to non-integrability when considered as a general-relativistic part of spacetime, and not just as a perturbation around a vacuum Kerr system. The rotation curves corroborate the aforementioned results, due to the formation of plateaus when the geodesics are inside the respective resonant island. We, finally, performed a tentative analysis of the widths of various $2/3$-resonant islands resulting by varying the BH spin and the compactness of the halo. We observe that the increment of both spin and compactness, typically, leads to larger, in span, resonant islands. 

Even though our work lies at the geodesic level, it has been shown that when radiation reaction is taken into account, the resulting non-Kerr EMRIs can potentially spend hundreds of revolutions inside a resonant island \cite{Apostolatos:2009vu,Lukes-Gerakopoulos:2010ipp}, where the periodicity of the central resonance is shared throughout it. The crossing of the secondary through successively-damped, ``resonantly-locked'', geodesics leads to a GW observable, namely a \emph{GW glitch} in the frequency evolution of the non-Kerr EMRI \cite{Destounis:2021mqv}. These glitches have a clear astrophysical interpretation \cite{Destounis:2021rko,Destounis:2023gpw,Destounis:2023khj}, can last for weeks \cite{Destounis:2021rko}, instead of minutes or hours as in Kerr EMRIs \cite{Ruangsri:2013hra}, and are not associated with the typical instrumental glitches that occur due to detector noise \cite{Cornish:2014kda,Coughlin:2019ref,Cabero:2019orq,Edwards:2020tlp,Cornish:2020dwh,Chatziioannou:2021ezd,Muratore:2025knh}. Thus, we can extrapolate that if our analysis is extended at the inspiral level, then similar glitches will appear in the GW frequency evolution of general-relativistic, matter-enriched rotating EMRIs.

In the dawn of GW astronomy, the astrophysical environments around realistic, spinning BHs, such as those residing in the core of galaxies, have encountered a significant interest from the GW community, especially for sources like EMRIs that are prime targets of the LISA mission. Our geodesic analysis demonstrates that the assumption of vacuum Kerr BHs with a perturbative, but not backreacting, addition of the astrophysical environment is not enough to describe the general-relativistic nature of the full system. By constructing a general-relativistic BH within a matter halo from first principles, using the Einstein cluster formalism \cite{Cardoso:2021wlq}, a complete understanding of environmental effects around BHs can be achieved at the level of the metric. Nevertheless, EMRIs evolve under radiation reaction, hence they need to be treated with fully-relativistic schemes at the dynamical level as well \cite{Vicente:2025gsg,Polcar:2025yto,Dyson:2025dlj} (instead of using Newtonian or post-Newtonian treatments). 

The relativistic treatment of inspiral dynamics has already been achieved for exact GR solutions that describe matter-enriched, static and spherically-symmetric BHs \cite{Cardoso:2022whc,Vicente:2022ivh,Datta:2025ruh}. The final missing piece to the puzzle is to generalize the fully-relativistic treatment of EMRIs of Refs. \cite{Cardoso:2022whc,Datta:2025ruh} to include the rotation of the primary. Nevertheless, it is easier said than done because the rotating astrophysical BHs used in this work do not have a closed analytic form, but rather are constructed numerically. On top of that, the generalization of the relativistic techniques prescribed in \cite{Cardoso:2022whc,Vicente:2022ivh,Vicente:2025gsg,Datta:2025ruh} needs to be accomplished, i.e., to transform from the Regge-Wheeler-Zerilli-matter equations (zero BH spin) to Teukolsky-matter relativistic equations (non-zero BH spin). 

Even if the above generalization proves to be a thorny and time-consuming process, we can still infer some rough detectability aspects, that are an outcome of the current geodesic work. A typical resonance in a vacuum Kerr EMRI produces dephasing of dozens of radians \cite{Flanagan:2010cd,Flanagan:2012kg,Ruangsri:2013hra}, without taking into account the dephasing that an astrophysical environments can further introduce. Imagining that a zero-volume single point of geodesic periodicity in phase space creates such an effect in Kerr EMRIs \cite{Berry:2016bit}, we could cautiously extrapolate that a finite volume of geodesic periodicity in phase-space, which is introduced by the formation of resonant islands, will produce even bigger resonant enhancement through unexpectedly extended resonance timescales. This could be in the form of beyond-Kerr-resonance dephasing of hundreds of radians, or even abrupt dephasing kicks when the island is crossed (see Refs. \cite{Destounis:2021mqv,Destounis:2021rko} for the resulting GW imprint of a resonant-island crossing). These effects alone might serve as conclusive evidence of the existence of astrophysical environments, and in general non-integrability. 

Thus, it is important that resonances, and in particular possible crossings through resonant islands, to be taken into consideration with novel state-of-the-art methods (see e.g. Ref. \cite{Pan:2023wau,Dyson:2025dlj}). Since GW modeling of generic Kerr EMRIs is a highly active subject of research, so does the GW modeling of resonant-island crossings should be explored for non-vacuum, general-relativistic spinning EMRIs. The combination of environmental effects, resonances, and astrophysical chaos, could form, in the future, the amalgamation of a well-explored branch of precise GW modeling for realistic EMRIs within astrophysical environments. One, though, needs to stay cautious for potential modifications of gravity, supermassive exotic compact object primaries, or triple systems \cite{Cardoso:2021vjq,Kuntz:2021hhm,Kuntz:2021ohi,Kuntz:2022onu,Camilloni:2023xvf,Camilloni:2023rra,Grilli:2024fds,Cocco:2025adu}, in order to avoid numerous degeneracies. 

\begin{acknowledgments}
The authors would like to warmly thank Vitor Cardoso for fruitful discussions.
K.D. acknowledges financial support provided by FCT–Fundação para a Ciência e a Tecnologia, I.P., under the Scientific Employment Stimulus – Individual Call – Grant No. 2023.07417.CEECIND/CP2830/CT0008. K.D. would also like to thank the Fundação para a Ciência e Tecnologia (FCT), Portugal, for the financial support to the Center for Astrophysics and Gravitation (CENTRA/IST/ULisboa) through grant No. UID/PRR/00099/2025 and grant No. UID/00099/2025.
P.F. is funded by the Deutsche Forschungsgemeinschaft (DFG, German Research Foundation) under Germany’s Excellence Strategy EXC 2181/1 - 390900948 (the Heidelberg STRUCTURES Excellence Cluster).
This project has received funding from the European Union’s Horizon MSCA-2022 research and innovation programme “Einstein Waves” under grant agreement No. 101131233.
\end{acknowledgments}

\bibliography{biblio}

\begin{thebibliography}{214}%
\makeatletter
\providecommand \@ifxundefined [1]{%
 \@ifx{#1\undefined}
}%
\providecommand \@ifnum [1]{%
 \ifnum #1\expandafter \@firstoftwo
 \else \expandafter \@secondoftwo
 \fi
}%
\providecommand \@ifx [1]{%
 \ifx #1\expandafter \@firstoftwo
 \else \expandafter \@secondoftwo
 \fi
}%
\providecommand \natexlab [1]{#1}%
\providecommand \enquote  [1]{``#1''}%
\providecommand \bibnamefont  [1]{#1}%
\providecommand \bibfnamefont [1]{#1}%
\providecommand \citenamefont [1]{#1}%
\providecommand \href@noop [0]{\@secondoftwo}%
\providecommand \href [0]{\begingroup \@sanitize@url \@href}%
\providecommand \@href[1]{\@@startlink{#1}\@@href}%
\providecommand \@@href[1]{\endgroup#1\@@endlink}%
\providecommand \@sanitize@url [0]{\catcode `\\12\catcode `\$12\catcode
  `\&12\catcode `\#12\catcode `\^12\catcode `\_12\catcode `\%12\relax}%
\providecommand \@@startlink[1]{}%
\providecommand \@@endlink[0]{}%
\providecommand \url  [0]{\begingroup\@sanitize@url \@url }%
\providecommand \@url [1]{\endgroup\@href {#1}{\urlprefix }}%
\providecommand \urlprefix  [0]{URL }%
\providecommand \Eprint [0]{\href }%
\providecommand \doibase [0]{https://doi.org/}%
\providecommand \selectlanguage [0]{\@gobble}%
\providecommand \bibinfo  [0]{\@secondoftwo}%
\providecommand \bibfield  [0]{\@secondoftwo}%
\providecommand \translation [1]{[#1]}%
\providecommand \BibitemOpen [0]{}%
\providecommand \bibitemStop [0]{}%
\providecommand \bibitemNoStop [0]{.\EOS\space}%
\providecommand \EOS [0]{\spacefactor3000\relax}%
\providecommand \BibitemShut  [1]{\csname bibitem#1\endcsname}%
\let\auto@bib@innerbib\@empty
\bibitem [{\citenamefont {Abbott}\ \emph {et~al.}(2023)\citenamefont {Abbott}
  \emph {et~al.}}]{KAGRA:2021vkt}%
  \BibitemOpen
  \bibfield  {author} {\bibinfo {author} {\bibfnamefont {R.}~\bibnamefont
  {Abbott}} \emph {et~al.} (\bibinfo {collaboration} {KAGRA, VIRGO, LIGO
  Scientific}),\ }\bibfield  {title} {\bibinfo {title} {{GWTC-3: Compact Binary
  Coalescences Observed by LIGO and Virgo during the Second Part of the Third
  Observing Run}},\ }\href {https://doi.org/10.1103/PhysRevX.13.041039}
  {\bibfield  {journal} {\bibinfo  {journal} {Phys. Rev. X}\ }\textbf {\bibinfo
  {volume} {13}},\ \bibinfo {pages} {041039} (\bibinfo {year} {2023})},\
  \Eprint {https://arxiv.org/abs/2111.03606} {arXiv:2111.03606 [gr-qc]}
  \BibitemShut {NoStop}%
\bibitem [{\citenamefont {Davis}\ \emph {et~al.}(2021)\citenamefont {Davis}
  \emph {et~al.}}]{LIGO:2021ppb}%
  \BibitemOpen
  \bibfield  {author} {\bibinfo {author} {\bibfnamefont {D.}~\bibnamefont
  {Davis}} \emph {et~al.} (\bibinfo {collaboration} {LIGO}),\ }\bibfield
  {title} {\bibinfo {title} {{LIGO detector characterization in the second and
  third observing runs}},\ }\href {https://doi.org/10.1088/1361-6382/abfd85}
  {\bibfield  {journal} {\bibinfo  {journal} {Class. Quant. Grav.}\ }\textbf
  {\bibinfo {volume} {38}},\ \bibinfo {pages} {135014} (\bibinfo {year}
  {2021})},\ \Eprint {https://arxiv.org/abs/2101.11673} {arXiv:2101.11673
  [astro-ph.IM]} \BibitemShut {NoStop}%
\bibitem [{\citenamefont {Barack}\ \emph {et~al.}(2019)\citenamefont {Barack}
  \emph {et~al.}}]{Barack:2018yly}%
  \BibitemOpen
  \bibfield  {author} {\bibinfo {author} {\bibfnamefont {L.}~\bibnamefont
  {Barack}} \emph {et~al.},\ }\bibfield  {title} {\bibinfo {title} {{Black
  holes, gravitational waves and fundamental physics: a roadmap}},\ }\href
  {https://doi.org/10.1088/1361-6382/ab0587} {\bibfield  {journal} {\bibinfo
  {journal} {Class. Quant. Grav.}\ }\textbf {\bibinfo {volume} {36}},\ \bibinfo
  {pages} {143001} (\bibinfo {year} {2019})},\ \Eprint
  {https://arxiv.org/abs/1806.05195} {arXiv:1806.05195 [gr-qc]} \BibitemShut
  {NoStop}%
\bibitem [{\citenamefont {Abbott}\ \emph {et~al.}(2016)\citenamefont {Abbott}
  \emph {et~al.}}]{LIGOScientific:2016aoc}%
  \BibitemOpen
  \bibfield  {author} {\bibinfo {author} {\bibfnamefont {B.~P.}\ \bibnamefont
  {Abbott}} \emph {et~al.} (\bibinfo {collaboration} {LIGO Scientific,
  Virgo}),\ }\bibfield  {title} {\bibinfo {title} {{Observation of
  Gravitational Waves from a Binary Black Hole Merger}},\ }\href
  {https://doi.org/10.1103/PhysRevLett.116.061102} {\bibfield  {journal}
  {\bibinfo  {journal} {Phys. Rev. Lett.}\ }\textbf {\bibinfo {volume} {116}},\
  \bibinfo {pages} {061102} (\bibinfo {year} {2016})},\ \Eprint
  {https://arxiv.org/abs/1602.03837} {arXiv:1602.03837 [gr-qc]} \BibitemShut
  {NoStop}%
\bibitem [{\citenamefont {Amaro-Seoane}\ \emph {et~al.}(2017)\citenamefont
  {Amaro-Seoane} \emph {et~al.}}]{LISA:2017pwj}%
  \BibitemOpen
  \bibfield  {author} {\bibinfo {author} {\bibfnamefont {P.}~\bibnamefont
  {Amaro-Seoane}} \emph {et~al.} (\bibinfo {collaboration} {LISA}),\
  }\href@noop {} {\bibinfo {title} {{Laser Interferometer Space Antenna}}}
  (\bibinfo {year} {2017}),\ \Eprint {https://arxiv.org/abs/1702.00786}
  {arXiv:1702.00786 [astro-ph.IM]} \BibitemShut {NoStop}%
\bibitem [{\citenamefont {Schutz}(1999)}]{Schutz:1999xj}%
  \BibitemOpen
  \bibfield  {author} {\bibinfo {author} {\bibfnamefont {B.~F.}\ \bibnamefont
  {Schutz}},\ }\bibfield  {title} {\bibinfo {title} {{Gravitational wave
  astronomy}},\ }\href {https://doi.org/10.1088/0264-9381/16/12A/307}
  {\bibfield  {journal} {\bibinfo  {journal} {Class. Quant. Grav.}\ }\textbf
  {\bibinfo {volume} {16}},\ \bibinfo {pages} {A131} (\bibinfo {year}
  {1999})},\ \Eprint {https://arxiv.org/abs/gr-qc/9911034}
  {arXiv:gr-qc/9911034} \BibitemShut {NoStop}%
\bibitem [{\citenamefont {Bailes}\ \emph {et~al.}(2021)\citenamefont {Bailes}
  \emph {et~al.}}]{Bailes:2021tot}%
  \BibitemOpen
  \bibfield  {author} {\bibinfo {author} {\bibfnamefont {M.}~\bibnamefont
  {Bailes}} \emph {et~al.},\ }\bibfield  {title} {\bibinfo {title}
  {{Gravitational-wave physics and astronomy in the 2020s and 2030s}},\ }\href
  {https://doi.org/10.1038/s42254-021-00303-8} {\bibfield  {journal} {\bibinfo
  {journal} {Nature Rev. Phys.}\ }\textbf {\bibinfo {volume} {3}},\ \bibinfo
  {pages} {344} (\bibinfo {year} {2021})}\BibitemShut {NoStop}%
\bibitem [{\citenamefont {Barausse}\ \emph {et~al.}(2020)\citenamefont
  {Barausse} \emph {et~al.}}]{Barausse:2020rsu}%
  \BibitemOpen
  \bibfield  {author} {\bibinfo {author} {\bibfnamefont {E.}~\bibnamefont
  {Barausse}} \emph {et~al.},\ }\bibfield  {title} {\bibinfo {title}
  {{Prospects for Fundamental Physics with LISA}},\ }\href
  {https://doi.org/10.1007/s10714-020-02691-1} {\bibfield  {journal} {\bibinfo
  {journal} {Gen. Rel. Grav.}\ }\textbf {\bibinfo {volume} {52}},\ \bibinfo
  {pages} {81} (\bibinfo {year} {2020})},\ \Eprint
  {https://arxiv.org/abs/2001.09793} {arXiv:2001.09793 [gr-qc]} \BibitemShut
  {NoStop}%
\bibitem [{\citenamefont {Arun}\ \emph {et~al.}(2022)\citenamefont {Arun} \emph
  {et~al.}}]{LISA:2022kgy}%
  \BibitemOpen
  \bibfield  {author} {\bibinfo {author} {\bibfnamefont {K.~G.}\ \bibnamefont
  {Arun}} \emph {et~al.} (\bibinfo {collaboration} {LISA}),\ }\bibfield
  {title} {\bibinfo {title} {{New horizons for fundamental physics with
  LISA}},\ }\href {https://doi.org/10.1007/s41114-022-00036-9} {\bibfield
  {journal} {\bibinfo  {journal} {Living Rev. Rel.}\ }\textbf {\bibinfo
  {volume} {25}},\ \bibinfo {pages} {4} (\bibinfo {year} {2022})},\ \Eprint
  {https://arxiv.org/abs/2205.01597} {arXiv:2205.01597 [gr-qc]} \BibitemShut
  {NoStop}%
\bibitem [{\citenamefont {Karnesis}\ \emph {et~al.}(2022)\citenamefont
  {Karnesis} \emph {et~al.}}]{Karnesis:2022vdp}%
  \BibitemOpen
  \bibfield  {author} {\bibinfo {author} {\bibfnamefont {N.}~\bibnamefont
  {Karnesis}} \emph {et~al.},\ }\href@noop {} {\bibinfo {title} {{The Laser
  Interferometer Space Antenna mission in Greece White Paper}}} (\bibinfo
  {year} {2022}),\ \Eprint {https://arxiv.org/abs/2209.04358} {arXiv:2209.04358
  [gr-qc]} \BibitemShut {NoStop}%
\bibitem [{\citenamefont {Amaro-Seoane}\ \emph {et~al.}(2022)\citenamefont
  {Amaro-Seoane} \emph {et~al.}}]{Amaro-Seoane:2022rxf}%
  \BibitemOpen
  \bibfield  {author} {\bibinfo {author} {\bibfnamefont {P.}~\bibnamefont
  {Amaro-Seoane}} \emph {et~al.},\ }\href@noop {} {\bibinfo {title}
  {{Astrophysics with the Laser Interferometer Space Antenna}}} (\bibinfo
  {year} {2022}),\ \Eprint {https://arxiv.org/abs/2203.06016} {arXiv:2203.06016
  [gr-qc]} \BibitemShut {NoStop}%
\bibitem [{\citenamefont {Seoane}\ \emph {et~al.}(2023)\citenamefont {Seoane}
  \emph {et~al.}}]{LISA:2022yao}%
  \BibitemOpen
  \bibfield  {author} {\bibinfo {author} {\bibfnamefont {P.~A.}\ \bibnamefont
  {Seoane}} \emph {et~al.} (\bibinfo {collaboration} {LISA}),\ }\bibfield
  {title} {\bibinfo {title} {{Astrophysics with the Laser Interferometer Space
  Antenna}},\ }\href {https://doi.org/10.1007/s41114-022-00041-y} {\bibfield
  {journal} {\bibinfo  {journal} {Living Rev. Rel.}\ }\textbf {\bibinfo
  {volume} {26}},\ \bibinfo {pages} {2} (\bibinfo {year} {2023})},\ \Eprint
  {https://arxiv.org/abs/2203.06016} {arXiv:2203.06016 [gr-qc]} \BibitemShut
  {NoStop}%
\bibitem [{\citenamefont {Ghez}\ \emph {et~al.}(1998)\citenamefont {Ghez},
  \citenamefont {Klein}, \citenamefont {Morris},\ and\ \citenamefont
  {Becklin}}]{Ghez:1998ph}%
  \BibitemOpen
  \bibfield  {author} {\bibinfo {author} {\bibfnamefont {A.~M.}\ \bibnamefont
  {Ghez}}, \bibinfo {author} {\bibfnamefont {B.~L.}\ \bibnamefont {Klein}},
  \bibinfo {author} {\bibfnamefont {M.}~\bibnamefont {Morris}},\ and\ \bibinfo
  {author} {\bibfnamefont {E.~E.}\ \bibnamefont {Becklin}},\ }\bibfield
  {title} {\bibinfo {title} {{High proper motion stars in the vicinity of Sgr
  A*: Evidence for a supermassive black hole at the center of our galaxy}},\
  }\href {https://doi.org/10.1086/306528} {\bibfield  {journal} {\bibinfo
  {journal} {Astrophys. J.}\ }\textbf {\bibinfo {volume} {509}},\ \bibinfo
  {pages} {678} (\bibinfo {year} {1998})},\ \Eprint
  {https://arxiv.org/abs/astro-ph/9807210} {arXiv:astro-ph/9807210}
  \BibitemShut {NoStop}%
\bibitem [{\citenamefont {{Genzel}}\ \emph {et~al.}(2010)\citenamefont
  {{Genzel}}, \citenamefont {{Eisenhauer}},\ and\ \citenamefont
  {{Gillessen}}}]{Genzel:2010}%
  \BibitemOpen
  \bibfield  {author} {\bibinfo {author} {\bibfnamefont {R.}~\bibnamefont
  {{Genzel}}}, \bibinfo {author} {\bibfnamefont {F.}~\bibnamefont
  {{Eisenhauer}}},\ and\ \bibinfo {author} {\bibfnamefont {S.}~\bibnamefont
  {{Gillessen}}},\ }\bibfield  {title} {\bibinfo {title} {{The Galactic Center
  massive black hole and nuclear star cluster}},\ }\href
  {https://doi.org/10.1103/RevModPhys.82.3121} {\bibfield  {journal} {\bibinfo
  {journal} {Reviews of Modern Physics}\ }\textbf {\bibinfo {volume} {82}},\
  \bibinfo {pages} {3121} (\bibinfo {year} {2010})},\ \Eprint
  {https://arxiv.org/abs/1006.0064} {arXiv:1006.0064 [astro-ph.GA]}
  \BibitemShut {NoStop}%
\bibitem [{\citenamefont {Amaro-Seoane}\ \emph {et~al.}(2007)\citenamefont
  {Amaro-Seoane}, \citenamefont {Gair}, \citenamefont {Freitag}, \citenamefont
  {Coleman~Miller}, \citenamefont {Mandel}, \citenamefont {Cutler},\ and\
  \citenamefont {Babak}}]{Amaro-Seoane:2007osp}%
  \BibitemOpen
  \bibfield  {author} {\bibinfo {author} {\bibfnamefont {P.}~\bibnamefont
  {Amaro-Seoane}}, \bibinfo {author} {\bibfnamefont {J.~R.}\ \bibnamefont
  {Gair}}, \bibinfo {author} {\bibfnamefont {M.}~\bibnamefont {Freitag}},
  \bibinfo {author} {\bibfnamefont {M.}~\bibnamefont {Coleman~Miller}},
  \bibinfo {author} {\bibfnamefont {I.}~\bibnamefont {Mandel}}, \bibinfo
  {author} {\bibfnamefont {C.~J.}\ \bibnamefont {Cutler}},\ and\ \bibinfo
  {author} {\bibfnamefont {S.}~\bibnamefont {Babak}},\ }\bibfield  {title}
  {\bibinfo {title} {{Astrophysics, detection and science applications of
  intermediate- and extreme mass-ratio inspirals}},\ }\href
  {https://doi.org/10.1088/0264-9381/24/17/R01} {\bibfield  {journal} {\bibinfo
   {journal} {Class. Quant. Grav.}\ }\textbf {\bibinfo {volume} {24}},\
  \bibinfo {pages} {R113} (\bibinfo {year} {2007})},\ \Eprint
  {https://arxiv.org/abs/astro-ph/0703495} {arXiv:astro-ph/0703495}
  \BibitemShut {NoStop}%
\bibitem [{\citenamefont {Gair}\ \emph {et~al.}(2017)\citenamefont {Gair},
  \citenamefont {Babak}, \citenamefont {Sesana}, \citenamefont {Amaro-Seoane},
  \citenamefont {Barausse}, \citenamefont {Berry}, \citenamefont {Berti},\ and\
  \citenamefont {Sopuerta}}]{Gair:2017ynp}%
  \BibitemOpen
  \bibfield  {author} {\bibinfo {author} {\bibfnamefont {J.~R.}\ \bibnamefont
  {Gair}}, \bibinfo {author} {\bibfnamefont {S.}~\bibnamefont {Babak}},
  \bibinfo {author} {\bibfnamefont {A.}~\bibnamefont {Sesana}}, \bibinfo
  {author} {\bibfnamefont {P.}~\bibnamefont {Amaro-Seoane}}, \bibinfo {author}
  {\bibfnamefont {E.}~\bibnamefont {Barausse}}, \bibinfo {author}
  {\bibfnamefont {C.~P.~L.}\ \bibnamefont {Berry}}, \bibinfo {author}
  {\bibfnamefont {E.}~\bibnamefont {Berti}},\ and\ \bibinfo {author}
  {\bibfnamefont {C.}~\bibnamefont {Sopuerta}},\ }\bibfield  {title} {\bibinfo
  {title} {{Prospects for observing extreme-mass-ratio inspirals with LISA}},\
  }\href {https://doi.org/10.1088/1742-6596/840/1/012021} {\bibfield  {journal}
  {\bibinfo  {journal} {J. Phys. Conf. Ser.}\ }\textbf {\bibinfo {volume}
  {840}},\ \bibinfo {pages} {012021} (\bibinfo {year} {2017})},\ \Eprint
  {https://arxiv.org/abs/1704.00009} {arXiv:1704.00009 [astro-ph.GA]}
  \BibitemShut {NoStop}%
\bibitem [{\citenamefont {Babak}\ \emph {et~al.}(2017)\citenamefont {Babak},
  \citenamefont {Gair}, \citenamefont {Sesana}, \citenamefont {Barausse},
  \citenamefont {Sopuerta}, \citenamefont {Berry}, \citenamefont {Berti},
  \citenamefont {Amaro-Seoane}, \citenamefont {Petiteau},\ and\ \citenamefont
  {Klein}}]{Babak:2017tow}%
  \BibitemOpen
  \bibfield  {author} {\bibinfo {author} {\bibfnamefont {S.}~\bibnamefont
  {Babak}}, \bibinfo {author} {\bibfnamefont {J.}~\bibnamefont {Gair}},
  \bibinfo {author} {\bibfnamefont {A.}~\bibnamefont {Sesana}}, \bibinfo
  {author} {\bibfnamefont {E.}~\bibnamefont {Barausse}}, \bibinfo {author}
  {\bibfnamefont {C.~F.}\ \bibnamefont {Sopuerta}}, \bibinfo {author}
  {\bibfnamefont {C.~P.~L.}\ \bibnamefont {Berry}}, \bibinfo {author}
  {\bibfnamefont {E.}~\bibnamefont {Berti}}, \bibinfo {author} {\bibfnamefont
  {P.}~\bibnamefont {Amaro-Seoane}}, \bibinfo {author} {\bibfnamefont
  {A.}~\bibnamefont {Petiteau}},\ and\ \bibinfo {author} {\bibfnamefont
  {A.}~\bibnamefont {Klein}},\ }\bibfield  {title} {\bibinfo {title} {{Science
  with the space-based interferometer LISA. V: Extreme mass-ratio inspirals}},\
  }\href {https://doi.org/10.1103/PhysRevD.95.103012} {\bibfield  {journal}
  {\bibinfo  {journal} {Phys. Rev. D}\ }\textbf {\bibinfo {volume} {95}},\
  \bibinfo {pages} {103012} (\bibinfo {year} {2017})},\ \Eprint
  {https://arxiv.org/abs/1703.09722} {arXiv:1703.09722 [gr-qc]} \BibitemShut
  {NoStop}%
\bibitem [{\citenamefont {Auclair}\ \emph {et~al.}(2023)\citenamefont {Auclair}
  \emph {et~al.}}]{LISACosmologyWorkingGroup:2022jok}%
  \BibitemOpen
  \bibfield  {author} {\bibinfo {author} {\bibfnamefont {P.}~\bibnamefont
  {Auclair}} \emph {et~al.} (\bibinfo {collaboration} {LISA Cosmology Working
  Group}),\ }\bibfield  {title} {\bibinfo {title} {{Cosmology with the Laser
  Interferometer Space Antenna}},\ }\href
  {https://doi.org/10.1007/s41114-023-00045-2} {\bibfield  {journal} {\bibinfo
  {journal} {Living Rev. Rel.}\ }\textbf {\bibinfo {volume} {26}},\ \bibinfo
  {pages} {5} (\bibinfo {year} {2023})},\ \Eprint
  {https://arxiv.org/abs/2204.05434} {arXiv:2204.05434 [astro-ph.CO]}
  \BibitemShut {NoStop}%
\bibitem [{\citenamefont {Raveh}\ and\ \citenamefont
  {Perets}(2021)}]{Raveh:2020jxg}%
  \BibitemOpen
  \bibfield  {author} {\bibinfo {author} {\bibfnamefont {Y.}~\bibnamefont
  {Raveh}}\ and\ \bibinfo {author} {\bibfnamefont {H.~B.}\ \bibnamefont
  {Perets}},\ }\bibfield  {title} {\bibinfo {title} {{Extreme mass-ratio
  gravitational-wave sources: Mass segregation and post binary tidal-disruption
  captures}},\ }\href {https://doi.org/10.1093/mnras/staa4001} {\bibfield
  {journal} {\bibinfo  {journal} {Mon. Not. Roy. Astron. Soc.}\ }\textbf
  {\bibinfo {volume} {501}},\ \bibinfo {pages} {5012} (\bibinfo {year}
  {2021})},\ \Eprint {https://arxiv.org/abs/2011.13952} {arXiv:2011.13952
  [astro-ph.GA]} \BibitemShut {NoStop}%
\bibitem [{\citenamefont {Amaro-Seoane}(2018)}]{Amaro-Seoane:2012lgq}%
  \BibitemOpen
  \bibfield  {author} {\bibinfo {author} {\bibfnamefont {P.}~\bibnamefont
  {Amaro-Seoane}},\ }\bibfield  {title} {\bibinfo {title} {{Relativistic
  dynamics and extreme mass ratio inspirals}},\ }\href
  {https://doi.org/10.1007/s41114-018-0013-8} {\bibfield  {journal} {\bibinfo
  {journal} {Living Rev. Rel.}\ }\textbf {\bibinfo {volume} {21}},\ \bibinfo
  {pages} {4} (\bibinfo {year} {2018})},\ \Eprint
  {https://arxiv.org/abs/1205.5240} {arXiv:1205.5240 [astro-ph.CO]}
  \BibitemShut {NoStop}%
\bibitem [{\citenamefont {Pan}\ and\ \citenamefont {Yang}(2021)}]{Pan:2021ksp}%
  \BibitemOpen
  \bibfield  {author} {\bibinfo {author} {\bibfnamefont {Z.}~\bibnamefont
  {Pan}}\ and\ \bibinfo {author} {\bibfnamefont {H.}~\bibnamefont {Yang}},\
  }\bibfield  {title} {\bibinfo {title} {{Formation Rate of Extreme Mass Ratio
  Inspirals in Active Galactic Nuclei}},\ }\href
  {https://doi.org/10.1103/PhysRevD.103.103018} {\bibfield  {journal} {\bibinfo
   {journal} {Phys. Rev. D}\ }\textbf {\bibinfo {volume} {103}},\ \bibinfo
  {pages} {103018} (\bibinfo {year} {2021})},\ \Eprint
  {https://arxiv.org/abs/2101.09146} {arXiv:2101.09146 [astro-ph.HE]}
  \BibitemShut {NoStop}%
\bibitem [{\citenamefont {Pan}\ \emph {et~al.}(2021)\citenamefont {Pan},
  \citenamefont {Lyu},\ and\ \citenamefont {Yang}}]{Pan:2021oob}%
  \BibitemOpen
  \bibfield  {author} {\bibinfo {author} {\bibfnamefont {Z.}~\bibnamefont
  {Pan}}, \bibinfo {author} {\bibfnamefont {Z.}~\bibnamefont {Lyu}},\ and\
  \bibinfo {author} {\bibfnamefont {H.}~\bibnamefont {Yang}},\ }\bibfield
  {title} {\bibinfo {title} {{Wet extreme mass ratio inspirals may be more
  common for spaceborne gravitational wave detection}},\ }\href
  {https://doi.org/10.1103/PhysRevD.104.063007} {\bibfield  {journal} {\bibinfo
   {journal} {Phys. Rev. D}\ }\textbf {\bibinfo {volume} {104}},\ \bibinfo
  {pages} {063007} (\bibinfo {year} {2021})},\ \Eprint
  {https://arxiv.org/abs/2104.01208} {arXiv:2104.01208 [astro-ph.HE]}
  \BibitemShut {NoStop}%
\bibitem [{\citenamefont {Chua}\ \emph {et~al.}(2021)\citenamefont {Chua},
  \citenamefont {Katz}, \citenamefont {Warburton},\ and\ \citenamefont
  {Hughes}}]{Chua:2020stf}%
  \BibitemOpen
  \bibfield  {author} {\bibinfo {author} {\bibfnamefont {A.~J.~K.}\
  \bibnamefont {Chua}}, \bibinfo {author} {\bibfnamefont {M.~L.}\ \bibnamefont
  {Katz}}, \bibinfo {author} {\bibfnamefont {N.}~\bibnamefont {Warburton}},\
  and\ \bibinfo {author} {\bibfnamefont {S.~A.}\ \bibnamefont {Hughes}},\
  }\bibfield  {title} {\bibinfo {title} {{Rapid generation of fully
  relativistic extreme-mass-ratio-inspiral waveform templates for LISA data
  analysis}},\ }\href {https://doi.org/10.1103/PhysRevLett.126.051102}
  {\bibfield  {journal} {\bibinfo  {journal} {Phys. Rev. Lett.}\ }\textbf
  {\bibinfo {volume} {126}},\ \bibinfo {pages} {051102} (\bibinfo {year}
  {2021})},\ \Eprint {https://arxiv.org/abs/2008.06071} {arXiv:2008.06071
  [gr-qc]} \BibitemShut {NoStop}%
\bibitem [{\citenamefont {Katz}\ \emph {et~al.}(2021)\citenamefont {Katz},
  \citenamefont {Chua}, \citenamefont {Speri}, \citenamefont {Warburton},\ and\
  \citenamefont {Hughes}}]{Katz:2021yft}%
  \BibitemOpen
  \bibfield  {author} {\bibinfo {author} {\bibfnamefont {M.~L.}\ \bibnamefont
  {Katz}}, \bibinfo {author} {\bibfnamefont {A.~J.~K.}\ \bibnamefont {Chua}},
  \bibinfo {author} {\bibfnamefont {L.}~\bibnamefont {Speri}}, \bibinfo
  {author} {\bibfnamefont {N.}~\bibnamefont {Warburton}},\ and\ \bibinfo
  {author} {\bibfnamefont {S.~A.}\ \bibnamefont {Hughes}},\ }\bibfield  {title}
  {\bibinfo {title} {{Fast extreme-mass-ratio-inspiral waveforms: New tools for
  millihertz gravitational-wave data analysis}},\ }\href
  {https://doi.org/10.1103/PhysRevD.104.064047} {\bibfield  {journal} {\bibinfo
   {journal} {Phys. Rev. D}\ }\textbf {\bibinfo {volume} {104}},\ \bibinfo
  {pages} {064047} (\bibinfo {year} {2021})},\ \Eprint
  {https://arxiv.org/abs/2104.04582} {arXiv:2104.04582 [gr-qc]} \BibitemShut
  {NoStop}%
\bibitem [{\citenamefont {Afshordi}\ \emph {et~al.}(2023)\citenamefont
  {Afshordi} \emph {et~al.}}]{LISAConsortiumWaveformWorkingGroup:2023arg}%
  \BibitemOpen
  \bibfield  {author} {\bibinfo {author} {\bibfnamefont {N.}~\bibnamefont
  {Afshordi}} \emph {et~al.} (\bibinfo {collaboration} {LISA Consortium
  Waveform Working Group}),\ }\href@noop {} {\bibinfo {title} {{Waveform
  Modelling for the Laser Interferometer Space Antenna}}} (\bibinfo {year}
  {2023}),\ \Eprint {https://arxiv.org/abs/2311.01300} {arXiv:2311.01300
  [gr-qc]} \BibitemShut {NoStop}%
\bibitem [{\citenamefont {Glampedakis}\ and\ \citenamefont
  {Babak}(2006)}]{Glampedakis:2005cf}%
  \BibitemOpen
  \bibfield  {author} {\bibinfo {author} {\bibfnamefont {K.}~\bibnamefont
  {Glampedakis}}\ and\ \bibinfo {author} {\bibfnamefont {S.}~\bibnamefont
  {Babak}},\ }\bibfield  {title} {\bibinfo {title} {{Mapping spacetimes with
  LISA: Inspiral of a test-body in a `quasi-Kerr' field}},\ }\href
  {https://doi.org/10.1088/0264-9381/23/12/013} {\bibfield  {journal} {\bibinfo
   {journal} {Class. Quant. Grav.}\ }\textbf {\bibinfo {volume} {23}},\
  \bibinfo {pages} {4167} (\bibinfo {year} {2006})},\ \Eprint
  {https://arxiv.org/abs/gr-qc/0510057} {arXiv:gr-qc/0510057} \BibitemShut
  {NoStop}%
\bibitem [{\citenamefont {Gair}\ \emph {et~al.}(2013)\citenamefont {Gair},
  \citenamefont {Vallisneri}, \citenamefont {Larson},\ and\ \citenamefont
  {Baker}}]{Gair:2012nm}%
  \BibitemOpen
  \bibfield  {author} {\bibinfo {author} {\bibfnamefont {J.~R.}\ \bibnamefont
  {Gair}}, \bibinfo {author} {\bibfnamefont {M.}~\bibnamefont {Vallisneri}},
  \bibinfo {author} {\bibfnamefont {S.~L.}\ \bibnamefont {Larson}},\ and\
  \bibinfo {author} {\bibfnamefont {J.~G.}\ \bibnamefont {Baker}},\ }\bibfield
  {title} {\bibinfo {title} {{Testing General Relativity with Low-Frequency,
  Space-Based Gravitational-Wave Detectors}},\ }\href
  {https://doi.org/10.12942/lrr-2013-7} {\bibfield  {journal} {\bibinfo
  {journal} {Living Rev. Rel.}\ }\textbf {\bibinfo {volume} {16}},\ \bibinfo
  {pages} {7} (\bibinfo {year} {2013})},\ \Eprint
  {https://arxiv.org/abs/1212.5575} {arXiv:1212.5575 [gr-qc]} \BibitemShut
  {NoStop}%
\bibitem [{\citenamefont {Yunes}\ \emph {et~al.}(2012)\citenamefont {Yunes},
  \citenamefont {Pani},\ and\ \citenamefont {Cardoso}}]{Yunes:2011aa}%
  \BibitemOpen
  \bibfield  {author} {\bibinfo {author} {\bibfnamefont {N.}~\bibnamefont
  {Yunes}}, \bibinfo {author} {\bibfnamefont {P.}~\bibnamefont {Pani}},\ and\
  \bibinfo {author} {\bibfnamefont {V.}~\bibnamefont {Cardoso}},\ }\bibfield
  {title} {\bibinfo {title} {{Gravitational Waves from Quasicircular Extreme
  Mass-Ratio Inspirals as Probes of Scalar-Tensor Theories}},\ }\href
  {https://doi.org/10.1103/PhysRevD.85.102003} {\bibfield  {journal} {\bibinfo
  {journal} {Phys. Rev. D}\ }\textbf {\bibinfo {volume} {85}},\ \bibinfo
  {pages} {102003} (\bibinfo {year} {2012})},\ \Eprint
  {https://arxiv.org/abs/1112.3351} {arXiv:1112.3351 [gr-qc]} \BibitemShut
  {NoStop}%
\bibitem [{\citenamefont {Cardoso}\ and\ \citenamefont
  {Gualtieri}(2016)}]{Cardoso:2016ryw}%
  \BibitemOpen
  \bibfield  {author} {\bibinfo {author} {\bibfnamefont {V.}~\bibnamefont
  {Cardoso}}\ and\ \bibinfo {author} {\bibfnamefont {L.}~\bibnamefont
  {Gualtieri}},\ }\bibfield  {title} {\bibinfo {title} {{Testing the black hole
  {\textquoteleft}no-hair{\textquoteright} hypothesis}},\ }\href
  {https://doi.org/10.1088/0264-9381/33/17/174001} {\bibfield  {journal}
  {\bibinfo  {journal} {Class. Quant. Grav.}\ }\textbf {\bibinfo {volume}
  {33}},\ \bibinfo {pages} {174001} (\bibinfo {year} {2016})},\ \Eprint
  {https://arxiv.org/abs/1607.03133} {arXiv:1607.03133 [gr-qc]} \BibitemShut
  {NoStop}%
\bibitem [{\citenamefont {Pani}\ \emph {et~al.}(2011)\citenamefont {Pani},
  \citenamefont {Cardoso},\ and\ \citenamefont {Gualtieri}}]{Pani:2011xj}%
  \BibitemOpen
  \bibfield  {author} {\bibinfo {author} {\bibfnamefont {P.}~\bibnamefont
  {Pani}}, \bibinfo {author} {\bibfnamefont {V.}~\bibnamefont {Cardoso}},\ and\
  \bibinfo {author} {\bibfnamefont {L.}~\bibnamefont {Gualtieri}},\ }\bibfield
  {title} {\bibinfo {title} {{Gravitational waves from extreme mass-ratio
  inspirals in Dynamical Chern-Simons gravity}},\ }\href
  {https://doi.org/10.1103/PhysRevD.83.104048} {\bibfield  {journal} {\bibinfo
  {journal} {Phys. Rev. D}\ }\textbf {\bibinfo {volume} {83}},\ \bibinfo
  {pages} {104048} (\bibinfo {year} {2011})},\ \Eprint
  {https://arxiv.org/abs/1104.1183} {arXiv:1104.1183 [gr-qc]} \BibitemShut
  {NoStop}%
\bibitem [{\citenamefont {Cardoso}\ \emph {et~al.}(2011)\citenamefont
  {Cardoso}, \citenamefont {Chakrabarti}, \citenamefont {Pani}, \citenamefont
  {Berti},\ and\ \citenamefont {Gualtieri}}]{Cardoso:2011xi}%
  \BibitemOpen
  \bibfield  {author} {\bibinfo {author} {\bibfnamefont {V.}~\bibnamefont
  {Cardoso}}, \bibinfo {author} {\bibfnamefont {S.}~\bibnamefont
  {Chakrabarti}}, \bibinfo {author} {\bibfnamefont {P.}~\bibnamefont {Pani}},
  \bibinfo {author} {\bibfnamefont {E.}~\bibnamefont {Berti}},\ and\ \bibinfo
  {author} {\bibfnamefont {L.}~\bibnamefont {Gualtieri}},\ }\bibfield  {title}
  {\bibinfo {title} {{Floating and sinking: The Imprint of massive scalars
  around rotating black holes}},\ }\href
  {https://doi.org/10.1103/PhysRevLett.107.241101} {\bibfield  {journal}
  {\bibinfo  {journal} {Phys. Rev. Lett.}\ }\textbf {\bibinfo {volume} {107}},\
  \bibinfo {pages} {241101} (\bibinfo {year} {2011})},\ \Eprint
  {https://arxiv.org/abs/1109.6021} {arXiv:1109.6021 [gr-qc]} \BibitemShut
  {NoStop}%
\bibitem [{\citenamefont {Canizares}\ \emph {et~al.}(2012)\citenamefont
  {Canizares}, \citenamefont {Gair},\ and\ \citenamefont
  {Sopuerta}}]{Canizares:2012is}%
  \BibitemOpen
  \bibfield  {author} {\bibinfo {author} {\bibfnamefont {P.}~\bibnamefont
  {Canizares}}, \bibinfo {author} {\bibfnamefont {J.~R.}\ \bibnamefont
  {Gair}},\ and\ \bibinfo {author} {\bibfnamefont {C.~F.}\ \bibnamefont
  {Sopuerta}},\ }\bibfield  {title} {\bibinfo {title} {{Testing Chern-Simons
  Modified Gravity with Gravitational-Wave Detections of Extreme-Mass-Ratio
  Binaries}},\ }\href {https://doi.org/10.1103/PhysRevD.86.044010} {\bibfield
  {journal} {\bibinfo  {journal} {Phys. Rev. D}\ }\textbf {\bibinfo {volume}
  {86}},\ \bibinfo {pages} {044010} (\bibinfo {year} {2012})},\ \Eprint
  {https://arxiv.org/abs/1205.1253} {arXiv:1205.1253 [gr-qc]} \BibitemShut
  {NoStop}%
\bibitem [{\citenamefont {Berti}\ \emph {et~al.}(2015)\citenamefont {Berti}
  \emph {et~al.}}]{Berti:2015itd}%
  \BibitemOpen
  \bibfield  {author} {\bibinfo {author} {\bibfnamefont {E.}~\bibnamefont
  {Berti}} \emph {et~al.},\ }\bibfield  {title} {\bibinfo {title} {{Testing
  General Relativity with Present and Future Astrophysical Observations}},\
  }\href {https://doi.org/10.1088/0264-9381/32/24/243001} {\bibfield  {journal}
  {\bibinfo  {journal} {Class. Quant. Grav.}\ }\textbf {\bibinfo {volume}
  {32}},\ \bibinfo {pages} {243001} (\bibinfo {year} {2015})},\ \Eprint
  {https://arxiv.org/abs/1501.07274} {arXiv:1501.07274 [gr-qc]} \BibitemShut
  {NoStop}%
\bibitem [{\citenamefont {Chua}\ \emph {et~al.}(2018)\citenamefont {Chua},
  \citenamefont {Hee}, \citenamefont {Handley}, \citenamefont {Higson},
  \citenamefont {Moore}, \citenamefont {Gair}, \citenamefont {Hobson},\ and\
  \citenamefont {Lasenby}}]{Chua:2018yng}%
  \BibitemOpen
  \bibfield  {author} {\bibinfo {author} {\bibfnamefont {A.~J.~K.}\
  \bibnamefont {Chua}}, \bibinfo {author} {\bibfnamefont {S.}~\bibnamefont
  {Hee}}, \bibinfo {author} {\bibfnamefont {W.~J.}\ \bibnamefont {Handley}},
  \bibinfo {author} {\bibfnamefont {E.}~\bibnamefont {Higson}}, \bibinfo
  {author} {\bibfnamefont {C.~J.}\ \bibnamefont {Moore}}, \bibinfo {author}
  {\bibfnamefont {J.~R.}\ \bibnamefont {Gair}}, \bibinfo {author}
  {\bibfnamefont {M.~P.}\ \bibnamefont {Hobson}},\ and\ \bibinfo {author}
  {\bibfnamefont {A.~N.}\ \bibnamefont {Lasenby}},\ }\bibfield  {title}
  {\bibinfo {title} {{Towards a framework for testing general relativity with
  extreme-mass-ratio-inspiral observations}},\ }\href
  {https://doi.org/10.1093/mnras/sty1079} {\bibfield  {journal} {\bibinfo
  {journal} {Mon. Not. Roy. Astron. Soc.}\ }\textbf {\bibinfo {volume} {478}},\
  \bibinfo {pages} {28} (\bibinfo {year} {2018})},\ \Eprint
  {https://arxiv.org/abs/1803.10210} {arXiv:1803.10210 [gr-qc]} \BibitemShut
  {NoStop}%
\bibitem [{\citenamefont {Berti}\ \emph {et~al.}(2019)\citenamefont {Berti}
  \emph {et~al.}}]{Berti:2019xgr}%
  \BibitemOpen
  \bibfield  {author} {\bibinfo {author} {\bibfnamefont {E.}~\bibnamefont
  {Berti}} \emph {et~al.},\ }\href@noop {} {\bibinfo {title} {{Tests of General
  Relativity and Fundamental Physics with Space-based Gravitational Wave
  Detectors}}} (\bibinfo {year} {2019}),\ \Eprint
  {https://arxiv.org/abs/1903.02781} {arXiv:1903.02781 [astro-ph.HE]}
  \BibitemShut {NoStop}%
\bibitem [{\citenamefont {Zhang}\ and\ \citenamefont
  {Yang}(2019)}]{Zhang:2018kib}%
  \BibitemOpen
  \bibfield  {author} {\bibinfo {author} {\bibfnamefont {J.}~\bibnamefont
  {Zhang}}\ and\ \bibinfo {author} {\bibfnamefont {H.}~\bibnamefont {Yang}},\
  }\bibfield  {title} {\bibinfo {title} {{Gravitational floating orbits around
  hairy black holes}},\ }\href {https://doi.org/10.1103/PhysRevD.99.064018}
  {\bibfield  {journal} {\bibinfo  {journal} {Phys. Rev. D}\ }\textbf {\bibinfo
  {volume} {99}},\ \bibinfo {pages} {064018} (\bibinfo {year} {2019})},\
  \Eprint {https://arxiv.org/abs/1808.02905} {arXiv:1808.02905 [gr-qc]}
  \BibitemShut {NoStop}%
\bibitem [{\citenamefont {Berry}\ \emph {et~al.}(2019)\citenamefont {Berry},
  \citenamefont {Hughes}, \citenamefont {Sopuerta}, \citenamefont {Chua},
  \citenamefont {Heffernan}, \citenamefont {Holley-Bockelmann}, \citenamefont
  {Mihaylov}, \citenamefont {Miller},\ and\ \citenamefont
  {Sesana}}]{Berry:2019wgg}%
  \BibitemOpen
  \bibfield  {author} {\bibinfo {author} {\bibfnamefont {C.~P.~L.}\
  \bibnamefont {Berry}}, \bibinfo {author} {\bibfnamefont {S.~A.}\ \bibnamefont
  {Hughes}}, \bibinfo {author} {\bibfnamefont {C.~F.}\ \bibnamefont
  {Sopuerta}}, \bibinfo {author} {\bibfnamefont {A.~J.~K.}\ \bibnamefont
  {Chua}}, \bibinfo {author} {\bibfnamefont {A.}~\bibnamefont {Heffernan}},
  \bibinfo {author} {\bibfnamefont {K.}~\bibnamefont {Holley-Bockelmann}},
  \bibinfo {author} {\bibfnamefont {D.~P.}\ \bibnamefont {Mihaylov}}, \bibinfo
  {author} {\bibfnamefont {M.~C.}\ \bibnamefont {Miller}},\ and\ \bibinfo
  {author} {\bibfnamefont {A.}~\bibnamefont {Sesana}},\ }\bibfield  {title}
  {\bibinfo {title} {{The unique potential of extreme mass-ratio inspirals for
  gravitational-wave astronomy}},\ }\href@noop {} {\bibfield  {journal}
  {\bibinfo  {journal} {Bull. Am. Astron. Soc.}\ }\textbf {\bibinfo {volume}
  {51}},\ \bibinfo {pages} {42} (\bibinfo {year} {2019})},\ \Eprint
  {https://arxiv.org/abs/1903.03686} {arXiv:1903.03686 [astro-ph.HE]}
  \BibitemShut {NoStop}%
\bibitem [{\citenamefont {C{\'a}rdenas-Avenda{\~n}o}\ and\ \citenamefont
  {Sopuerta}(2024)}]{Cardenas-Avendano:2024mqp}%
  \BibitemOpen
  \bibfield  {author} {\bibinfo {author} {\bibfnamefont {A.}~\bibnamefont
  {C{\'a}rdenas-Avenda{\~n}o}}\ and\ \bibinfo {author} {\bibfnamefont {C.~F.}\
  \bibnamefont {Sopuerta}},\ }\bibinfo {title} {{Testing Gravity with
  Extreme-Mass-Ratio Inspirals}}\ (\bibinfo {year} {2024})\ \Eprint
  {https://arxiv.org/abs/2401.08085} {arXiv:2401.08085 [gr-qc]} \BibitemShut
  {NoStop}%
\bibitem [{\citenamefont {Berti}(2024)}]{Berti:2024orb}%
  \BibitemOpen
  \bibfield  {author} {\bibinfo {author} {\bibfnamefont {E.}~\bibnamefont
  {Berti}},\ }\bibfield  {title} {\bibinfo {title} {{Tests of general
  relativity with future detectors}},\ }\href
  {https://doi.org/10.1007/s10714-024-03332-7} {\bibfield  {journal} {\bibinfo
  {journal} {Gen. Rel. Grav.}\ }\textbf {\bibinfo {volume} {56}},\ \bibinfo
  {pages} {145} (\bibinfo {year} {2024})}\BibitemShut {NoStop}%
\bibitem [{\citenamefont {Datta}\ and\ \citenamefont
  {Bose}(2019)}]{Datta:2019euh}%
  \BibitemOpen
  \bibfield  {author} {\bibinfo {author} {\bibfnamefont {S.}~\bibnamefont
  {Datta}}\ and\ \bibinfo {author} {\bibfnamefont {S.}~\bibnamefont {Bose}},\
  }\bibfield  {title} {\bibinfo {title} {{Probing the nature of central objects
  in extreme-mass-ratio inspirals with gravitational waves}},\ }\href
  {https://doi.org/10.1103/PhysRevD.99.084001} {\bibfield  {journal} {\bibinfo
  {journal} {Phys. Rev. D}\ }\textbf {\bibinfo {volume} {99}},\ \bibinfo
  {pages} {084001} (\bibinfo {year} {2019})},\ \Eprint
  {https://arxiv.org/abs/1902.01723} {arXiv:1902.01723 [gr-qc]} \BibitemShut
  {NoStop}%
\bibitem [{\citenamefont {Hannuksela}\ \emph {et~al.}(2019)\citenamefont
  {Hannuksela}, \citenamefont {Wong}, \citenamefont {Brito}, \citenamefont
  {Berti},\ and\ \citenamefont {Li}}]{Hannuksela:2018izj}%
  \BibitemOpen
  \bibfield  {author} {\bibinfo {author} {\bibfnamefont {O.~A.}\ \bibnamefont
  {Hannuksela}}, \bibinfo {author} {\bibfnamefont {K.~W.~K.}\ \bibnamefont
  {Wong}}, \bibinfo {author} {\bibfnamefont {R.}~\bibnamefont {Brito}},
  \bibinfo {author} {\bibfnamefont {E.}~\bibnamefont {Berti}},\ and\ \bibinfo
  {author} {\bibfnamefont {T.~G.~F.}\ \bibnamefont {Li}},\ }\bibfield  {title}
  {\bibinfo {title} {{Probing the existence of ultralight bosons with a single
  gravitational-wave measurement}},\ }\href
  {https://doi.org/10.1038/s41550-019-0712-4} {\bibfield  {journal} {\bibinfo
  {journal} {Nature Astron.}\ }\textbf {\bibinfo {volume} {3}},\ \bibinfo
  {pages} {447} (\bibinfo {year} {2019})},\ \Eprint
  {https://arxiv.org/abs/1804.09659} {arXiv:1804.09659 [astro-ph.HE]}
  \BibitemShut {NoStop}%
\bibitem [{\citenamefont {Maselli}\ \emph {et~al.}(2020)\citenamefont
  {Maselli}, \citenamefont {Franchini}, \citenamefont {Gualtieri},\ and\
  \citenamefont {Sotiriou}}]{Maselli:2020zgv}%
  \BibitemOpen
  \bibfield  {author} {\bibinfo {author} {\bibfnamefont {A.}~\bibnamefont
  {Maselli}}, \bibinfo {author} {\bibfnamefont {N.}~\bibnamefont {Franchini}},
  \bibinfo {author} {\bibfnamefont {L.}~\bibnamefont {Gualtieri}},\ and\
  \bibinfo {author} {\bibfnamefont {T.~P.}\ \bibnamefont {Sotiriou}},\
  }\bibfield  {title} {\bibinfo {title} {{Detecting scalar fields with Extreme
  Mass Ratio Inspirals}},\ }\href
  {https://doi.org/10.1103/PhysRevLett.125.141101} {\bibfield  {journal}
  {\bibinfo  {journal} {Phys. Rev. Lett.}\ }\textbf {\bibinfo {volume} {125}},\
  \bibinfo {pages} {141101} (\bibinfo {year} {2020})},\ \Eprint
  {https://arxiv.org/abs/2004.11895} {arXiv:2004.11895 [gr-qc]} \BibitemShut
  {NoStop}%
\bibitem [{\citenamefont {Maselli}\ \emph {et~al.}(2022)\citenamefont
  {Maselli}, \citenamefont {Franchini}, \citenamefont {Gualtieri},
  \citenamefont {Sotiriou}, \citenamefont {Barsanti},\ and\ \citenamefont
  {Pani}}]{Maselli:2021men}%
  \BibitemOpen
  \bibfield  {author} {\bibinfo {author} {\bibfnamefont {A.}~\bibnamefont
  {Maselli}}, \bibinfo {author} {\bibfnamefont {N.}~\bibnamefont {Franchini}},
  \bibinfo {author} {\bibfnamefont {L.}~\bibnamefont {Gualtieri}}, \bibinfo
  {author} {\bibfnamefont {T.~P.}\ \bibnamefont {Sotiriou}}, \bibinfo {author}
  {\bibfnamefont {S.}~\bibnamefont {Barsanti}},\ and\ \bibinfo {author}
  {\bibfnamefont {P.}~\bibnamefont {Pani}},\ }\bibfield  {title} {\bibinfo
  {title} {{Detecting fundamental fields with LISA observations of
  gravitational waves from extreme mass-ratio inspirals}},\ }\href
  {https://doi.org/10.1038/s41550-021-01589-5} {\bibfield  {journal} {\bibinfo
  {journal} {Nature Astron.}\ }\textbf {\bibinfo {volume} {6}},\ \bibinfo
  {pages} {464} (\bibinfo {year} {2022})},\ \Eprint
  {https://arxiv.org/abs/2106.11325} {arXiv:2106.11325 [gr-qc]} \BibitemShut
  {NoStop}%
\bibitem [{\citenamefont {Barsanti}\ \emph
  {et~al.}(2022{\natexlab{a}})\citenamefont {Barsanti}, \citenamefont
  {Maselli}, \citenamefont {Sotiriou},\ and\ \citenamefont
  {Gualtieri}}]{Barsanti:2022vvl}%
  \BibitemOpen
  \bibfield  {author} {\bibinfo {author} {\bibfnamefont {S.}~\bibnamefont
  {Barsanti}}, \bibinfo {author} {\bibfnamefont {A.}~\bibnamefont {Maselli}},
  \bibinfo {author} {\bibfnamefont {T.~P.}\ \bibnamefont {Sotiriou}},\ and\
  \bibinfo {author} {\bibfnamefont {L.}~\bibnamefont {Gualtieri}},\ }\href@noop
  {} {\bibinfo {title} {{Detecting massive scalar fields with Extreme
  Mass-Ratio Inspirals}}} (\bibinfo {year} {2022}{\natexlab{a}}),\ \Eprint
  {https://arxiv.org/abs/2212.03888} {arXiv:2212.03888 [gr-qc]} \BibitemShut
  {NoStop}%
\bibitem [{\citenamefont {Barsanti}\ \emph
  {et~al.}(2022{\natexlab{b}})\citenamefont {Barsanti}, \citenamefont
  {Franchini}, \citenamefont {Gualtieri}, \citenamefont {Maselli},\ and\
  \citenamefont {Sotiriou}}]{Barsanti:2022ana}%
  \BibitemOpen
  \bibfield  {author} {\bibinfo {author} {\bibfnamefont {S.}~\bibnamefont
  {Barsanti}}, \bibinfo {author} {\bibfnamefont {N.}~\bibnamefont {Franchini}},
  \bibinfo {author} {\bibfnamefont {L.}~\bibnamefont {Gualtieri}}, \bibinfo
  {author} {\bibfnamefont {A.}~\bibnamefont {Maselli}},\ and\ \bibinfo {author}
  {\bibfnamefont {T.~P.}\ \bibnamefont {Sotiriou}},\ }\bibfield  {title}
  {\bibinfo {title} {{Extreme mass-ratio inspirals as probes of scalar fields:
  Eccentric equatorial orbits around Kerr black holes}},\ }\href
  {https://doi.org/10.1103/PhysRevD.106.044029} {\bibfield  {journal} {\bibinfo
   {journal} {Phys. Rev. D}\ }\textbf {\bibinfo {volume} {106}},\ \bibinfo
  {pages} {044029} (\bibinfo {year} {2022}{\natexlab{b}})},\ \Eprint
  {https://arxiv.org/abs/2203.05003} {arXiv:2203.05003 [gr-qc]} \BibitemShut
  {NoStop}%
\bibitem [{\citenamefont {Mitra}\ \emph {et~al.}(2024)\citenamefont {Mitra},
  \citenamefont {Chakraborty}, \citenamefont {Vicente},\ and\ \citenamefont
  {Feng}}]{Mitra:2023sny}%
  \BibitemOpen
  \bibfield  {author} {\bibinfo {author} {\bibfnamefont {S.}~\bibnamefont
  {Mitra}}, \bibinfo {author} {\bibfnamefont {S.}~\bibnamefont {Chakraborty}},
  \bibinfo {author} {\bibfnamefont {R.}~\bibnamefont {Vicente}},\ and\ \bibinfo
  {author} {\bibfnamefont {J.~C.}\ \bibnamefont {Feng}},\ }\bibfield  {title}
  {\bibinfo {title} {{Probing the quantum nature of black holes with ultralight
  boson environments}},\ }\href {https://doi.org/10.1103/PhysRevD.110.084012}
  {\bibfield  {journal} {\bibinfo  {journal} {Phys. Rev. D}\ }\textbf {\bibinfo
  {volume} {110}},\ \bibinfo {pages} {084012} (\bibinfo {year} {2024})},\
  \Eprint {https://arxiv.org/abs/2312.06783} {arXiv:2312.06783 [gr-qc]}
  \BibitemShut {NoStop}%
\bibitem [{\citenamefont {Zhang}\ and\ \citenamefont
  {Gong}(2024)}]{Zhang:2024ogc}%
  \BibitemOpen
  \bibfield  {author} {\bibinfo {author} {\bibfnamefont {C.}~\bibnamefont
  {Zhang}}\ and\ \bibinfo {author} {\bibfnamefont {Y.}~\bibnamefont {Gong}},\
  }\bibfield  {title} {\bibinfo {title} {{Probing new fundamental fields with
  extreme mass ratio inspirals}},\ }\href
  {https://doi.org/10.1103/PhysRevD.110.104052} {\bibfield  {journal} {\bibinfo
   {journal} {Phys. Rev. D}\ }\textbf {\bibinfo {volume} {110}},\ \bibinfo
  {pages} {104052} (\bibinfo {year} {2024})},\ \Eprint
  {https://arxiv.org/abs/2407.07449} {arXiv:2407.07449 [gr-qc]} \BibitemShut
  {NoStop}%
\bibitem [{\citenamefont {Kuntz}\ \emph {et~al.}(2020)\citenamefont {Kuntz},
  \citenamefont {Penco},\ and\ \citenamefont {Piazza}}]{Kuntz:2020yow}%
  \BibitemOpen
  \bibfield  {author} {\bibinfo {author} {\bibfnamefont {A.}~\bibnamefont
  {Kuntz}}, \bibinfo {author} {\bibfnamefont {R.}~\bibnamefont {Penco}},\ and\
  \bibinfo {author} {\bibfnamefont {F.}~\bibnamefont {Piazza}},\ }\bibfield
  {title} {\bibinfo {title} {{Extreme Mass Ratio Inspirals with Scalar Hair}},\
  }\href {https://doi.org/10.1088/1475-7516/2020/08/023} {\bibfield  {journal}
  {\bibinfo  {journal} {JCAP}\ }\textbf {\bibinfo {volume} {08}},\ \bibinfo
  {pages} {023}},\ \Eprint {https://arxiv.org/abs/2004.10772} {arXiv:2004.10772
  [gr-qc]} \BibitemShut {NoStop}%
\bibitem [{\citenamefont {Lestingi}\ \emph {et~al.}(2024)\citenamefont
  {Lestingi}, \citenamefont {Cannizzaro},\ and\ \citenamefont
  {Pani}}]{Lestingi:2023ovn}%
  \BibitemOpen
  \bibfield  {author} {\bibinfo {author} {\bibfnamefont {J.}~\bibnamefont
  {Lestingi}}, \bibinfo {author} {\bibfnamefont {E.}~\bibnamefont
  {Cannizzaro}},\ and\ \bibinfo {author} {\bibfnamefont {P.}~\bibnamefont
  {Pani}},\ }\bibfield  {title} {\bibinfo {title} {{Extreme mass-ratio
  inspirals as probes of fundamental dipoles}},\ }\href
  {https://doi.org/10.1103/PhysRevD.109.044052} {\bibfield  {journal} {\bibinfo
   {journal} {Phys. Rev. D}\ }\textbf {\bibinfo {volume} {109}},\ \bibinfo
  {pages} {044052} (\bibinfo {year} {2024})},\ \Eprint
  {https://arxiv.org/abs/2310.07772} {arXiv:2310.07772 [gr-qc]} \BibitemShut
  {NoStop}%
\bibitem [{\citenamefont {Della~Rocca}\ \emph {et~al.}(2024)\citenamefont
  {Della~Rocca}, \citenamefont {Barsanti}, \citenamefont {Gualtieri},\ and\
  \citenamefont {Maselli}}]{DellaRocca:2024pnm}%
  \BibitemOpen
  \bibfield  {author} {\bibinfo {author} {\bibfnamefont {M.}~\bibnamefont
  {Della~Rocca}}, \bibinfo {author} {\bibfnamefont {S.}~\bibnamefont
  {Barsanti}}, \bibinfo {author} {\bibfnamefont {L.}~\bibnamefont
  {Gualtieri}},\ and\ \bibinfo {author} {\bibfnamefont {A.}~\bibnamefont
  {Maselli}},\ }\bibfield  {title} {\bibinfo {title} {{Extreme mass-ratio
  inspirals as probes of scalar fields: Inclined circular orbits around Kerr
  black holes}},\ }\href {https://doi.org/10.1103/PhysRevD.109.104079}
  {\bibfield  {journal} {\bibinfo  {journal} {Phys. Rev. D}\ }\textbf {\bibinfo
  {volume} {109}},\ \bibinfo {pages} {104079} (\bibinfo {year} {2024})},\
  \Eprint {https://arxiv.org/abs/2401.09542} {arXiv:2401.09542 [gr-qc]}
  \BibitemShut {NoStop}%
\bibitem [{\citenamefont {Zi}\ and\ \citenamefont {Kumar}(2025)}]{Zi:2025lio}%
  \BibitemOpen
  \bibfield  {author} {\bibinfo {author} {\bibfnamefont {T.}~\bibnamefont
  {Zi}}\ and\ \bibinfo {author} {\bibfnamefont {S.}~\bibnamefont {Kumar}},\
  }\href@noop {} {\bibinfo {title} {{Probing scalar field with generic extreme
  mass-ratio inspirals around Kerr black holes}}} (\bibinfo {year} {2025}),\
  \Eprint {https://arxiv.org/abs/2508.00516} {arXiv:2508.00516 [gr-qc]}
  \BibitemShut {NoStop}%
\bibitem [{\citenamefont {Gondolo}\ and\ \citenamefont
  {Silk}(1999)}]{Gondolo:1999ef}%
  \BibitemOpen
  \bibfield  {author} {\bibinfo {author} {\bibfnamefont {P.}~\bibnamefont
  {Gondolo}}\ and\ \bibinfo {author} {\bibfnamefont {J.}~\bibnamefont {Silk}},\
  }\bibfield  {title} {\bibinfo {title} {{Dark matter annihilation at the
  galactic center}},\ }\href {https://doi.org/10.1103/PhysRevLett.83.1719}
  {\bibfield  {journal} {\bibinfo  {journal} {Phys. Rev. Lett.}\ }\textbf
  {\bibinfo {volume} {83}},\ \bibinfo {pages} {1719} (\bibinfo {year}
  {1999})},\ \Eprint {https://arxiv.org/abs/astro-ph/9906391}
  {arXiv:astro-ph/9906391} \BibitemShut {NoStop}%
\bibitem [{\citenamefont {Sigl}\ \emph {et~al.}(2007)\citenamefont {Sigl},
  \citenamefont {Schnittman},\ and\ \citenamefont {Buonanno}}]{Sigl:2006cg}%
  \BibitemOpen
  \bibfield  {author} {\bibinfo {author} {\bibfnamefont {G.}~\bibnamefont
  {Sigl}}, \bibinfo {author} {\bibfnamefont {J.}~\bibnamefont {Schnittman}},\
  and\ \bibinfo {author} {\bibfnamefont {A.}~\bibnamefont {Buonanno}},\
  }\bibfield  {title} {\bibinfo {title} {{Gravitational-wave background from
  compact objects embedded in AGN accretion disks}},\ }\href
  {https://doi.org/10.1103/PhysRevD.75.024034} {\bibfield  {journal} {\bibinfo
  {journal} {Phys. Rev. D}\ }\textbf {\bibinfo {volume} {75}},\ \bibinfo
  {pages} {024034} (\bibinfo {year} {2007})},\ \Eprint
  {https://arxiv.org/abs/astro-ph/0610680} {arXiv:astro-ph/0610680}
  \BibitemShut {NoStop}%
\bibitem [{\citenamefont {Preto}\ and\ \citenamefont
  {Amaro-Seoane}(2010)}]{Preto:2009kd}%
  \BibitemOpen
  \bibfield  {author} {\bibinfo {author} {\bibfnamefont {M.}~\bibnamefont
  {Preto}}\ and\ \bibinfo {author} {\bibfnamefont {P.}~\bibnamefont
  {Amaro-Seoane}},\ }\bibfield  {title} {\bibinfo {title} {{On strong mass
  segregation around a massive black hole: Implications for lower-frequency
  gravitational-wave astrophysics}},\ }\href
  {https://doi.org/10.1088/2041-8205/708/1/L42} {\bibfield  {journal} {\bibinfo
   {journal} {Astrophys. J. Lett.}\ }\textbf {\bibinfo {volume} {708}},\
  \bibinfo {pages} {L42} (\bibinfo {year} {2010})},\ \Eprint
  {https://arxiv.org/abs/0910.3206} {arXiv:0910.3206 [astro-ph.GA]}
  \BibitemShut {NoStop}%
\bibitem [{\citenamefont {Yunes}\ \emph {et~al.}(2011)\citenamefont {Yunes},
  \citenamefont {Kocsis}, \citenamefont {Loeb},\ and\ \citenamefont
  {Haiman}}]{Yunes:2011ws}%
  \BibitemOpen
  \bibfield  {author} {\bibinfo {author} {\bibfnamefont {N.}~\bibnamefont
  {Yunes}}, \bibinfo {author} {\bibfnamefont {B.}~\bibnamefont {Kocsis}},
  \bibinfo {author} {\bibfnamefont {A.}~\bibnamefont {Loeb}},\ and\ \bibinfo
  {author} {\bibfnamefont {Z.}~\bibnamefont {Haiman}},\ }\bibfield  {title}
  {\bibinfo {title} {{Imprint of Accretion Disk-Induced Migration on
  Gravitational Waves from Extreme Mass Ratio Inspirals}},\ }\href
  {https://doi.org/10.1103/PhysRevLett.107.171103} {\bibfield  {journal}
  {\bibinfo  {journal} {Phys. Rev. Lett.}\ }\textbf {\bibinfo {volume} {107}},\
  \bibinfo {pages} {171103} (\bibinfo {year} {2011})},\ \Eprint
  {https://arxiv.org/abs/1103.4609} {arXiv:1103.4609 [astro-ph.CO]}
  \BibitemShut {NoStop}%
\bibitem [{\citenamefont {Kocsis}\ \emph {et~al.}(2011)\citenamefont {Kocsis},
  \citenamefont {Yunes},\ and\ \citenamefont {Loeb}}]{Kocsis:2011dr}%
  \BibitemOpen
  \bibfield  {author} {\bibinfo {author} {\bibfnamefont {B.}~\bibnamefont
  {Kocsis}}, \bibinfo {author} {\bibfnamefont {N.}~\bibnamefont {Yunes}},\ and\
  \bibinfo {author} {\bibfnamefont {A.}~\bibnamefont {Loeb}},\ }\bibfield
  {title} {\bibinfo {title} {{Observable Signatures of EMRI Black Hole Binaries
  Embedded in Thin Accretion Disks}},\ }\href
  {https://doi.org/10.1103/PhysRevD.86.049907} {\bibfield  {journal} {\bibinfo
  {journal} {Phys. Rev. D}\ }\textbf {\bibinfo {volume} {84}},\ \bibinfo
  {pages} {024032} (\bibinfo {year} {2011})},\ \Eprint
  {https://arxiv.org/abs/1104.2322} {arXiv:1104.2322 [astro-ph.GA]}
  \BibitemShut {NoStop}%
\bibitem [{\citenamefont {Macedo}\ \emph {et~al.}(2013)\citenamefont {Macedo},
  \citenamefont {Pani}, \citenamefont {Cardoso},\ and\ \citenamefont
  {Crispino}}]{Macedo:2013qea}%
  \BibitemOpen
  \bibfield  {author} {\bibinfo {author} {\bibfnamefont {C.~F.~B.}\
  \bibnamefont {Macedo}}, \bibinfo {author} {\bibfnamefont {P.}~\bibnamefont
  {Pani}}, \bibinfo {author} {\bibfnamefont {V.}~\bibnamefont {Cardoso}},\ and\
  \bibinfo {author} {\bibfnamefont {L.~C.~B.}\ \bibnamefont {Crispino}},\
  }\bibfield  {title} {\bibinfo {title} {{Into the lair: gravitational-wave
  signatures of dark matter}},\ }\href
  {https://doi.org/10.1088/0004-637X/774/1/48} {\bibfield  {journal} {\bibinfo
  {journal} {Astrophys. J.}\ }\textbf {\bibinfo {volume} {774}},\ \bibinfo
  {pages} {48} (\bibinfo {year} {2013})},\ \Eprint
  {https://arxiv.org/abs/1302.2646} {arXiv:1302.2646 [gr-qc]} \BibitemShut
  {NoStop}%
\bibitem [{\citenamefont {Sadeghian}\ \emph {et~al.}(2013)\citenamefont
  {Sadeghian}, \citenamefont {Ferrer},\ and\ \citenamefont
  {Will}}]{Sadeghian:2013laa}%
  \BibitemOpen
  \bibfield  {author} {\bibinfo {author} {\bibfnamefont {L.}~\bibnamefont
  {Sadeghian}}, \bibinfo {author} {\bibfnamefont {F.}~\bibnamefont {Ferrer}},\
  and\ \bibinfo {author} {\bibfnamefont {C.~M.}\ \bibnamefont {Will}},\
  }\bibfield  {title} {\bibinfo {title} {{Dark matter distributions around
  massive black holes: A general relativistic analysis}},\ }\href
  {https://doi.org/10.1103/PhysRevD.88.063522} {\bibfield  {journal} {\bibinfo
  {journal} {Phys. Rev. D}\ }\textbf {\bibinfo {volume} {88}},\ \bibinfo
  {pages} {063522} (\bibinfo {year} {2013})},\ \Eprint
  {https://arxiv.org/abs/1305.2619} {arXiv:1305.2619 [astro-ph.GA]}
  \BibitemShut {NoStop}%
\bibitem [{\citenamefont {Ferrer}\ \emph {et~al.}(2017)\citenamefont {Ferrer},
  \citenamefont {da~Rosa},\ and\ \citenamefont {Will}}]{Ferrer:2017xwm}%
  \BibitemOpen
  \bibfield  {author} {\bibinfo {author} {\bibfnamefont {F.}~\bibnamefont
  {Ferrer}}, \bibinfo {author} {\bibfnamefont {A.~M.}\ \bibnamefont
  {da~Rosa}},\ and\ \bibinfo {author} {\bibfnamefont {C.~M.}\ \bibnamefont
  {Will}},\ }\bibfield  {title} {\bibinfo {title} {{Dark matter spikes in the
  vicinity of Kerr black holes}},\ }\href
  {https://doi.org/10.1103/PhysRevD.96.083014} {\bibfield  {journal} {\bibinfo
  {journal} {Phys. Rev. D}\ }\textbf {\bibinfo {volume} {96}},\ \bibinfo
  {pages} {083014} (\bibinfo {year} {2017})},\ \Eprint
  {https://arxiv.org/abs/1707.06302} {arXiv:1707.06302 [astro-ph.CO]}
  \BibitemShut {NoStop}%
\bibitem [{\citenamefont {Hannuksela}\ \emph {et~al.}(2020)\citenamefont
  {Hannuksela}, \citenamefont {Ng},\ and\ \citenamefont
  {Li}}]{Hannuksela:2019vip}%
  \BibitemOpen
  \bibfield  {author} {\bibinfo {author} {\bibfnamefont {O.~A.}\ \bibnamefont
  {Hannuksela}}, \bibinfo {author} {\bibfnamefont {K.~C.~Y.}\ \bibnamefont
  {Ng}},\ and\ \bibinfo {author} {\bibfnamefont {T.~G.~F.}\ \bibnamefont
  {Li}},\ }\bibfield  {title} {\bibinfo {title} {{Extreme dark matter tests
  with extreme mass ratio inspirals}},\ }\href
  {https://doi.org/10.1103/PhysRevD.102.103022} {\bibfield  {journal} {\bibinfo
   {journal} {Phys. Rev. D}\ }\textbf {\bibinfo {volume} {102}},\ \bibinfo
  {pages} {103022} (\bibinfo {year} {2020})},\ \Eprint
  {https://arxiv.org/abs/1906.11845} {arXiv:1906.11845 [astro-ph.CO]}
  \BibitemShut {NoStop}%
\bibitem [{\citenamefont {Cardoso}\ and\ \citenamefont
  {Maselli}(2020)}]{Cardoso:2019rou}%
  \BibitemOpen
  \bibfield  {author} {\bibinfo {author} {\bibfnamefont {V.}~\bibnamefont
  {Cardoso}}\ and\ \bibinfo {author} {\bibfnamefont {A.}~\bibnamefont
  {Maselli}},\ }\bibfield  {title} {\bibinfo {title} {{Constraints on the
  astrophysical environment of binaries with gravitational-wave
  observations}},\ }\href {https://doi.org/10.1051/0004-6361/202037654}
  {\bibfield  {journal} {\bibinfo  {journal} {Astron. Astrophys.}\ }\textbf
  {\bibinfo {volume} {644}},\ \bibinfo {pages} {A147} (\bibinfo {year}
  {2020})},\ \Eprint {https://arxiv.org/abs/1909.05870} {arXiv:1909.05870
  [astro-ph.HE]} \BibitemShut {NoStop}%
\bibitem [{\citenamefont {Polcar}\ \emph {et~al.}(2022)\citenamefont {Polcar},
  \citenamefont {Lukes-Gerakopoulos},\ and\ \citenamefont
  {Witzany}}]{Polcar:2022bwv}%
  \BibitemOpen
  \bibfield  {author} {\bibinfo {author} {\bibfnamefont {L.}~\bibnamefont
  {Polcar}}, \bibinfo {author} {\bibfnamefont {G.}~\bibnamefont
  {Lukes-Gerakopoulos}},\ and\ \bibinfo {author} {\bibfnamefont
  {V.}~\bibnamefont {Witzany}},\ }\bibfield  {title} {\bibinfo {title}
  {{Extreme mass ratio inspirals into black holes surrounded by matter}},\
  }\href {https://doi.org/10.1103/PhysRevD.106.044069} {\bibfield  {journal}
  {\bibinfo  {journal} {Phys. Rev. D}\ }\textbf {\bibinfo {volume} {106}},\
  \bibinfo {pages} {044069} (\bibinfo {year} {2022})},\ \Eprint
  {https://arxiv.org/abs/2205.08516} {arXiv:2205.08516 [gr-qc]} \BibitemShut
  {NoStop}%
\bibitem [{\citenamefont {Speri}\ \emph {et~al.}(2022)\citenamefont {Speri},
  \citenamefont {Antonelli}, \citenamefont {Sberna}, \citenamefont {Babak},
  \citenamefont {Barausse}, \citenamefont {Gair},\ and\ \citenamefont
  {Katz}}]{Speri:2022upm}%
  \BibitemOpen
  \bibfield  {author} {\bibinfo {author} {\bibfnamefont {L.}~\bibnamefont
  {Speri}}, \bibinfo {author} {\bibfnamefont {A.}~\bibnamefont {Antonelli}},
  \bibinfo {author} {\bibfnamefont {L.}~\bibnamefont {Sberna}}, \bibinfo
  {author} {\bibfnamefont {S.}~\bibnamefont {Babak}}, \bibinfo {author}
  {\bibfnamefont {E.}~\bibnamefont {Barausse}}, \bibinfo {author}
  {\bibfnamefont {J.~R.}\ \bibnamefont {Gair}},\ and\ \bibinfo {author}
  {\bibfnamefont {M.~L.}\ \bibnamefont {Katz}},\ }\href@noop {} {\bibinfo
  {title} {{Measuring accretion-disk effects with gravitational waves from
  extreme mass ratio inspirals}}} (\bibinfo {year} {2022}),\ \Eprint
  {https://arxiv.org/abs/2207.10086} {arXiv:2207.10086 [gr-qc]} \BibitemShut
  {NoStop}%
\bibitem [{\citenamefont {Duque}\ \emph {et~al.}(2024)\citenamefont {Duque},
  \citenamefont {Macedo}, \citenamefont {Vicente},\ and\ \citenamefont
  {Cardoso}}]{Duque:2023seg}%
  \BibitemOpen
  \bibfield  {author} {\bibinfo {author} {\bibfnamefont {F.}~\bibnamefont
  {Duque}}, \bibinfo {author} {\bibfnamefont {C.~F.~B.}\ \bibnamefont
  {Macedo}}, \bibinfo {author} {\bibfnamefont {R.}~\bibnamefont {Vicente}},\
  and\ \bibinfo {author} {\bibfnamefont {V.}~\bibnamefont {Cardoso}},\
  }\bibfield  {title} {\bibinfo {title} {{Extreme-Mass-Ratio Inspirals in
  Ultralight Dark Matter}},\ }\href
  {https://doi.org/10.1103/PhysRevLett.133.121404} {\bibfield  {journal}
  {\bibinfo  {journal} {Phys. Rev. Lett.}\ }\textbf {\bibinfo {volume} {133}},\
  \bibinfo {pages} {121404} (\bibinfo {year} {2024})},\ \Eprint
  {https://arxiv.org/abs/2312.06767} {arXiv:2312.06767 [gr-qc]} \BibitemShut
  {NoStop}%
\bibitem [{\citenamefont {Brito}\ and\ \citenamefont
  {Shah}(2023)}]{Brito:2023pyl}%
  \BibitemOpen
  \bibfield  {author} {\bibinfo {author} {\bibfnamefont {R.}~\bibnamefont
  {Brito}}\ and\ \bibinfo {author} {\bibfnamefont {S.}~\bibnamefont {Shah}},\
  }\bibfield  {title} {\bibinfo {title} {{Extreme mass-ratio inspirals into
  black holes surrounded by scalar clouds}},\ }\href
  {https://doi.org/10.1103/PhysRevD.108.084019} {\bibfield  {journal} {\bibinfo
   {journal} {Phys. Rev. D}\ }\textbf {\bibinfo {volume} {108}},\ \bibinfo
  {pages} {084019} (\bibinfo {year} {2023})},\ \bibinfo {note} {[Erratum:
  Phys.Rev.D 110, 109902 (2024)]},\ \Eprint {https://arxiv.org/abs/2307.16093}
  {arXiv:2307.16093 [gr-qc]} \BibitemShut {NoStop}%
\bibitem [{\citenamefont {Traykova}\ \emph {et~al.}(2021)\citenamefont
  {Traykova}, \citenamefont {Clough}, \citenamefont {Helfer}, \citenamefont
  {Berti}, \citenamefont {Ferreira},\ and\ \citenamefont
  {Hui}}]{Traykova:2021dua}%
  \BibitemOpen
  \bibfield  {author} {\bibinfo {author} {\bibfnamefont {D.}~\bibnamefont
  {Traykova}}, \bibinfo {author} {\bibfnamefont {K.}~\bibnamefont {Clough}},
  \bibinfo {author} {\bibfnamefont {T.}~\bibnamefont {Helfer}}, \bibinfo
  {author} {\bibfnamefont {E.}~\bibnamefont {Berti}}, \bibinfo {author}
  {\bibfnamefont {P.~G.}\ \bibnamefont {Ferreira}},\ and\ \bibinfo {author}
  {\bibfnamefont {L.}~\bibnamefont {Hui}},\ }\bibfield  {title} {\bibinfo
  {title} {{Dynamical friction from scalar dark matter in the relativistic
  regime}},\ }\href {https://doi.org/10.1103/PhysRevD.104.103014} {\bibfield
  {journal} {\bibinfo  {journal} {Phys. Rev. D}\ }\textbf {\bibinfo {volume}
  {104}},\ \bibinfo {pages} {103014} (\bibinfo {year} {2021})},\ \Eprint
  {https://arxiv.org/abs/2106.08280} {arXiv:2106.08280 [gr-qc]} \BibitemShut
  {NoStop}%
\bibitem [{\citenamefont {Traykova}\ \emph {et~al.}(2023)\citenamefont
  {Traykova}, \citenamefont {Vicente}, \citenamefont {Clough}, \citenamefont
  {Helfer}, \citenamefont {Berti}, \citenamefont {Ferreira},\ and\
  \citenamefont {Hui}}]{Traykova:2023qyv}%
  \BibitemOpen
  \bibfield  {author} {\bibinfo {author} {\bibfnamefont {D.}~\bibnamefont
  {Traykova}}, \bibinfo {author} {\bibfnamefont {R.}~\bibnamefont {Vicente}},
  \bibinfo {author} {\bibfnamefont {K.}~\bibnamefont {Clough}}, \bibinfo
  {author} {\bibfnamefont {T.}~\bibnamefont {Helfer}}, \bibinfo {author}
  {\bibfnamefont {E.}~\bibnamefont {Berti}}, \bibinfo {author} {\bibfnamefont
  {P.~G.}\ \bibnamefont {Ferreira}},\ and\ \bibinfo {author} {\bibfnamefont
  {L.}~\bibnamefont {Hui}},\ }\bibfield  {title} {\bibinfo {title}
  {{Relativistic drag forces on black holes from scalar dark matter clouds of
  all sizes}},\ }\href {https://doi.org/10.1103/PhysRevD.108.L121502}
  {\bibfield  {journal} {\bibinfo  {journal} {Phys. Rev. D}\ }\textbf {\bibinfo
  {volume} {108}},\ \bibinfo {pages} {L121502} (\bibinfo {year} {2023})},\
  \Eprint {https://arxiv.org/abs/2305.10492} {arXiv:2305.10492 [gr-qc]}
  \BibitemShut {NoStop}%
\bibitem [{\citenamefont {Chatterjee}\ \emph {et~al.}(2023)\citenamefont
  {Chatterjee}, \citenamefont {Mondal},\ and\ \citenamefont
  {Basu}}]{Chatterjee:2023cry}%
  \BibitemOpen
  \bibfield  {author} {\bibinfo {author} {\bibfnamefont {S.}~\bibnamefont
  {Chatterjee}}, \bibinfo {author} {\bibfnamefont {S.}~\bibnamefont {Mondal}},\
  and\ \bibinfo {author} {\bibfnamefont {P.}~\bibnamefont {Basu}},\ }\bibfield
  {title} {\bibinfo {title} {{Detectability of gas-rich
  E/IMRI{\textquoteright}s in LISA band: observable signature of transonic
  accretion flow}},\ }\href {https://doi.org/10.1093/mnras/stad3132} {\bibfield
   {journal} {\bibinfo  {journal} {Mon. Not. Roy. Astron. Soc.}\ }\textbf
  {\bibinfo {volume} {526}},\ \bibinfo {pages} {5612} (\bibinfo {year}
  {2023})},\ \Eprint {https://arxiv.org/abs/2307.12144} {arXiv:2307.12144
  [astro-ph.HE]} \BibitemShut {NoStop}%
\bibitem [{\citenamefont {Santini}\ \emph {et~al.}(2023)\citenamefont
  {Santini}, \citenamefont {Gerosa}, \citenamefont {Cotesta},\ and\
  \citenamefont {Berti}}]{Santini:2023ukl}%
  \BibitemOpen
  \bibfield  {author} {\bibinfo {author} {\bibfnamefont {A.}~\bibnamefont
  {Santini}}, \bibinfo {author} {\bibfnamefont {D.}~\bibnamefont {Gerosa}},
  \bibinfo {author} {\bibfnamefont {R.}~\bibnamefont {Cotesta}},\ and\ \bibinfo
  {author} {\bibfnamefont {E.}~\bibnamefont {Berti}},\ }\bibfield  {title}
  {\bibinfo {title} {{Black-hole mergers in disklike environments could explain
  the observed q-{\ensuremath{\chi}}eff correlation}},\ }\href
  {https://doi.org/10.1103/PhysRevD.108.083033} {\bibfield  {journal} {\bibinfo
   {journal} {Phys. Rev. D}\ }\textbf {\bibinfo {volume} {108}},\ \bibinfo
  {pages} {083033} (\bibinfo {year} {2023})},\ \Eprint
  {https://arxiv.org/abs/2308.12998} {arXiv:2308.12998 [astro-ph.HE]}
  \BibitemShut {NoStop}%
\bibitem [{\citenamefont {Roy}\ and\ \citenamefont
  {Vicente}(2025)}]{Roy:2024rhe}%
  \BibitemOpen
  \bibfield  {author} {\bibinfo {author} {\bibfnamefont {S.}~\bibnamefont
  {Roy}}\ and\ \bibinfo {author} {\bibfnamefont {R.}~\bibnamefont {Vicente}},\
  }\bibfield  {title} {\bibinfo {title} {{Compact binary coalescences in dense
  gaseous environments can pose as ones in vacuum}},\ }\href
  {https://doi.org/10.1103/PhysRevD.111.084037} {\bibfield  {journal} {\bibinfo
   {journal} {Phys. Rev. D}\ }\textbf {\bibinfo {volume} {111}},\ \bibinfo
  {pages} {084037} (\bibinfo {year} {2025})},\ \Eprint
  {https://arxiv.org/abs/2410.16388} {arXiv:2410.16388 [gr-qc]} \BibitemShut
  {NoStop}%
\bibitem [{\citenamefont {Duque}\ \emph {et~al.}(2025)\citenamefont {Duque},
  \citenamefont {Kejriwal}, \citenamefont {Sberna}, \citenamefont {Speri},\
  and\ \citenamefont {Gair}}]{Duque:2024mfw}%
  \BibitemOpen
  \bibfield  {author} {\bibinfo {author} {\bibfnamefont {F.}~\bibnamefont
  {Duque}}, \bibinfo {author} {\bibfnamefont {S.}~\bibnamefont {Kejriwal}},
  \bibinfo {author} {\bibfnamefont {L.}~\bibnamefont {Sberna}}, \bibinfo
  {author} {\bibfnamefont {L.}~\bibnamefont {Speri}},\ and\ \bibinfo {author}
  {\bibfnamefont {J.}~\bibnamefont {Gair}},\ }\bibfield  {title} {\bibinfo
  {title} {{Constraining accretion physics with gravitational waves from
  eccentric extreme-mass-ratio inspirals}},\ }\href
  {https://doi.org/10.1103/PhysRevD.111.084006} {\bibfield  {journal} {\bibinfo
   {journal} {Phys. Rev. D}\ }\textbf {\bibinfo {volume} {111}},\ \bibinfo
  {pages} {084006} (\bibinfo {year} {2025})},\ \Eprint
  {https://arxiv.org/abs/2411.03436} {arXiv:2411.03436 [gr-qc]} \BibitemShut
  {NoStop}%
\bibitem [{\citenamefont {Lyu}\ \emph {et~al.}(2024)\citenamefont {Lyu},
  \citenamefont {Pan}, \citenamefont {Mao}, \citenamefont {Jiang},\ and\
  \citenamefont {Yang}}]{Lyu:2024gnk}%
  \BibitemOpen
  \bibfield  {author} {\bibinfo {author} {\bibfnamefont {Z.}~\bibnamefont
  {Lyu}}, \bibinfo {author} {\bibfnamefont {Z.}~\bibnamefont {Pan}}, \bibinfo
  {author} {\bibfnamefont {J.}~\bibnamefont {Mao}}, \bibinfo {author}
  {\bibfnamefont {N.}~\bibnamefont {Jiang}},\ and\ \bibinfo {author}
  {\bibfnamefont {H.}~\bibnamefont {Yang}},\ }\href@noop {} {\bibinfo {title}
  {{Science Opportunities of Wet Extreme Mass-Ratio Inspirals}}} (\bibinfo
  {year} {2024}),\ \Eprint {https://arxiv.org/abs/2501.03252} {arXiv:2501.03252
  [astro-ph.HE]} \BibitemShut {NoStop}%
\bibitem [{\citenamefont {Dyson}\ \emph {et~al.}(2025)\citenamefont {Dyson},
  \citenamefont {Spieksma}, \citenamefont {Brito}, \citenamefont {van~de
  Meent},\ and\ \citenamefont {Dolan}}]{Dyson:2025dlj}%
  \BibitemOpen
  \bibfield  {author} {\bibinfo {author} {\bibfnamefont {C.}~\bibnamefont
  {Dyson}}, \bibinfo {author} {\bibfnamefont {T.~F.~M.}\ \bibnamefont
  {Spieksma}}, \bibinfo {author} {\bibfnamefont {R.}~\bibnamefont {Brito}},
  \bibinfo {author} {\bibfnamefont {M.}~\bibnamefont {van~de Meent}},\ and\
  \bibinfo {author} {\bibfnamefont {S.}~\bibnamefont {Dolan}},\ }\bibfield
  {title} {\bibinfo {title} {{Environmental Effects in Extreme-Mass-Ratio
  Inspirals: Perturbations to the Environment in Kerr Spacetimes}},\ }\href
  {https://doi.org/10.1103/PhysRevLett.134.211403} {\bibfield  {journal}
  {\bibinfo  {journal} {Phys. Rev. Lett.}\ }\textbf {\bibinfo {volume} {134}},\
  \bibinfo {pages} {211403} (\bibinfo {year} {2025})},\ \Eprint
  {https://arxiv.org/abs/2501.09806} {arXiv:2501.09806 [gr-qc]} \BibitemShut
  {NoStop}%
\bibitem [{\citenamefont {Mitra}\ \emph {et~al.}(2025)\citenamefont {Mitra},
  \citenamefont {Speeney}, \citenamefont {Chakraborty},\ and\ \citenamefont
  {Berti}}]{Mitra:2025tag}%
  \BibitemOpen
  \bibfield  {author} {\bibinfo {author} {\bibfnamefont {S.}~\bibnamefont
  {Mitra}}, \bibinfo {author} {\bibfnamefont {N.}~\bibnamefont {Speeney}},
  \bibinfo {author} {\bibfnamefont {S.}~\bibnamefont {Chakraborty}},\ and\
  \bibinfo {author} {\bibfnamefont {E.}~\bibnamefont {Berti}},\ }\href@noop {}
  {\bibinfo {title} {{Extreme mass ratio inspirals in rotating dark matter
  spikes}}} (\bibinfo {year} {2025}),\ \Eprint
  {https://arxiv.org/abs/2505.04697} {arXiv:2505.04697 [gr-qc]} \BibitemShut
  {NoStop}%
\bibitem [{\citenamefont {Blas}\ \emph {et~al.}(2025)\citenamefont {Blas},
  \citenamefont {Gasparotto},\ and\ \citenamefont {Vicente}}]{Blas:2024duy}%
  \BibitemOpen
  \bibfield  {author} {\bibinfo {author} {\bibfnamefont {D.}~\bibnamefont
  {Blas}}, \bibinfo {author} {\bibfnamefont {S.}~\bibnamefont {Gasparotto}},\
  and\ \bibinfo {author} {\bibfnamefont {R.}~\bibnamefont {Vicente}},\
  }\bibfield  {title} {\bibinfo {title} {{Searching for ultralight dark matter
  through frequency modulation of gravitational waves}},\ }\href
  {https://doi.org/10.1103/PhysRevD.111.042008} {\bibfield  {journal} {\bibinfo
   {journal} {Phys. Rev. D}\ }\textbf {\bibinfo {volume} {111}},\ \bibinfo
  {pages} {042008} (\bibinfo {year} {2025})},\ \Eprint
  {https://arxiv.org/abs/2410.07330} {arXiv:2410.07330 [hep-ph]} \BibitemShut
  {NoStop}%
\bibitem [{\citenamefont {Vicente}\ \emph {et~al.}(2025)\citenamefont
  {Vicente}, \citenamefont {Karydas},\ and\ \citenamefont
  {Bertone}}]{Vicente:2025gsg}%
  \BibitemOpen
  \bibfield  {author} {\bibinfo {author} {\bibfnamefont {R.}~\bibnamefont
  {Vicente}}, \bibinfo {author} {\bibfnamefont {T.~K.}\ \bibnamefont
  {Karydas}},\ and\ \bibinfo {author} {\bibfnamefont {G.}~\bibnamefont
  {Bertone}},\ }\href@noop {} {\bibinfo {title} {{A fully relativistic
  treatment of EMRIs in collisionless environments}}} (\bibinfo {year}
  {2025}),\ \Eprint {https://arxiv.org/abs/2505.09715} {arXiv:2505.09715
  [gr-qc]} \BibitemShut {NoStop}%
\bibitem [{\citenamefont {Polcar}\ and\ \citenamefont
  {Witzany}(2025)}]{Polcar:2025yto}%
  \BibitemOpen
  \bibfield  {author} {\bibinfo {author} {\bibfnamefont {L.}~\bibnamefont
  {Polcar}}\ and\ \bibinfo {author} {\bibfnamefont {V.}~\bibnamefont
  {Witzany}},\ }\href@noop {} {\bibinfo {title} {{Towards relativistic
  inspirals into black holes surrounded by matter}}} (\bibinfo {year} {2025}),\
  \Eprint {https://arxiv.org/abs/2507.15720} {arXiv:2507.15720 [gr-qc]}
  \BibitemShut {NoStop}%
\bibitem [{\citenamefont {Rahman}\ \emph {et~al.}(2024)\citenamefont {Rahman},
  \citenamefont {Kumar},\ and\ \citenamefont {Bhattacharyya}}]{Rahman:2023sof}%
  \BibitemOpen
  \bibfield  {author} {\bibinfo {author} {\bibfnamefont {M.}~\bibnamefont
  {Rahman}}, \bibinfo {author} {\bibfnamefont {S.}~\bibnamefont {Kumar}},\ and\
  \bibinfo {author} {\bibfnamefont {A.}~\bibnamefont {Bhattacharyya}},\
  }\bibfield  {title} {\bibinfo {title} {{Probing astrophysical environment
  with eccentric extreme mass-ratio inspirals}},\ }\href
  {https://doi.org/10.1088/1475-7516/2024/01/035} {\bibfield  {journal}
  {\bibinfo  {journal} {JCAP}\ }\textbf {\bibinfo {volume} {01}},\ \bibinfo
  {pages} {035}},\ \Eprint {https://arxiv.org/abs/2306.14971} {arXiv:2306.14971
  [gr-qc]} \BibitemShut {NoStop}%
\bibitem [{\citenamefont {Rahman}\ and\ \citenamefont
  {Takahashi}(2025)}]{Rahman:2025mip}%
  \BibitemOpen
  \bibfield  {author} {\bibinfo {author} {\bibfnamefont {M.}~\bibnamefont
  {Rahman}}\ and\ \bibinfo {author} {\bibfnamefont {T.}~\bibnamefont
  {Takahashi}},\ }\href@noop {} {\bibinfo {title} {{Post-adiabatic waveforms
  from extreme mass ratio inspirals in the presence of dark matter}}} (\bibinfo
  {year} {2025}),\ \Eprint {https://arxiv.org/abs/2507.06923} {arXiv:2507.06923
  [gr-qc]} \BibitemShut {NoStop}%
\bibitem [{\citenamefont {Caneva~Santoro}\ \emph {et~al.}(2024)\citenamefont
  {Caneva~Santoro}, \citenamefont {Roy}, \citenamefont {Vicente}, \citenamefont
  {Haney}, \citenamefont {Piccinni}, \citenamefont {Del~Pozzo},\ and\
  \citenamefont {Martinez}}]{CanevaSantoro:2023aol}%
  \BibitemOpen
  \bibfield  {author} {\bibinfo {author} {\bibfnamefont {G.}~\bibnamefont
  {Caneva~Santoro}}, \bibinfo {author} {\bibfnamefont {S.}~\bibnamefont {Roy}},
  \bibinfo {author} {\bibfnamefont {R.}~\bibnamefont {Vicente}}, \bibinfo
  {author} {\bibfnamefont {M.}~\bibnamefont {Haney}}, \bibinfo {author}
  {\bibfnamefont {O.~J.}\ \bibnamefont {Piccinni}}, \bibinfo {author}
  {\bibfnamefont {W.}~\bibnamefont {Del~Pozzo}},\ and\ \bibinfo {author}
  {\bibfnamefont {M.}~\bibnamefont {Martinez}},\ }\bibfield  {title} {\bibinfo
  {title} {{First Constraints on Compact Binary Environments from LIGO-Virgo
  Data}},\ }\href {https://doi.org/10.1103/PhysRevLett.132.251401} {\bibfield
  {journal} {\bibinfo  {journal} {Phys. Rev. Lett.}\ }\textbf {\bibinfo
  {volume} {132}},\ \bibinfo {pages} {251401} (\bibinfo {year} {2024})},\
  \Eprint {https://arxiv.org/abs/2309.05061} {arXiv:2309.05061 [gr-qc]}
  \BibitemShut {NoStop}%
\bibitem [{\citenamefont {Zwick}\ \emph
  {et~al.}(2025{\natexlab{a}})\citenamefont {Zwick}, \citenamefont
  {Tak{\'a}tsy}, \citenamefont {Saini}, \citenamefont {Hendriks}, \citenamefont
  {Samsing}, \citenamefont {Tiede}, \citenamefont {Rowan},\ and\ \citenamefont
  {Trani}}]{Zwick:2025wkt}%
  \BibitemOpen
  \bibfield  {author} {\bibinfo {author} {\bibfnamefont {L.}~\bibnamefont
  {Zwick}}, \bibinfo {author} {\bibfnamefont {J.}~\bibnamefont {Tak{\'a}tsy}},
  \bibinfo {author} {\bibfnamefont {P.}~\bibnamefont {Saini}}, \bibinfo
  {author} {\bibfnamefont {K.}~\bibnamefont {Hendriks}}, \bibinfo {author}
  {\bibfnamefont {J.}~\bibnamefont {Samsing}}, \bibinfo {author} {\bibfnamefont
  {C.}~\bibnamefont {Tiede}}, \bibinfo {author} {\bibfnamefont
  {C.}~\bibnamefont {Rowan}},\ and\ \bibinfo {author} {\bibfnamefont {A.~A.}\
  \bibnamefont {Trani}},\ }\href@noop {} {\bibinfo {title} {{Environmental
  effects in stellar mass gravitational wave sources I: Expected fraction of
  signals with significant dephasing in the dynamical and AGN channels}}}
  (\bibinfo {year} {2025}{\natexlab{a}}),\ \Eprint
  {https://arxiv.org/abs/2503.24084} {arXiv:2503.24084 [astro-ph.HE]}
  \BibitemShut {NoStop}%
\bibitem [{\citenamefont {Zwick}\ \emph
  {et~al.}(2025{\natexlab{b}})\citenamefont {Zwick}, \citenamefont {Hendriks},
  \citenamefont {O'Neill}, \citenamefont {Tak{\'a}tsy}, \citenamefont
  {Kirkeberg}, \citenamefont {Tiede}, \citenamefont {Stegmann}, \citenamefont
  {Samsing},\ and\ \citenamefont {D'Orazio}}]{Zwick:2025ine}%
  \BibitemOpen
  \bibfield  {author} {\bibinfo {author} {\bibfnamefont {L.}~\bibnamefont
  {Zwick}}, \bibinfo {author} {\bibfnamefont {K.}~\bibnamefont {Hendriks}},
  \bibinfo {author} {\bibfnamefont {D.}~\bibnamefont {O'Neill}}, \bibinfo
  {author} {\bibfnamefont {J.}~\bibnamefont {Tak{\'a}tsy}}, \bibinfo {author}
  {\bibfnamefont {P.}~\bibnamefont {Kirkeberg}}, \bibinfo {author}
  {\bibfnamefont {C.}~\bibnamefont {Tiede}}, \bibinfo {author} {\bibfnamefont
  {J.}~\bibnamefont {Stegmann}}, \bibinfo {author} {\bibfnamefont
  {J.}~\bibnamefont {Samsing}},\ and\ \bibinfo {author} {\bibfnamefont {D.~J.}\
  \bibnamefont {D'Orazio}},\ }\href@noop {} {\bibinfo {title} {{Dissecting
  environmental effects with eccentric gravitational wave sources}}} (\bibinfo
  {year} {2025}{\natexlab{b}}),\ \Eprint {https://arxiv.org/abs/2506.09140}
  {arXiv:2506.09140 [astro-ph.HE]} \BibitemShut {NoStop}%
\bibitem [{\citenamefont {Destounis}\ and\ \citenamefont
  {Duque}(2023)}]{Destounis:2023ruj}%
  \BibitemOpen
  \bibfield  {author} {\bibinfo {author} {\bibfnamefont {K.}~\bibnamefont
  {Destounis}}\ and\ \bibinfo {author} {\bibfnamefont {F.}~\bibnamefont
  {Duque}},\ }\bibfield  {title} {\bibinfo {title} {{Black-hole spectroscopy:
  quasinormal modes, ringdown stability and the pseudospectrum}}\ }(\bibinfo
  {year} {2023})\ \Eprint {https://arxiv.org/abs/2308.16227} {arXiv:2308.16227
  [gr-qc]} \BibitemShut {NoStop}%
\bibitem [{\citenamefont {Jaramillo}\ \emph {et~al.}(2021)\citenamefont
  {Jaramillo}, \citenamefont {Panosso~Macedo},\ and\ \citenamefont
  {Al~Sheikh}}]{Jaramillo:2020tuu}%
  \BibitemOpen
  \bibfield  {author} {\bibinfo {author} {\bibfnamefont {J.~L.}\ \bibnamefont
  {Jaramillo}}, \bibinfo {author} {\bibfnamefont {R.}~\bibnamefont
  {Panosso~Macedo}},\ and\ \bibinfo {author} {\bibfnamefont {L.}~\bibnamefont
  {Al~Sheikh}},\ }\bibfield  {title} {\bibinfo {title} {{Pseudospectrum and
  Black Hole Quasinormal Mode Instability}},\ }\href
  {https://doi.org/10.1103/PhysRevX.11.031003} {\bibfield  {journal} {\bibinfo
  {journal} {Phys. Rev. X}\ }\textbf {\bibinfo {volume} {11}},\ \bibinfo
  {pages} {031003} (\bibinfo {year} {2021})},\ \Eprint
  {https://arxiv.org/abs/2004.06434} {arXiv:2004.06434 [gr-qc]} \BibitemShut
  {NoStop}%
\bibitem [{\citenamefont {Jaramillo}\ \emph {et~al.}(2022)\citenamefont
  {Jaramillo}, \citenamefont {Panosso~Macedo},\ and\ \citenamefont
  {Sheikh}}]{Jaramillo:2021tmt}%
  \BibitemOpen
  \bibfield  {author} {\bibinfo {author} {\bibfnamefont {J.~L.}\ \bibnamefont
  {Jaramillo}}, \bibinfo {author} {\bibfnamefont {R.}~\bibnamefont
  {Panosso~Macedo}},\ and\ \bibinfo {author} {\bibfnamefont {L.~A.}\
  \bibnamefont {Sheikh}},\ }\bibfield  {title} {\bibinfo {title}
  {{Gravitational Wave Signatures of Black Hole Quasinormal Mode
  Instability}},\ }\href {https://doi.org/10.1103/PhysRevLett.128.211102}
  {\bibfield  {journal} {\bibinfo  {journal} {Phys. Rev. Lett.}\ }\textbf
  {\bibinfo {volume} {128}},\ \bibinfo {pages} {211102} (\bibinfo {year}
  {2022})},\ \Eprint {https://arxiv.org/abs/2105.03451} {arXiv:2105.03451
  [gr-qc]} \BibitemShut {NoStop}%
\bibitem [{\citenamefont {Cheung}\ \emph {et~al.}(2022)\citenamefont {Cheung},
  \citenamefont {Destounis}, \citenamefont {Macedo}, \citenamefont {Berti},\
  and\ \citenamefont {Cardoso}}]{Cheung:2021bol}%
  \BibitemOpen
  \bibfield  {author} {\bibinfo {author} {\bibfnamefont {M.~H.-Y.}\
  \bibnamefont {Cheung}}, \bibinfo {author} {\bibfnamefont {K.}~\bibnamefont
  {Destounis}}, \bibinfo {author} {\bibfnamefont {R.~P.}\ \bibnamefont
  {Macedo}}, \bibinfo {author} {\bibfnamefont {E.}~\bibnamefont {Berti}},\ and\
  \bibinfo {author} {\bibfnamefont {V.}~\bibnamefont {Cardoso}},\ }\bibfield
  {title} {\bibinfo {title} {{Destabilizing the Fundamental Mode of Black
  Holes: The Elephant and the Flea}},\ }\href
  {https://doi.org/10.1103/PhysRevLett.128.111103} {\bibfield  {journal}
  {\bibinfo  {journal} {Phys. Rev. Lett.}\ }\textbf {\bibinfo {volume} {128}},\
  \bibinfo {pages} {111103} (\bibinfo {year} {2022})},\ \Eprint
  {https://arxiv.org/abs/2111.05415} {arXiv:2111.05415 [gr-qc]} \BibitemShut
  {NoStop}%
\bibitem [{\citenamefont {Destounis}\ \emph
  {et~al.}(2021{\natexlab{a}})\citenamefont {Destounis}, \citenamefont
  {Macedo}, \citenamefont {Berti}, \citenamefont {Cardoso},\ and\ \citenamefont
  {Jaramillo}}]{Destounis:2021lum}%
  \BibitemOpen
  \bibfield  {author} {\bibinfo {author} {\bibfnamefont {K.}~\bibnamefont
  {Destounis}}, \bibinfo {author} {\bibfnamefont {R.~P.}\ \bibnamefont
  {Macedo}}, \bibinfo {author} {\bibfnamefont {E.}~\bibnamefont {Berti}},
  \bibinfo {author} {\bibfnamefont {V.}~\bibnamefont {Cardoso}},\ and\ \bibinfo
  {author} {\bibfnamefont {J.~L.}\ \bibnamefont {Jaramillo}},\ }\bibfield
  {title} {\bibinfo {title} {{Pseudospectrum of Reissner-Nordstr\"om black
  holes: Quasinormal mode instability and universality}},\ }\href
  {https://doi.org/10.1103/PhysRevD.104.084091} {\bibfield  {journal} {\bibinfo
   {journal} {Phys. Rev. D}\ }\textbf {\bibinfo {volume} {104}},\ \bibinfo
  {pages} {084091} (\bibinfo {year} {2021}{\natexlab{a}})},\ \Eprint
  {https://arxiv.org/abs/2107.09673} {arXiv:2107.09673 [gr-qc]} \BibitemShut
  {NoStop}%
\bibitem [{\citenamefont {Boyanov}\ \emph {et~al.}(2023)\citenamefont
  {Boyanov}, \citenamefont {Destounis}, \citenamefont {Panosso~Macedo},
  \citenamefont {Cardoso},\ and\ \citenamefont {Jaramillo}}]{Boyanov:2022ark}%
  \BibitemOpen
  \bibfield  {author} {\bibinfo {author} {\bibfnamefont {V.}~\bibnamefont
  {Boyanov}}, \bibinfo {author} {\bibfnamefont {K.}~\bibnamefont {Destounis}},
  \bibinfo {author} {\bibfnamefont {R.}~\bibnamefont {Panosso~Macedo}},
  \bibinfo {author} {\bibfnamefont {V.}~\bibnamefont {Cardoso}},\ and\ \bibinfo
  {author} {\bibfnamefont {J.~L.}\ \bibnamefont {Jaramillo}},\ }\bibfield
  {title} {\bibinfo {title} {{Pseudospectrum of horizonless compact objects: A
  bootstrap instability mechanism}},\ }\href
  {https://doi.org/10.1103/PhysRevD.107.064012} {\bibfield  {journal} {\bibinfo
   {journal} {Phys. Rev. D}\ }\textbf {\bibinfo {volume} {107}},\ \bibinfo
  {pages} {064012} (\bibinfo {year} {2023})},\ \Eprint
  {https://arxiv.org/abs/2209.12950} {arXiv:2209.12950 [gr-qc]} \BibitemShut
  {NoStop}%
\bibitem [{\citenamefont {Sarkar}\ \emph {et~al.}(2023)\citenamefont {Sarkar},
  \citenamefont {Rahman},\ and\ \citenamefont {Chakraborty}}]{Sarkar:2023rhp}%
  \BibitemOpen
  \bibfield  {author} {\bibinfo {author} {\bibfnamefont {S.}~\bibnamefont
  {Sarkar}}, \bibinfo {author} {\bibfnamefont {M.}~\bibnamefont {Rahman}},\
  and\ \bibinfo {author} {\bibfnamefont {S.}~\bibnamefont {Chakraborty}},\
  }\bibfield  {title} {\bibinfo {title} {{Perturbing the perturbed: Stability
  of quasinormal modes in presence of a positive cosmological constant}},\
  }\href {https://doi.org/10.1103/PhysRevD.108.104002} {\bibfield  {journal}
  {\bibinfo  {journal} {Phys. Rev. D}\ }\textbf {\bibinfo {volume} {108}},\
  \bibinfo {pages} {104002} (\bibinfo {year} {2023})},\ \Eprint
  {https://arxiv.org/abs/2304.06829} {arXiv:2304.06829 [gr-qc]} \BibitemShut
  {NoStop}%
\bibitem [{\citenamefont {Destounis}\ \emph {et~al.}(2024)\citenamefont
  {Destounis}, \citenamefont {Boyanov},\ and\ \citenamefont
  {Panosso~Macedo}}]{Destounis:2023nmb}%
  \BibitemOpen
  \bibfield  {author} {\bibinfo {author} {\bibfnamefont {K.}~\bibnamefont
  {Destounis}}, \bibinfo {author} {\bibfnamefont {V.}~\bibnamefont {Boyanov}},\
  and\ \bibinfo {author} {\bibfnamefont {R.}~\bibnamefont {Panosso~Macedo}},\
  }\bibfield  {title} {\bibinfo {title} {{Pseudospectrum of de Sitter black
  holes}},\ }\href {https://doi.org/10.1103/PhysRevD.109.044023} {\bibfield
  {journal} {\bibinfo  {journal} {Phys. Rev. D}\ }\textbf {\bibinfo {volume}
  {109}},\ \bibinfo {pages} {044023} (\bibinfo {year} {2024})},\ \Eprint
  {https://arxiv.org/abs/2312.11630} {arXiv:2312.11630 [gr-qc]} \BibitemShut
  {NoStop}%
\bibitem [{\citenamefont {Boyanov}\ \emph {et~al.}(2024)\citenamefont
  {Boyanov}, \citenamefont {Cardoso}, \citenamefont {Destounis}, \citenamefont
  {Jaramillo},\ and\ \citenamefont {Panosso~Macedo}}]{Boyanov:2023qqf}%
  \BibitemOpen
  \bibfield  {author} {\bibinfo {author} {\bibfnamefont {V.}~\bibnamefont
  {Boyanov}}, \bibinfo {author} {\bibfnamefont {V.}~\bibnamefont {Cardoso}},
  \bibinfo {author} {\bibfnamefont {K.}~\bibnamefont {Destounis}}, \bibinfo
  {author} {\bibfnamefont {J.~L.}\ \bibnamefont {Jaramillo}},\ and\ \bibinfo
  {author} {\bibfnamefont {R.}~\bibnamefont {Panosso~Macedo}},\ }\bibfield
  {title} {\bibinfo {title} {{Structural aspects of the anti{\textendash}de
  Sitter black hole pseudospectrum}},\ }\href
  {https://doi.org/10.1103/PhysRevD.109.064068} {\bibfield  {journal} {\bibinfo
   {journal} {Phys. Rev. D}\ }\textbf {\bibinfo {volume} {109}},\ \bibinfo
  {pages} {064068} (\bibinfo {year} {2024})},\ \Eprint
  {https://arxiv.org/abs/2312.11998} {arXiv:2312.11998 [gr-qc]} \BibitemShut
  {NoStop}%
\bibitem [{\citenamefont {Are{\'a}n}\ \emph {et~al.}(2023)\citenamefont
  {Are{\'a}n}, \citenamefont {Fari{\~n}a},\ and\ \citenamefont
  {Landsteiner}}]{Arean:2023ejh}%
  \BibitemOpen
  \bibfield  {author} {\bibinfo {author} {\bibfnamefont {D.}~\bibnamefont
  {Are{\'a}n}}, \bibinfo {author} {\bibfnamefont {D.~G.}\ \bibnamefont
  {Fari{\~n}a}},\ and\ \bibinfo {author} {\bibfnamefont {K.}~\bibnamefont
  {Landsteiner}},\ }\bibfield  {title} {\bibinfo {title} {{Pseudospectra of
  holographic quasinormal modes}},\ }\href
  {https://doi.org/10.1007/JHEP12(2023)187} {\bibfield  {journal} {\bibinfo
  {journal} {JHEP}\ }\textbf {\bibinfo {volume} {12}},\ \bibinfo {pages}
  {187}},\ \Eprint {https://arxiv.org/abs/2307.08751} {arXiv:2307.08751
  [hep-th]} \BibitemShut {NoStop}%
\bibitem [{\citenamefont {Cownden}\ \emph {et~al.}(2024)\citenamefont
  {Cownden}, \citenamefont {Pantelidou},\ and\ \citenamefont
  {Zilh{\~a}o}}]{Cownden:2023dam}%
  \BibitemOpen
  \bibfield  {author} {\bibinfo {author} {\bibfnamefont {B.}~\bibnamefont
  {Cownden}}, \bibinfo {author} {\bibfnamefont {C.}~\bibnamefont
  {Pantelidou}},\ and\ \bibinfo {author} {\bibfnamefont {M.}~\bibnamefont
  {Zilh{\~a}o}},\ }\bibfield  {title} {\bibinfo {title} {{The pseudospectra of
  black holes in AdS}},\ }\href {https://doi.org/10.1007/JHEP05(2024)202}
  {\bibfield  {journal} {\bibinfo  {journal} {JHEP}\ }\textbf {\bibinfo
  {volume} {05}},\ \bibinfo {pages} {202}},\ \Eprint
  {https://arxiv.org/abs/2312.08352} {arXiv:2312.08352 [gr-qc]} \BibitemShut
  {NoStop}%
\bibitem [{\citenamefont {Courty}\ \emph {et~al.}(2023)\citenamefont {Courty},
  \citenamefont {Destounis},\ and\ \citenamefont {Pani}}]{Courty:2023rxk}%
  \BibitemOpen
  \bibfield  {author} {\bibinfo {author} {\bibfnamefont {A.}~\bibnamefont
  {Courty}}, \bibinfo {author} {\bibfnamefont {K.}~\bibnamefont {Destounis}},\
  and\ \bibinfo {author} {\bibfnamefont {P.}~\bibnamefont {Pani}},\ }\bibfield
  {title} {\bibinfo {title} {{Spectral instability of quasinormal modes and
  strong cosmic censorship}},\ }\href
  {https://doi.org/10.1103/PhysRevD.108.104027} {\bibfield  {journal} {\bibinfo
   {journal} {Phys. Rev. D}\ }\textbf {\bibinfo {volume} {108}},\ \bibinfo
  {pages} {104027} (\bibinfo {year} {2023})},\ \Eprint
  {https://arxiv.org/abs/2307.11155} {arXiv:2307.11155 [gr-qc]} \BibitemShut
  {NoStop}%
\bibitem [{\citenamefont {Rosato}\ \emph {et~al.}(2024)\citenamefont {Rosato},
  \citenamefont {Destounis},\ and\ \citenamefont {Pani}}]{Rosato:2024arw}%
  \BibitemOpen
  \bibfield  {author} {\bibinfo {author} {\bibfnamefont {R.~F.}\ \bibnamefont
  {Rosato}}, \bibinfo {author} {\bibfnamefont {K.}~\bibnamefont {Destounis}},\
  and\ \bibinfo {author} {\bibfnamefont {P.}~\bibnamefont {Pani}},\ }\bibfield
  {title} {\bibinfo {title} {{Ringdown stability: Graybody factors as stable
  gravitational-wave observables}},\ }\href
  {https://doi.org/10.1103/PhysRevD.110.L121501} {\bibfield  {journal}
  {\bibinfo  {journal} {Phys. Rev. D}\ }\textbf {\bibinfo {volume} {110}},\
  \bibinfo {pages} {L121501} (\bibinfo {year} {2024})},\ \Eprint
  {https://arxiv.org/abs/2406.01692} {arXiv:2406.01692 [gr-qc]} \BibitemShut
  {NoStop}%
\bibitem [{\citenamefont {Boyanov}(2024)}]{Boyanov:2024fgc}%
  \BibitemOpen
  \bibfield  {author} {\bibinfo {author} {\bibfnamefont {V.}~\bibnamefont
  {Boyanov}},\ }\bibfield  {title} {\bibinfo {title} {{On destabilising
  quasi-normal modes with a radially concentrated perturbation}},\ }\href
  {https://doi.org/10.3389/fphy.2024.1511757} {\bibfield  {journal} {\bibinfo
  {journal} {Front. in Phys.}\ }\textbf {\bibinfo {volume} {12}},\ \bibinfo
  {pages} {1511757} (\bibinfo {year} {2024})},\ \Eprint
  {https://arxiv.org/abs/2410.11547} {arXiv:2410.11547 [gr-qc]} \BibitemShut
  {NoStop}%
\bibitem [{\citenamefont {Cai}\ \emph {et~al.}(2025)\citenamefont {Cai},
  \citenamefont {Cao}, \citenamefont {Chen}, \citenamefont {Guo}, \citenamefont
  {Wu},\ and\ \citenamefont {Zhou}}]{Cai:2025irl}%
  \BibitemOpen
  \bibfield  {author} {\bibinfo {author} {\bibfnamefont {R.-G.}\ \bibnamefont
  {Cai}}, \bibinfo {author} {\bibfnamefont {L.-M.}\ \bibnamefont {Cao}},
  \bibinfo {author} {\bibfnamefont {J.-N.}\ \bibnamefont {Chen}}, \bibinfo
  {author} {\bibfnamefont {Z.-K.}\ \bibnamefont {Guo}}, \bibinfo {author}
  {\bibfnamefont {L.-B.}\ \bibnamefont {Wu}},\ and\ \bibinfo {author}
  {\bibfnamefont {Y.-S.}\ \bibnamefont {Zhou}},\ }\bibfield  {title} {\bibinfo
  {title} {{Pseudospectrum for the Kerr black hole with spin s=0 case}},\
  }\href {https://doi.org/10.1103/PhysRevD.111.084011} {\bibfield  {journal}
  {\bibinfo  {journal} {Phys. Rev. D}\ }\textbf {\bibinfo {volume} {111}},\
  \bibinfo {pages} {084011} (\bibinfo {year} {2025})},\ \Eprint
  {https://arxiv.org/abs/2501.02522} {arXiv:2501.02522 [gr-qc]} \BibitemShut
  {NoStop}%
\bibitem [{\citenamefont {Spieksma}\ \emph {et~al.}(2025)\citenamefont
  {Spieksma}, \citenamefont {Cardoso}, \citenamefont {Carullo}, \citenamefont
  {Della~Rocca},\ and\ \citenamefont {Duque}}]{Spieksma:2024voy}%
  \BibitemOpen
  \bibfield  {author} {\bibinfo {author} {\bibfnamefont {T.~F.~M.}\
  \bibnamefont {Spieksma}}, \bibinfo {author} {\bibfnamefont {V.}~\bibnamefont
  {Cardoso}}, \bibinfo {author} {\bibfnamefont {G.}~\bibnamefont {Carullo}},
  \bibinfo {author} {\bibfnamefont {M.}~\bibnamefont {Della~Rocca}},\ and\
  \bibinfo {author} {\bibfnamefont {F.}~\bibnamefont {Duque}},\ }\bibfield
  {title} {\bibinfo {title} {{Black Hole Spectroscopy in Environments:
  Detectability Prospects}},\ }\href
  {https://doi.org/10.1103/PhysRevLett.134.081402} {\bibfield  {journal}
  {\bibinfo  {journal} {Phys. Rev. Lett.}\ }\textbf {\bibinfo {volume} {134}},\
  \bibinfo {pages} {081402} (\bibinfo {year} {2025})},\ \Eprint
  {https://arxiv.org/abs/2409.05950} {arXiv:2409.05950 [gr-qc]} \BibitemShut
  {NoStop}%
\bibitem [{\citenamefont {Datta}(2024)}]{Datta:2023zmd}%
  \BibitemOpen
  \bibfield  {author} {\bibinfo {author} {\bibfnamefont {S.}~\bibnamefont
  {Datta}},\ }\bibfield  {title} {\bibinfo {title} {{Black holes immersed in
  dark matter: Energy condition and sound speed}},\ }\href
  {https://doi.org/10.1103/PhysRevD.109.104042} {\bibfield  {journal} {\bibinfo
   {journal} {Phys. Rev. D}\ }\textbf {\bibinfo {volume} {109}},\ \bibinfo
  {pages} {104042} (\bibinfo {year} {2024})},\ \Eprint
  {https://arxiv.org/abs/2312.01277} {arXiv:2312.01277 [gr-qc]} \BibitemShut
  {NoStop}%
\bibitem [{\citenamefont {Datta}\ and\ \citenamefont
  {Maselli}(2025)}]{Datta:2025ruh}%
  \BibitemOpen
  \bibfield  {author} {\bibinfo {author} {\bibfnamefont {S.}~\bibnamefont
  {Datta}}\ and\ \bibinfo {author} {\bibfnamefont {A.}~\bibnamefont
  {Maselli}},\ }\href@noop {} {\bibinfo {title} {{A multi-parameter expansion
  for the evolution of asymmetric binaries in astrophysical environments}}}
  (\bibinfo {year} {2025}),\ \Eprint {https://arxiv.org/abs/2507.04471}
  {arXiv:2507.04471 [gr-qc]} \BibitemShut {NoStop}%
\bibitem [{\citenamefont {Pease}(1916)}]{Pease:1916}%
  \BibitemOpen
  \bibfield  {author} {\bibinfo {author} {\bibfnamefont {F.~G.}\ \bibnamefont
  {Pease}},\ }\bibfield  {title} {\bibinfo {title} {The rotation and radial
  velocity of the spiral nebula ngcc4594},\ }\href
  {https://doi.org/10.1088/122533a} {\bibfield  {journal} {\bibinfo  {journal}
  {Publications of the Astronomical Society of the Pacific}\ }\textbf {\bibinfo
  {volume} {28}},\ \bibinfo {pages} {191} (\bibinfo {year} {1916})}\BibitemShut
  {NoStop}%
\bibitem [{\citenamefont {Pease}(1918)}]{Pease:1918}%
  \BibitemOpen
  \bibfield  {author} {\bibinfo {author} {\bibfnamefont {F.~G.}\ \bibnamefont
  {Pease}},\ }\bibfield  {title} {\bibinfo {title} {The rotation and radial
  velocity of the central part of the andromeda nebula},\ }\href
  {https://doi.org/10.1073/pnas.4.1.21} {\bibfield  {journal} {\bibinfo
  {journal} {Proceedings of the National Academy of Sciences}\ }\textbf
  {\bibinfo {volume} {4}},\ \bibinfo {pages} {21} (\bibinfo {year} {1918})},\
  \Eprint
  {https://arxiv.org/abs/https://www.pnas.org/doi/pdf/10.1073/pnas.4.1.21}
  {https://www.pnas.org/doi/pdf/10.1073/pnas.4.1.21} \BibitemShut {NoStop}%
\bibitem [{\citenamefont {{Schwarzschild}}(1954)}]{Schwarzschild:1954}%
  \BibitemOpen
  \bibfield  {author} {\bibinfo {author} {\bibfnamefont {M.}~\bibnamefont
  {{Schwarzschild}}},\ }\bibfield  {title} {\bibinfo {title} {{Mass
  distribution and mass-luminosity ratio in galaxies}},\ }\href
  {https://doi.org/10.1086/107013} {\bibfield  {journal} {\bibinfo  {journal}
  {The Astronomical Journal}\ }\textbf {\bibinfo {volume} {59}},\ \bibinfo
  {pages} {273} (\bibinfo {year} {1954})}\BibitemShut {NoStop}%
\bibitem [{\citenamefont {Lindblad}(1959)}]{Lindblad:1959}%
  \BibitemOpen
  \bibfield  {author} {\bibinfo {author} {\bibfnamefont {B.}~\bibnamefont
  {Lindblad}},\ }\bibinfo {title} {Galactic dynamics},\ in\ \href
  {https://doi.org/10.1007/978-3-642-45932-0_2} {\emph {\bibinfo {booktitle}
  {Astrophysik IV: Sternsysteme / Astrophysics IV: Stellar Systems}}},\
  \bibinfo {editor} {edited by\ \bibinfo {editor} {\bibfnamefont
  {S.}~\bibnamefont {Fl{\"u}gge}}}\ (\bibinfo  {publisher} {Springer Berlin
  Heidelberg},\ \bibinfo {address} {Berlin, Heidelberg},\ \bibinfo {year}
  {1959})\ pp.\ \bibinfo {pages} {21--99}\BibitemShut {NoStop}%
\bibitem [{\citenamefont {{de Vaucouleurs}}(1959)}]{deVaucouleurs:1959}%
  \BibitemOpen
  \bibfield  {author} {\bibinfo {author} {\bibfnamefont {G.}~\bibnamefont {{de
  Vaucouleurs}}},\ }\bibfield  {title} {\bibinfo {title} {{Classification and
  Morphology of External Galaxies.}},\ }\href
  {https://doi.org/10.1007/978-3-642-45932-0_7} {\bibfield  {journal} {\bibinfo
   {journal} {Handbuch der Physik}\ }\textbf {\bibinfo {volume} {53}},\
  \bibinfo {pages} {275} (\bibinfo {year} {1959})}\BibitemShut {NoStop}%
\bibitem [{\citenamefont {{de Vaucouleurs}}\ and\ \citenamefont
  {Freeman}(1972)}]{deVaucouleurs:1972}%
  \BibitemOpen
  \bibfield  {author} {\bibinfo {author} {\bibfnamefont {G.}~\bibnamefont {{de
  Vaucouleurs}}}\ and\ \bibinfo {author} {\bibfnamefont {K.}~\bibnamefont
  {Freeman}},\ }\bibfield  {title} {\bibinfo {title} {Structure and dynamics of
  barred spiral galaxies, in particular of the magellanic type},\ }\href
  {https://doi.org/https://doi.org/10.1016/0083-6656(72)90026-8} {\bibfield
  {journal} {\bibinfo  {journal} {Vistas in Astronomy}\ }\textbf {\bibinfo
  {volume} {14}},\ \bibinfo {pages} {163} (\bibinfo {year} {1972})}\BibitemShut
  {NoStop}%
\bibitem [{\citenamefont {{Burbidge}}(1975)}]{Burbidge:1975}%
  \BibitemOpen
  \bibfield  {author} {\bibinfo {author} {\bibfnamefont {G.}~\bibnamefont
  {{Burbidge}}},\ }\bibfield  {title} {\bibinfo {title} {{On the masses and
  relative velocities of galaxies.}},\ }\href {https://doi.org/10.1086/181731}
  {\bibfield  {journal} {\bibinfo  {journal} {The Astrophysical Journal
  Letters}\ }\textbf {\bibinfo {volume} {196}},\ \bibinfo {pages} {L7}
  (\bibinfo {year} {1975})}\BibitemShut {NoStop}%
\bibitem [{\citenamefont {{van der Kruit}}\ and\ \citenamefont
  {{Bosma}}(1978)}]{vanderKruit:1978}%
  \BibitemOpen
  \bibfield  {author} {\bibinfo {author} {\bibfnamefont {P.~C.}\ \bibnamefont
  {{van der Kruit}}}\ and\ \bibinfo {author} {\bibfnamefont {A.}~\bibnamefont
  {{Bosma}}},\ }\bibfield  {title} {\bibinfo {title} {{The rotation curves and
  orientation parameters of the spiral galaxies NGC 2715, 5033 and 5055.}},\
  }\href@noop {} {\bibfield  {journal} {\bibinfo  {journal} {Astronomy and
  Astrophysics}\ }\textbf {\bibinfo {volume} {34}},\ \bibinfo {pages} {259}
  (\bibinfo {year} {1978})}\BibitemShut {NoStop}%
\bibitem [{\citenamefont {Sofue}\ and\ \citenamefont
  {Rubin}(2001)}]{Sofue:2000jx}%
  \BibitemOpen
  \bibfield  {author} {\bibinfo {author} {\bibfnamefont {Y.}~\bibnamefont
  {Sofue}}\ and\ \bibinfo {author} {\bibfnamefont {V.}~\bibnamefont {Rubin}},\
  }\bibfield  {title} {\bibinfo {title} {{Rotation curves of spiral
  galaxies}},\ }\href {https://doi.org/10.1146/annurev.astro.39.1.137}
  {\bibfield  {journal} {\bibinfo  {journal} {Ann. Rev. Astron. Astrophys.}\
  }\textbf {\bibinfo {volume} {39}},\ \bibinfo {pages} {137} (\bibinfo {year}
  {2001})},\ \Eprint {https://arxiv.org/abs/astro-ph/0010594}
  {arXiv:astro-ph/0010594} \BibitemShut {NoStop}%
\bibitem [{\citenamefont {{Canizares}}\ \emph {et~al.}(1979)\citenamefont
  {{Canizares}}, \citenamefont {{Clark}}, \citenamefont {{Markert}},
  \citenamefont {{Berg}}, \citenamefont {{Smedira}}, \citenamefont {{Bardas}},
  \citenamefont {{Schnopper}},\ and\ \citenamefont
  {{Kalata}}}]{Canizares:1979}%
  \BibitemOpen
  \bibfield  {author} {\bibinfo {author} {\bibfnamefont {C.~R.}\ \bibnamefont
  {{Canizares}}}, \bibinfo {author} {\bibfnamefont {G.~W.}\ \bibnamefont
  {{Clark}}}, \bibinfo {author} {\bibfnamefont {T.~H.}\ \bibnamefont
  {{Markert}}}, \bibinfo {author} {\bibfnamefont {C.}~\bibnamefont {{Berg}}},
  \bibinfo {author} {\bibfnamefont {M.}~\bibnamefont {{Smedira}}}, \bibinfo
  {author} {\bibfnamefont {D.}~\bibnamefont {{Bardas}}}, \bibinfo {author}
  {\bibfnamefont {H.}~\bibnamefont {{Schnopper}}},\ and\ \bibinfo {author}
  {\bibfnamefont {K.}~\bibnamefont {{Kalata}}},\ }\bibfield  {title} {\bibinfo
  {title} {{High-resolution X-ray spectroscopy of M87 with the Einstein
  Observatory: the detection of an O VIII emission line.}},\ }\href
  {https://doi.org/10.1086/183104} {\bibfield  {journal} {\bibinfo  {journal}
  {The Astrophysical Journal Letters}\ }\textbf {\bibinfo {volume} {234}},\
  \bibinfo {pages} {L33} (\bibinfo {year} {1979})}\BibitemShut {NoStop}%
\bibitem [{\citenamefont {{Canizares}}\ \emph {et~al.}(1982)\citenamefont
  {{Canizares}}, \citenamefont {{Clark}}, \citenamefont {{Jernigan}},\ and\
  \citenamefont {{Markert}}}]{Canizares:1982}%
  \BibitemOpen
  \bibfield  {author} {\bibinfo {author} {\bibfnamefont {C.~R.}\ \bibnamefont
  {{Canizares}}}, \bibinfo {author} {\bibfnamefont {G.~W.}\ \bibnamefont
  {{Clark}}}, \bibinfo {author} {\bibfnamefont {J.~G.}\ \bibnamefont
  {{Jernigan}}},\ and\ \bibinfo {author} {\bibfnamefont {T.~H.}\ \bibnamefont
  {{Markert}}},\ }\bibfield  {title} {\bibinfo {title} {{X-ray spectroscopy of
  the galaxy M 87: radiative accretion of the hot plasma halo.}},\ }\href
  {https://doi.org/10.1086/160393} {\bibfield  {journal} {\bibinfo  {journal}
  {\apj}\ }\textbf {\bibinfo {volume} {262}},\ \bibinfo {pages} {33} (\bibinfo
  {year} {1982})}\BibitemShut {NoStop}%
\bibitem [{\citenamefont {{Boughn}}\ and\ \citenamefont
  {{Saulson}}(1983)}]{Boughn:1983}%
  \BibitemOpen
  \bibfield  {author} {\bibinfo {author} {\bibfnamefont {S.~P.}\ \bibnamefont
  {{Boughn}}}\ and\ \bibinfo {author} {\bibfnamefont {P.~R.}\ \bibnamefont
  {{Saulson}}},\ }\bibfield  {title} {\bibinfo {title} {{Infrared photometry of
  the halo of M 87.}},\ }\href {https://doi.org/10.1086/183957} {\bibfield
  {journal} {\bibinfo  {journal} {The Astrophysical Journal Letters}\ }\textbf
  {\bibinfo {volume} {265}},\ \bibinfo {pages} {L55} (\bibinfo {year}
  {1983})}\BibitemShut {NoStop}%
\bibitem [{\citenamefont {Akiyama}\ \emph {et~al.}(2019)\citenamefont {Akiyama}
  \emph {et~al.}}]{EventHorizonTelescope:2019dse}%
  \BibitemOpen
  \bibfield  {author} {\bibinfo {author} {\bibfnamefont {K.}~\bibnamefont
  {Akiyama}} \emph {et~al.} (\bibinfo {collaboration} {Event Horizon
  Telescope}),\ }\bibfield  {title} {\bibinfo {title} {{First M87 Event Horizon
  Telescope Results. I. The Shadow of the Supermassive Black Hole}},\ }\href
  {https://doi.org/10.3847/2041-8213/ab0ec7} {\bibfield  {journal} {\bibinfo
  {journal} {Astrophys. J. Lett.}\ }\textbf {\bibinfo {volume} {875}},\
  \bibinfo {pages} {L1} (\bibinfo {year} {2019})},\ \Eprint
  {https://arxiv.org/abs/1906.11238} {arXiv:1906.11238 [astro-ph.GA]}
  \BibitemShut {NoStop}%
\bibitem [{\citenamefont {Akiyama}\ \emph {et~al.}(2022)\citenamefont {Akiyama}
  \emph {et~al.}}]{EventHorizonTelescope:2022wkp}%
  \BibitemOpen
  \bibfield  {author} {\bibinfo {author} {\bibfnamefont {K.}~\bibnamefont
  {Akiyama}} \emph {et~al.} (\bibinfo {collaboration} {Event Horizon
  Telescope}),\ }\bibfield  {title} {\bibinfo {title} {{First Sagittarius A*
  Event Horizon Telescope Results. I. The Shadow of the Supermassive Black Hole
  in the Center of the Milky Way}},\ }\href
  {https://doi.org/10.3847/2041-8213/ac6674} {\bibfield  {journal} {\bibinfo
  {journal} {Astrophys. J. Lett.}\ }\textbf {\bibinfo {volume} {930}},\
  \bibinfo {pages} {L12} (\bibinfo {year} {2022})}\BibitemShut {NoStop}%
\bibitem [{\citenamefont {Taylor}\ and\ \citenamefont
  {Poisson}(2008)}]{Taylor:2008xy}%
  \BibitemOpen
  \bibfield  {author} {\bibinfo {author} {\bibfnamefont {S.}~\bibnamefont
  {Taylor}}\ and\ \bibinfo {author} {\bibfnamefont {E.}~\bibnamefont
  {Poisson}},\ }\bibfield  {title} {\bibinfo {title} {{Nonrotating black hole
  in a post-Newtonian tidal environment}},\ }\href
  {https://doi.org/10.1103/PhysRevD.78.084016} {\bibfield  {journal} {\bibinfo
  {journal} {Phys. Rev. D}\ }\textbf {\bibinfo {volume} {78}},\ \bibinfo
  {pages} {084016} (\bibinfo {year} {2008})},\ \Eprint
  {https://arxiv.org/abs/0806.3052} {arXiv:0806.3052 [gr-qc]} \BibitemShut
  {NoStop}%
\bibitem [{\citenamefont {Poisson}\ and\ \citenamefont
  {Corrigan}(2018)}]{Poisson:2018qqd}%
  \BibitemOpen
  \bibfield  {author} {\bibinfo {author} {\bibfnamefont {E.}~\bibnamefont
  {Poisson}}\ and\ \bibinfo {author} {\bibfnamefont {E.}~\bibnamefont
  {Corrigan}},\ }\bibfield  {title} {\bibinfo {title} {{Nonrotating black hole
  in a post-Newtonian tidal environment II}},\ }\href
  {https://doi.org/10.1103/PhysRevD.97.124048} {\bibfield  {journal} {\bibinfo
  {journal} {Phys. Rev. D}\ }\textbf {\bibinfo {volume} {97}},\ \bibinfo
  {pages} {124048} (\bibinfo {year} {2018})},\ \Eprint
  {https://arxiv.org/abs/1804.01848} {arXiv:1804.01848 [gr-qc]} \BibitemShut
  {NoStop}%
\bibitem [{\citenamefont {Eda}\ \emph {et~al.}(2013)\citenamefont {Eda},
  \citenamefont {Itoh}, \citenamefont {Kuroyanagi},\ and\ \citenamefont
  {Silk}}]{Eda:2013gg}%
  \BibitemOpen
  \bibfield  {author} {\bibinfo {author} {\bibfnamefont {K.}~\bibnamefont
  {Eda}}, \bibinfo {author} {\bibfnamefont {Y.}~\bibnamefont {Itoh}}, \bibinfo
  {author} {\bibfnamefont {S.}~\bibnamefont {Kuroyanagi}},\ and\ \bibinfo
  {author} {\bibfnamefont {J.}~\bibnamefont {Silk}},\ }\bibfield  {title}
  {\bibinfo {title} {{New Probe of Dark-Matter Properties: Gravitational Waves
  from an Intermediate-Mass Black Hole Embedded in a Dark-Matter Minispike}},\
  }\href {https://doi.org/10.1103/PhysRevLett.110.221101} {\bibfield  {journal}
  {\bibinfo  {journal} {Phys. Rev. Lett.}\ }\textbf {\bibinfo {volume} {110}},\
  \bibinfo {pages} {221101} (\bibinfo {year} {2013})},\ \Eprint
  {https://arxiv.org/abs/1301.5971} {arXiv:1301.5971 [gr-qc]} \BibitemShut
  {NoStop}%
\bibitem [{\citenamefont {Tamanini}\ \emph {et~al.}(2020)\citenamefont
  {Tamanini}, \citenamefont {Klein}, \citenamefont {Bonvin}, \citenamefont
  {Barausse},\ and\ \citenamefont {Caprini}}]{Tamanini:2019usx}%
  \BibitemOpen
  \bibfield  {author} {\bibinfo {author} {\bibfnamefont {N.}~\bibnamefont
  {Tamanini}}, \bibinfo {author} {\bibfnamefont {A.}~\bibnamefont {Klein}},
  \bibinfo {author} {\bibfnamefont {C.}~\bibnamefont {Bonvin}}, \bibinfo
  {author} {\bibfnamefont {E.}~\bibnamefont {Barausse}},\ and\ \bibinfo
  {author} {\bibfnamefont {C.}~\bibnamefont {Caprini}},\ }\bibfield  {title}
  {\bibinfo {title} {{Peculiar acceleration of stellar-origin black hole
  binaries: Measurement and biases with LISA}},\ }\href
  {https://doi.org/10.1103/PhysRevD.101.063002} {\bibfield  {journal} {\bibinfo
   {journal} {Phys. Rev. D}\ }\textbf {\bibinfo {volume} {101}},\ \bibinfo
  {pages} {063002} (\bibinfo {year} {2020})},\ \Eprint
  {https://arxiv.org/abs/1907.02018} {arXiv:1907.02018 [astro-ph.IM]}
  \BibitemShut {NoStop}%
\bibitem [{\citenamefont {Kavanagh}\ \emph {et~al.}(2020)\citenamefont
  {Kavanagh}, \citenamefont {Nichols}, \citenamefont {Bertone},\ and\
  \citenamefont {Gaggero}}]{Kavanagh:2020cfn}%
  \BibitemOpen
  \bibfield  {author} {\bibinfo {author} {\bibfnamefont {B.~J.}\ \bibnamefont
  {Kavanagh}}, \bibinfo {author} {\bibfnamefont {D.~A.}\ \bibnamefont
  {Nichols}}, \bibinfo {author} {\bibfnamefont {G.}~\bibnamefont {Bertone}},\
  and\ \bibinfo {author} {\bibfnamefont {D.}~\bibnamefont {Gaggero}},\
  }\bibfield  {title} {\bibinfo {title} {{Detecting dark matter around black
  holes with gravitational waves: Effects of dark-matter dynamics on the
  gravitational waveform}},\ }\href
  {https://doi.org/10.1103/PhysRevD.102.083006} {\bibfield  {journal} {\bibinfo
   {journal} {Phys. Rev. D}\ }\textbf {\bibinfo {volume} {102}},\ \bibinfo
  {pages} {083006} (\bibinfo {year} {2020})},\ \Eprint
  {https://arxiv.org/abs/2002.12811} {arXiv:2002.12811 [gr-qc]} \BibitemShut
  {NoStop}%
\bibitem [{\citenamefont {Ylla}\ \emph {et~al.}(2025)\citenamefont {Ylla},
  \citenamefont {Koga},\ and\ \citenamefont {Yoo}}]{Ylla:2025hia}%
  \BibitemOpen
  \bibfield  {author} {\bibinfo {author} {\bibfnamefont {A.~U.~P.}\
  \bibnamefont {Ylla}}, \bibinfo {author} {\bibfnamefont {Y.}~\bibnamefont
  {Koga}},\ and\ \bibinfo {author} {\bibfnamefont {C.-M.}\ \bibnamefont
  {Yoo}},\ }\href {https://doi.org/10.1103/681r-bw7q} {\bibinfo {title} {{Test
  particle motion around a black hole dressed with a spherically symmetric
  stationary fluid}}} (\bibinfo {year} {2025}),\ \Eprint
  {https://arxiv.org/abs/2504.21755} {arXiv:2504.21755 [gr-qc]} \BibitemShut
  {NoStop}%
\bibitem [{\citenamefont {Cardoso}\ \emph
  {et~al.}(2022{\natexlab{a}})\citenamefont {Cardoso}, \citenamefont
  {Destounis}, \citenamefont {Duque}, \citenamefont {Macedo},\ and\
  \citenamefont {Maselli}}]{Cardoso:2021wlq}%
  \BibitemOpen
  \bibfield  {author} {\bibinfo {author} {\bibfnamefont {V.}~\bibnamefont
  {Cardoso}}, \bibinfo {author} {\bibfnamefont {K.}~\bibnamefont {Destounis}},
  \bibinfo {author} {\bibfnamefont {F.}~\bibnamefont {Duque}}, \bibinfo
  {author} {\bibfnamefont {R.~P.}\ \bibnamefont {Macedo}},\ and\ \bibinfo
  {author} {\bibfnamefont {A.}~\bibnamefont {Maselli}},\ }\bibfield  {title}
  {\bibinfo {title} {{Black holes in galaxies: Environmental impact on
  gravitational-wave generation and propagation}},\ }\href
  {https://doi.org/10.1103/PhysRevD.105.L061501} {\bibfield  {journal}
  {\bibinfo  {journal} {Phys. Rev. D}\ }\textbf {\bibinfo {volume} {105}},\
  \bibinfo {pages} {L061501} (\bibinfo {year} {2022}{\natexlab{a}})},\ \Eprint
  {https://arxiv.org/abs/2109.00005} {arXiv:2109.00005 [gr-qc]} \BibitemShut
  {NoStop}%
\bibitem [{\citenamefont {Hernquist}(1990)}]{Hernquist:1990be}%
  \BibitemOpen
  \bibfield  {author} {\bibinfo {author} {\bibfnamefont {L.}~\bibnamefont
  {Hernquist}},\ }\bibfield  {title} {\bibinfo {title} {{An Analytical Model
  for Spherical Galaxies and Bulges}},\ }\href {https://doi.org/10.1086/168845}
  {\bibfield  {journal} {\bibinfo  {journal} {Astrophys. J.}\ }\textbf
  {\bibinfo {volume} {356}},\ \bibinfo {pages} {359} (\bibinfo {year}
  {1990})}\BibitemShut {NoStop}%
\bibitem [{\citenamefont {Stuchl\'\i{}k}\ and\ \citenamefont
  {Vrba}(2021)}]{Stuchlik:2021gwg}%
  \BibitemOpen
  \bibfield  {author} {\bibinfo {author} {\bibfnamefont {Z.}~\bibnamefont
  {Stuchl\'\i{}k}}\ and\ \bibinfo {author} {\bibfnamefont {J.}~\bibnamefont
  {Vrba}},\ }\bibfield  {title} {\bibinfo {title} {{Supermassive black holes
  surrounded by dark matter modeled as anisotropic fluid: epicyclic
  oscillations and their fitting to observed QPOs}},\ }\href
  {https://doi.org/10.1088/1475-7516/2021/11/059} {\bibfield  {journal}
  {\bibinfo  {journal} {JCAP}\ }\textbf {\bibinfo {volume} {11}}\bibfield
  {number} {\bibinfo  {number} { (11)},\ \bibinfo {pages} {059}},\ }\Eprint
  {https://arxiv.org/abs/2110.07411} {arXiv:2110.07411 [gr-qc]} \BibitemShut
  {NoStop}%
\bibitem [{\citenamefont {Konoplya}\ and\ \citenamefont
  {Zhidenko}(2022)}]{Konoplya:2022hbl}%
  \BibitemOpen
  \bibfield  {author} {\bibinfo {author} {\bibfnamefont {R.~A.}\ \bibnamefont
  {Konoplya}}\ and\ \bibinfo {author} {\bibfnamefont {A.}~\bibnamefont
  {Zhidenko}},\ }\bibfield  {title} {\bibinfo {title} {{Solutions of the
  Einstein Equations for a Black Hole Surrounded by a Galactic Halo}},\ }\href
  {https://doi.org/10.3847/1538-4357/ac76bc} {\bibfield  {journal} {\bibinfo
  {journal} {Astrophys. J.}\ }\textbf {\bibinfo {volume} {933}},\ \bibinfo
  {pages} {166} (\bibinfo {year} {2022})},\ \Eprint
  {https://arxiv.org/abs/2202.02205} {arXiv:2202.02205 [gr-qc]} \BibitemShut
  {NoStop}%
\bibitem [{\citenamefont {Jusufi}(2022)}]{Jusufi:2022jxu}%
  \BibitemOpen
  \bibfield  {author} {\bibinfo {author} {\bibfnamefont {K.}~\bibnamefont
  {Jusufi}},\ }\href@noop {} {\bibinfo {title} {{Black holes surrounded by
  Einstein clusters as models of dark matter fluid}}} (\bibinfo {year}
  {2022}),\ \Eprint {https://arxiv.org/abs/2202.00010} {arXiv:2202.00010
  [gr-qc]} \BibitemShut {NoStop}%
\bibitem [{\citenamefont {Pezzella}\ \emph {et~al.}(2025)\citenamefont
  {Pezzella}, \citenamefont {Destounis}, \citenamefont {Maselli},\ and\
  \citenamefont {Cardoso}}]{Pezzella:2024tkf}%
  \BibitemOpen
  \bibfield  {author} {\bibinfo {author} {\bibfnamefont {L.}~\bibnamefont
  {Pezzella}}, \bibinfo {author} {\bibfnamefont {K.}~\bibnamefont {Destounis}},
  \bibinfo {author} {\bibfnamefont {A.}~\bibnamefont {Maselli}},\ and\ \bibinfo
  {author} {\bibfnamefont {V.}~\bibnamefont {Cardoso}},\ }\bibfield  {title}
  {\bibinfo {title} {{Quasinormal modes of black holes embedded in halos of
  matter}},\ }\href {https://doi.org/10.1103/PhysRevD.111.064026} {\bibfield
  {journal} {\bibinfo  {journal} {Phys. Rev. D}\ }\textbf {\bibinfo {volume}
  {111}},\ \bibinfo {pages} {064026} (\bibinfo {year} {2025})},\ \Eprint
  {https://arxiv.org/abs/2412.18651} {arXiv:2412.18651 [gr-qc]} \BibitemShut
  {NoStop}%
\bibitem [{\citenamefont {Cardoso}\ \emph
  {et~al.}(2022{\natexlab{b}})\citenamefont {Cardoso}, \citenamefont
  {Destounis}, \citenamefont {Duque}, \citenamefont {Panosso~Macedo},\ and\
  \citenamefont {Maselli}}]{Cardoso:2022whc}%
  \BibitemOpen
  \bibfield  {author} {\bibinfo {author} {\bibfnamefont {V.}~\bibnamefont
  {Cardoso}}, \bibinfo {author} {\bibfnamefont {K.}~\bibnamefont {Destounis}},
  \bibinfo {author} {\bibfnamefont {F.}~\bibnamefont {Duque}}, \bibinfo
  {author} {\bibfnamefont {R.}~\bibnamefont {Panosso~Macedo}},\ and\ \bibinfo
  {author} {\bibfnamefont {A.}~\bibnamefont {Maselli}},\ }\href@noop {}
  {\bibinfo {title} {{Gravitational waves from extreme-mass-ratio systems in
  astrophysical environments}}} (\bibinfo {year} {2022}{\natexlab{b}}),\
  \Eprint {https://arxiv.org/abs/2210.01133} {arXiv:2210.01133 [gr-qc]}
  \BibitemShut {NoStop}%
\bibitem [{\citenamefont {Speeney}\ \emph {et~al.}(2024)\citenamefont
  {Speeney}, \citenamefont {Berti}, \citenamefont {Cardoso},\ and\
  \citenamefont {Maselli}}]{Speeney:2024mas}%
  \BibitemOpen
  \bibfield  {author} {\bibinfo {author} {\bibfnamefont {N.}~\bibnamefont
  {Speeney}}, \bibinfo {author} {\bibfnamefont {E.}~\bibnamefont {Berti}},
  \bibinfo {author} {\bibfnamefont {V.}~\bibnamefont {Cardoso}},\ and\ \bibinfo
  {author} {\bibfnamefont {A.}~\bibnamefont {Maselli}},\ }\bibfield  {title}
  {\bibinfo {title} {{Black holes surrounded by generic matter distributions:
  Polar perturbations and energy flux}},\ }\href
  {https://doi.org/10.1103/PhysRevD.109.084068} {\bibfield  {journal} {\bibinfo
   {journal} {Phys. Rev. D}\ }\textbf {\bibinfo {volume} {109}},\ \bibinfo
  {pages} {084068} (\bibinfo {year} {2024})},\ \Eprint
  {https://arxiv.org/abs/2401.00932} {arXiv:2401.00932 [gr-qc]} \BibitemShut
  {NoStop}%
\bibitem [{\citenamefont {Figueiredo}\ \emph {et~al.}(2023)\citenamefont
  {Figueiredo}, \citenamefont {Maselli},\ and\ \citenamefont
  {Cardoso}}]{Figueiredo:2023gas}%
  \BibitemOpen
  \bibfield  {author} {\bibinfo {author} {\bibfnamefont {E.}~\bibnamefont
  {Figueiredo}}, \bibinfo {author} {\bibfnamefont {A.}~\bibnamefont
  {Maselli}},\ and\ \bibinfo {author} {\bibfnamefont {V.}~\bibnamefont
  {Cardoso}},\ }\href@noop {} {\bibinfo {title} {{Black holes surrounded by
  generic dark matter profiles: appearance and gravitational-wave emission}}}
  (\bibinfo {year} {2023}),\ \Eprint {https://arxiv.org/abs/2303.08183}
  {arXiv:2303.08183 [gr-qc]} \BibitemShut {NoStop}%
\bibitem [{\citenamefont {Gliorio}\ \emph {et~al.}(2025)\citenamefont
  {Gliorio}, \citenamefont {Berti}, \citenamefont {Maselli},\ and\
  \citenamefont {Speeney}}]{Gliorio:2025cbh}%
  \BibitemOpen
  \bibfield  {author} {\bibinfo {author} {\bibfnamefont {S.}~\bibnamefont
  {Gliorio}}, \bibinfo {author} {\bibfnamefont {E.}~\bibnamefont {Berti}},
  \bibinfo {author} {\bibfnamefont {A.}~\bibnamefont {Maselli}},\ and\ \bibinfo
  {author} {\bibfnamefont {N.}~\bibnamefont {Speeney}},\ }\href@noop {}
  {\bibinfo {title} {{Extreme mass ratio inspirals in dark matter halos:
  dynamics and distinguishability of halo models}}} (\bibinfo {year} {2025}),\
  \Eprint {https://arxiv.org/abs/2503.16649} {arXiv:2503.16649 [gr-qc]}
  \BibitemShut {NoStop}%
\bibitem [{\citenamefont {Fernandes}\ and\ \citenamefont
  {Cardoso}(2025{\natexlab{a}})}]{Fernandes:2025osu}%
  \BibitemOpen
  \bibfield  {author} {\bibinfo {author} {\bibfnamefont {P.~G.~S.}\
  \bibnamefont {Fernandes}}\ and\ \bibinfo {author} {\bibfnamefont
  {V.}~\bibnamefont {Cardoso}},\ }\href@noop {} {\bibinfo {title} {{Spinning
  black holes in astrophysical environments}}} (\bibinfo {year}
  {2025}{\natexlab{a}}),\ \Eprint {https://arxiv.org/abs/2507.04389}
  {arXiv:2507.04389 [gr-qc]} \BibitemShut {NoStop}%
\bibitem [{\citenamefont {Carter}(1968)}]{Carter:1968rr}%
  \BibitemOpen
  \bibfield  {author} {\bibinfo {author} {\bibfnamefont {B.}~\bibnamefont
  {Carter}},\ }\bibfield  {title} {\bibinfo {title} {{Global structure of the
  Kerr family of gravitational fields}},\ }\href
  {https://doi.org/10.1103/PhysRev.174.1559} {\bibfield  {journal} {\bibinfo
  {journal} {Phys. Rev.}\ }\textbf {\bibinfo {volume} {174}},\ \bibinfo {pages}
  {1559} (\bibinfo {year} {1968})}\BibitemShut {NoStop}%
\bibitem [{\citenamefont {Contopoulos}(2003)}]{Contopoulos_book}%
  \BibitemOpen
  \bibfield  {author} {\bibinfo {author} {\bibfnamefont {G.}~\bibnamefont
  {Contopoulos}},\ }\href {https://doi.org/10.1063/1.1634536} {\emph {\bibinfo
  {title} {Order and Chaos in Dynamical Astronomy}}}\ (\bibinfo  {publisher}
  {Springer-Verlag},\ \bibinfo {address} {New York},\ \bibinfo {year}
  {2003})\BibitemShut {NoStop}%
\bibitem [{\citenamefont {Flanagan}\ and\ \citenamefont
  {Hinderer}(2012)}]{Flanagan:2010cd}%
  \BibitemOpen
  \bibfield  {author} {\bibinfo {author} {\bibfnamefont {E.~E.}\ \bibnamefont
  {Flanagan}}\ and\ \bibinfo {author} {\bibfnamefont {T.}~\bibnamefont
  {Hinderer}},\ }\bibfield  {title} {\bibinfo {title} {{Transient resonances in
  the inspirals of point particles into black holes}},\ }\href
  {https://doi.org/10.1103/PhysRevLett.109.071102} {\bibfield  {journal}
  {\bibinfo  {journal} {Phys. Rev. Lett.}\ }\textbf {\bibinfo {volume} {109}},\
  \bibinfo {pages} {071102} (\bibinfo {year} {2012})},\ \Eprint
  {https://arxiv.org/abs/1009.4923} {arXiv:1009.4923 [gr-qc]} \BibitemShut
  {NoStop}%
\bibitem [{\citenamefont {Ruangsri}\ and\ \citenamefont
  {Hughes}(2014)}]{Ruangsri:2013hra}%
  \BibitemOpen
  \bibfield  {author} {\bibinfo {author} {\bibfnamefont {U.}~\bibnamefont
  {Ruangsri}}\ and\ \bibinfo {author} {\bibfnamefont {S.~A.}\ \bibnamefont
  {Hughes}},\ }\bibfield  {title} {\bibinfo {title} {{Census of transient
  orbital resonances encountered during binary inspiral}},\ }\href
  {https://doi.org/10.1103/PhysRevD.89.084036} {\bibfield  {journal} {\bibinfo
  {journal} {Phys. Rev. D}\ }\textbf {\bibinfo {volume} {89}},\ \bibinfo
  {pages} {084036} (\bibinfo {year} {2014})},\ \Eprint
  {https://arxiv.org/abs/1307.6483} {arXiv:1307.6483 [gr-qc]} \BibitemShut
  {NoStop}%
\bibitem [{\citenamefont {Berry}\ \emph {et~al.}(2016)\citenamefont {Berry},
  \citenamefont {Cole}, \citenamefont {Ca\~nizares},\ and\ \citenamefont
  {Gair}}]{Berry:2016bit}%
  \BibitemOpen
  \bibfield  {author} {\bibinfo {author} {\bibfnamefont {C.~P.~L.}\
  \bibnamefont {Berry}}, \bibinfo {author} {\bibfnamefont {R.~H.}\ \bibnamefont
  {Cole}}, \bibinfo {author} {\bibfnamefont {P.}~\bibnamefont {Ca\~nizares}},\
  and\ \bibinfo {author} {\bibfnamefont {J.~R.}\ \bibnamefont {Gair}},\
  }\bibfield  {title} {\bibinfo {title} {{Importance of transient resonances in
  extreme-mass-ratio inspirals}},\ }\href
  {https://doi.org/10.1103/PhysRevD.94.124042} {\bibfield  {journal} {\bibinfo
  {journal} {Phys. Rev. D}\ }\textbf {\bibinfo {volume} {94}},\ \bibinfo
  {pages} {124042} (\bibinfo {year} {2016})},\ \Eprint
  {https://arxiv.org/abs/1608.08951} {arXiv:1608.08951 [gr-qc]} \BibitemShut
  {NoStop}%
\bibitem [{\citenamefont {Speri}\ and\ \citenamefont
  {Gair}(2021)}]{Speri:2021psr}%
  \BibitemOpen
  \bibfield  {author} {\bibinfo {author} {\bibfnamefont {L.}~\bibnamefont
  {Speri}}\ and\ \bibinfo {author} {\bibfnamefont {J.~R.}\ \bibnamefont
  {Gair}},\ }\bibfield  {title} {\bibinfo {title} {{Assessing the impact of
  transient orbital resonances}},\ }\href
  {https://doi.org/10.1103/PhysRevD.103.124032} {\bibfield  {journal} {\bibinfo
   {journal} {Phys. Rev. D}\ }\textbf {\bibinfo {volume} {103}},\ \bibinfo
  {pages} {124032} (\bibinfo {year} {2021})},\ \Eprint
  {https://arxiv.org/abs/2103.06306} {arXiv:2103.06306 [gr-qc]} \BibitemShut
  {NoStop}%
\bibitem [{\citenamefont {Gupta}\ \emph {et~al.}(2022)\citenamefont {Gupta},
  \citenamefont {Speri}, \citenamefont {Bonga}, \citenamefont {Chua},\ and\
  \citenamefont {Tanaka}}]{Gupta:2022fbe}%
  \BibitemOpen
  \bibfield  {author} {\bibinfo {author} {\bibfnamefont {P.}~\bibnamefont
  {Gupta}}, \bibinfo {author} {\bibfnamefont {L.}~\bibnamefont {Speri}},
  \bibinfo {author} {\bibfnamefont {B.}~\bibnamefont {Bonga}}, \bibinfo
  {author} {\bibfnamefont {A.~J.~K.}\ \bibnamefont {Chua}},\ and\ \bibinfo
  {author} {\bibfnamefont {T.}~\bibnamefont {Tanaka}},\ }\bibfield  {title}
  {\bibinfo {title} {{Modeling transient resonances in extreme-mass-ratio
  inspirals}},\ }\href {https://doi.org/10.1103/PhysRevD.106.104001} {\bibfield
   {journal} {\bibinfo  {journal} {Phys. Rev. D}\ }\textbf {\bibinfo {volume}
  {106}},\ \bibinfo {pages} {104001} (\bibinfo {year} {2022})},\ \Eprint
  {https://arxiv.org/abs/2205.04808} {arXiv:2205.04808 [gr-qc]} \BibitemShut
  {NoStop}%
\bibitem [{\citenamefont {Apostolatos}\ \emph {et~al.}(2009)\citenamefont
  {Apostolatos}, \citenamefont {Lukes-Gerakopoulos},\ and\ \citenamefont
  {Contopoulos}}]{Apostolatos:2009vu}%
  \BibitemOpen
  \bibfield  {author} {\bibinfo {author} {\bibfnamefont {T.~A.}\ \bibnamefont
  {Apostolatos}}, \bibinfo {author} {\bibfnamefont {G.}~\bibnamefont
  {Lukes-Gerakopoulos}},\ and\ \bibinfo {author} {\bibfnamefont
  {G.}~\bibnamefont {Contopoulos}},\ }\bibfield  {title} {\bibinfo {title}
  {{How to Observe a Non-Kerr Spacetime Using Gravitational Waves}},\ }\href
  {https://doi.org/10.1103/PhysRevLett.103.111101} {\bibfield  {journal}
  {\bibinfo  {journal} {Phys. Rev. Lett.}\ }\textbf {\bibinfo {volume} {103}},\
  \bibinfo {pages} {111101} (\bibinfo {year} {2009})},\ \Eprint
  {https://arxiv.org/abs/0906.0093} {arXiv:0906.0093 [gr-qc]} \BibitemShut
  {NoStop}%
\bibitem [{\citenamefont {Lukes-Gerakopoulos}\ \emph
  {et~al.}(2010)\citenamefont {Lukes-Gerakopoulos}, \citenamefont
  {Apostolatos},\ and\ \citenamefont
  {Contopoulos}}]{Lukes-Gerakopoulos:2010ipp}%
  \BibitemOpen
  \bibfield  {author} {\bibinfo {author} {\bibfnamefont {G.}~\bibnamefont
  {Lukes-Gerakopoulos}}, \bibinfo {author} {\bibfnamefont {T.~A.}\ \bibnamefont
  {Apostolatos}},\ and\ \bibinfo {author} {\bibfnamefont {G.}~\bibnamefont
  {Contopoulos}},\ }\bibfield  {title} {\bibinfo {title} {{Observable signature
  of a background deviating from the Kerr metric}},\ }\href
  {https://doi.org/10.1103/PhysRevD.81.124005} {\bibfield  {journal} {\bibinfo
  {journal} {Phys. Rev. D}\ }\textbf {\bibinfo {volume} {81}},\ \bibinfo
  {pages} {124005} (\bibinfo {year} {2010})},\ \Eprint
  {https://arxiv.org/abs/1003.3120} {arXiv:1003.3120 [gr-qc]} \BibitemShut
  {NoStop}%
\bibitem [{\citenamefont {Destounis}\ \emph
  {et~al.}(2021{\natexlab{b}})\citenamefont {Destounis}, \citenamefont
  {Suvorov},\ and\ \citenamefont {Kokkotas}}]{Destounis:2021mqv}%
  \BibitemOpen
  \bibfield  {author} {\bibinfo {author} {\bibfnamefont {K.}~\bibnamefont
  {Destounis}}, \bibinfo {author} {\bibfnamefont {A.~G.}\ \bibnamefont
  {Suvorov}},\ and\ \bibinfo {author} {\bibfnamefont {K.~D.}\ \bibnamefont
  {Kokkotas}},\ }\bibfield  {title} {\bibinfo {title} {{Gravitational-wave
  glitches in chaotic extreme-mass-ratio inspirals}},\ }\href
  {https://doi.org/10.1103/PhysRevLett.126.141102} {\bibfield  {journal}
  {\bibinfo  {journal} {Phys. Rev. Lett.}\ }\textbf {\bibinfo {volume} {126}},\
  \bibinfo {pages} {141102} (\bibinfo {year} {2021}{\natexlab{b}})},\ \Eprint
  {https://arxiv.org/abs/2103.05643} {arXiv:2103.05643 [gr-qc]} \BibitemShut
  {NoStop}%
\bibitem [{\citenamefont {Destounis}\ and\ \citenamefont
  {Kokkotas}(2021)}]{Destounis:2021rko}%
  \BibitemOpen
  \bibfield  {author} {\bibinfo {author} {\bibfnamefont {K.}~\bibnamefont
  {Destounis}}\ and\ \bibinfo {author} {\bibfnamefont {K.~D.}\ \bibnamefont
  {Kokkotas}},\ }\bibfield  {title} {\bibinfo {title} {{Gravitational-wave
  glitches: Resonant islands and frequency jumps in nonintegrable
  extreme-mass-ratio inspirals}},\ }\href
  {https://doi.org/10.1103/PhysRevD.104.064023} {\bibfield  {journal} {\bibinfo
   {journal} {Phys. Rev. D}\ }\textbf {\bibinfo {volume} {104}},\ \bibinfo
  {pages} {064023} (\bibinfo {year} {2021})},\ \Eprint
  {https://arxiv.org/abs/2108.02782} {arXiv:2108.02782 [gr-qc]} \BibitemShut
  {NoStop}%
\bibitem [{\citenamefont {Destounis}\ \emph {et~al.}(2020)\citenamefont
  {Destounis}, \citenamefont {Suvorov},\ and\ \citenamefont
  {Kokkotas}}]{Destounis:2020kss}%
  \BibitemOpen
  \bibfield  {author} {\bibinfo {author} {\bibfnamefont {K.}~\bibnamefont
  {Destounis}}, \bibinfo {author} {\bibfnamefont {A.~G.}\ \bibnamefont
  {Suvorov}},\ and\ \bibinfo {author} {\bibfnamefont {K.~D.}\ \bibnamefont
  {Kokkotas}},\ }\bibfield  {title} {\bibinfo {title} {{Testing spacetime
  symmetry through gravitational waves from extreme-mass-ratio inspirals}},\
  }\href {https://doi.org/10.1103/PhysRevD.102.064041} {\bibfield  {journal}
  {\bibinfo  {journal} {Phys. Rev. D}\ }\textbf {\bibinfo {volume} {102}},\
  \bibinfo {pages} {064041} (\bibinfo {year} {2020})},\ \Eprint
  {https://arxiv.org/abs/2009.00028} {arXiv:2009.00028 [gr-qc]} \BibitemShut
  {NoStop}%
\bibitem [{\citenamefont {Eleni}\ \emph {et~al.}(2024)\citenamefont {Eleni},
  \citenamefont {Destounis}, \citenamefont {Apostolatos},\ and\ \citenamefont
  {Kokkotas}}]{Eleni:2024fgs}%
  \BibitemOpen
  \bibfield  {author} {\bibinfo {author} {\bibfnamefont {A.}~\bibnamefont
  {Eleni}}, \bibinfo {author} {\bibfnamefont {K.}~\bibnamefont {Destounis}},
  \bibinfo {author} {\bibfnamefont {T.~A.}\ \bibnamefont {Apostolatos}},\ and\
  \bibinfo {author} {\bibfnamefont {K.~D.}\ \bibnamefont {Kokkotas}},\
  }\bibfield  {title} {\bibinfo {title} {{Resonant excitation of eccentricity
  in spherical extreme-mass-ratio inspirals}},\ }\href
  {https://doi.org/10.1103/PhysRevD.110.124004} {\bibfield  {journal} {\bibinfo
   {journal} {Phys. Rev. D}\ }\textbf {\bibinfo {volume} {110}},\ \bibinfo
  {pages} {124004} (\bibinfo {year} {2024})},\ \Eprint
  {https://arxiv.org/abs/2408.02004} {arXiv:2408.02004 [gr-qc]} \BibitemShut
  {NoStop}%
\bibitem [{\citenamefont {Chen}\ \emph {et~al.}(2022)\citenamefont {Chen},
  \citenamefont {Lin},\ and\ \citenamefont {Patel}}]{Chen:2022znf}%
  \BibitemOpen
  \bibfield  {author} {\bibinfo {author} {\bibfnamefont {C.-Y.}\ \bibnamefont
  {Chen}}, \bibinfo {author} {\bibfnamefont {F.-L.}\ \bibnamefont {Lin}},\ and\
  \bibinfo {author} {\bibfnamefont {A.}~\bibnamefont {Patel}},\ }\bibfield
  {title} {\bibinfo {title} {{Resonant islands of effective-one-body
  dynamics}},\ }\href {https://doi.org/10.1103/PhysRevD.106.084064} {\bibfield
  {journal} {\bibinfo  {journal} {Phys. Rev. D}\ }\textbf {\bibinfo {volume}
  {106}},\ \bibinfo {pages} {084064} (\bibinfo {year} {2022})},\ \Eprint
  {https://arxiv.org/abs/2206.10966} {arXiv:2206.10966 [gr-qc]} \BibitemShut
  {NoStop}%
\bibitem [{\citenamefont {Contopoulos}\ \emph {et~al.}(2011)\citenamefont
  {Contopoulos}, \citenamefont {Lukes-Gerakopoulos},\ and\ \citenamefont
  {Apostolatos}}]{Contopoulos:2011dz}%
  \BibitemOpen
  \bibfield  {author} {\bibinfo {author} {\bibfnamefont {G.}~\bibnamefont
  {Contopoulos}}, \bibinfo {author} {\bibfnamefont {G.}~\bibnamefont
  {Lukes-Gerakopoulos}},\ and\ \bibinfo {author} {\bibfnamefont {T.~A.}\
  \bibnamefont {Apostolatos}},\ }\bibfield  {title} {\bibinfo {title} {{Orbits
  in a non-Kerr Dynamical System}},\ }\href
  {https://doi.org/10.1142/S0218127411029768} {\bibfield  {journal} {\bibinfo
  {journal} {Int. J. Bifurc. Chaos}\ }\textbf {\bibinfo {volume} {21}},\
  \bibinfo {pages} {2261} (\bibinfo {year} {2011})},\ \Eprint
  {https://arxiv.org/abs/1108.5057} {arXiv:1108.5057 [gr-qc]} \BibitemShut
  {NoStop}%
\bibitem [{\citenamefont {Lukes-Gerakopoulos}\ and\ \citenamefont
  {Witzany}(2021)}]{Lukes-Gerakopoulos:2021ybx}%
  \BibitemOpen
  \bibfield  {author} {\bibinfo {author} {\bibfnamefont {G.}~\bibnamefont
  {Lukes-Gerakopoulos}}\ and\ \bibinfo {author} {\bibfnamefont
  {V.}~\bibnamefont {Witzany}},\ }\bibfield  {title} {\bibinfo {title}
  {{Non-linear effects in EMRI dynamics and their imprints on gravitational
  waves}}\ }\href {https://doi.org/10.1007/978-981-15-4702-7-42-1}
  {10.1007/978-981-15-4702-7-42-1} (\bibinfo {year} {2021}),\ \Eprint
  {https://arxiv.org/abs/2103.06724} {arXiv:2103.06724 [gr-qc]} \BibitemShut
  {NoStop}%
\bibitem [{\citenamefont {Destounis}\ \emph
  {et~al.}(2023{\natexlab{a}})\citenamefont {Destounis}, \citenamefont {Huez},\
  and\ \citenamefont {Kokkotas}}]{Destounis:2023gpw}%
  \BibitemOpen
  \bibfield  {author} {\bibinfo {author} {\bibfnamefont {K.}~\bibnamefont
  {Destounis}}, \bibinfo {author} {\bibfnamefont {G.}~\bibnamefont {Huez}},\
  and\ \bibinfo {author} {\bibfnamefont {K.~D.}\ \bibnamefont {Kokkotas}},\
  }\href@noop {} {\bibinfo {title} {{Geodesics and gravitational waves in
  chaotic extreme-mass-ratio inspirals: The curious case of Zipoy-Voorhees
  black-hole mimickers}}} (\bibinfo {year} {2023}{\natexlab{a}}),\ \Eprint
  {https://arxiv.org/abs/2301.11483} {arXiv:2301.11483 [gr-qc]} \BibitemShut
  {NoStop}%
\bibitem [{\citenamefont {Destounis}\ \emph
  {et~al.}(2023{\natexlab{b}})\citenamefont {Destounis}, \citenamefont
  {Angeloni}, \citenamefont {Vaglio},\ and\ \citenamefont
  {Pani}}]{Destounis:2023khj}%
  \BibitemOpen
  \bibfield  {author} {\bibinfo {author} {\bibfnamefont {K.}~\bibnamefont
  {Destounis}}, \bibinfo {author} {\bibfnamefont {F.}~\bibnamefont {Angeloni}},
  \bibinfo {author} {\bibfnamefont {M.}~\bibnamefont {Vaglio}},\ and\ \bibinfo
  {author} {\bibfnamefont {P.}~\bibnamefont {Pani}},\ }\bibfield  {title}
  {\bibinfo {title} {{Extreme-mass-ratio inspirals into rotating boson stars:
  Nonintegrability, chaos, and transient resonances}},\ }\href
  {https://doi.org/10.1103/PhysRevD.108.084062} {\bibfield  {journal} {\bibinfo
   {journal} {Phys. Rev. D}\ }\textbf {\bibinfo {volume} {108}},\ \bibinfo
  {pages} {084062} (\bibinfo {year} {2023}{\natexlab{b}})},\ \Eprint
  {https://arxiv.org/abs/2305.05691} {arXiv:2305.05691 [gr-qc]} \BibitemShut
  {NoStop}%
\bibitem [{\citenamefont {Chen}\ \emph {et~al.}(2023)\citenamefont {Chen},
  \citenamefont {Chiang},\ and\ \citenamefont {Patel}}]{Chen:2023gwm}%
  \BibitemOpen
  \bibfield  {author} {\bibinfo {author} {\bibfnamefont {C.-Y.}\ \bibnamefont
  {Chen}}, \bibinfo {author} {\bibfnamefont {H.-W.}\ \bibnamefont {Chiang}},\
  and\ \bibinfo {author} {\bibfnamefont {A.}~\bibnamefont {Patel}},\ }\bibfield
   {title} {\bibinfo {title} {{Resonant orbits of rotating black holes beyond
  circularity: Discontinuity along a parameter shift}},\ }\href
  {https://doi.org/10.1103/PhysRevD.108.064016} {\bibfield  {journal} {\bibinfo
   {journal} {Phys. Rev. D}\ }\textbf {\bibinfo {volume} {108}},\ \bibinfo
  {pages} {064016} (\bibinfo {year} {2023})},\ \Eprint
  {https://arxiv.org/abs/2306.08356} {arXiv:2306.08356 [gr-qc]} \BibitemShut
  {NoStop}%
\bibitem [{\citenamefont {Mukherjee}\ \emph {et~al.}(2022)\citenamefont
  {Mukherjee}, \citenamefont {Kopacek},\ and\ \citenamefont
  {Lukes-Gerakopoulos}}]{Mukherjee:2022dju}%
  \BibitemOpen
  \bibfield  {author} {\bibinfo {author} {\bibfnamefont {S.}~\bibnamefont
  {Mukherjee}}, \bibinfo {author} {\bibfnamefont {O.}~\bibnamefont {Kopacek}},\
  and\ \bibinfo {author} {\bibfnamefont {G.}~\bibnamefont
  {Lukes-Gerakopoulos}},\ }\href@noop {} {\bibinfo {title} {{Resonance crossing
  of a charged body in a magnetized Kerr background: an analogue of extreme
  mass ratio inspiral}}} (\bibinfo {year} {2022}),\ \Eprint
  {https://arxiv.org/abs/2206.10302} {arXiv:2206.10302 [gr-qc]} \BibitemShut
  {NoStop}%
\bibitem [{\citenamefont {Eleni}\ and\ \citenamefont
  {Apostolatos}(2023)}]{Eleni:2023mjx}%
  \BibitemOpen
  \bibfield  {author} {\bibinfo {author} {\bibfnamefont {A.}~\bibnamefont
  {Eleni}}\ and\ \bibinfo {author} {\bibfnamefont {T.~A.}\ \bibnamefont
  {Apostolatos}},\ }\bibfield  {title} {\bibinfo {title} {{Enhanced plateau
  effect at resonance in realistic nonintegrable extreme-mass-ratio
  inspirals}},\ }\href {https://doi.org/10.1103/PhysRevD.108.124044} {\bibfield
   {journal} {\bibinfo  {journal} {Phys. Rev. D}\ }\textbf {\bibinfo {volume}
  {108}},\ \bibinfo {pages} {124044} (\bibinfo {year} {2023})},\ \Eprint
  {https://arxiv.org/abs/2306.17762} {arXiv:2306.17762 [gr-qc]} \BibitemShut
  {NoStop}%
\bibitem [{\citenamefont {Zelenka}\ \emph {et~al.}(2020)\citenamefont
  {Zelenka}, \citenamefont {Lukes-Gerakopoulos}, \citenamefont {Witzany},\ and\
  \citenamefont {Kop\'a\v{c}ek}}]{Zelenka:2019nyp}%
  \BibitemOpen
  \bibfield  {author} {\bibinfo {author} {\bibfnamefont {O.}~\bibnamefont
  {Zelenka}}, \bibinfo {author} {\bibfnamefont {G.}~\bibnamefont
  {Lukes-Gerakopoulos}}, \bibinfo {author} {\bibfnamefont {V.}~\bibnamefont
  {Witzany}},\ and\ \bibinfo {author} {\bibfnamefont {O.}~\bibnamefont
  {Kop\'a\v{c}ek}},\ }\bibfield  {title} {\bibinfo {title} {{Growth of
  resonances and chaos for a spinning test particle in the Schwarzschild
  background}},\ }\href {https://doi.org/10.1103/PhysRevD.101.024037}
  {\bibfield  {journal} {\bibinfo  {journal} {Phys. Rev. D}\ }\textbf {\bibinfo
  {volume} {101}},\ \bibinfo {pages} {024037} (\bibinfo {year} {2020})},\
  \Eprint {https://arxiv.org/abs/1911.00414} {arXiv:1911.00414 [gr-qc]}
  \BibitemShut {NoStop}%
\bibitem [{\citenamefont {De~Falco}\ and\ \citenamefont
  {Borrelli}(2021{\natexlab{a}})}]{DeFalco:2020yys}%
  \BibitemOpen
  \bibfield  {author} {\bibinfo {author} {\bibfnamefont {V.}~\bibnamefont
  {De~Falco}}\ and\ \bibinfo {author} {\bibfnamefont {W.}~\bibnamefont
  {Borrelli}},\ }\bibfield  {title} {\bibinfo {title} {{Detection of chaos in
  the general relativistic Poynting-Robertson effect: Kerr equatorial plane}},\
  }\href {https://doi.org/10.1103/PhysRevD.103.064014} {\bibfield  {journal}
  {\bibinfo  {journal} {Phys. Rev. D}\ }\textbf {\bibinfo {volume} {103}},\
  \bibinfo {pages} {064014} (\bibinfo {year} {2021}{\natexlab{a}})},\ \Eprint
  {https://arxiv.org/abs/2001.04979} {arXiv:2001.04979 [gr-qc]} \BibitemShut
  {NoStop}%
\bibitem [{\citenamefont {De~Falco}\ and\ \citenamefont
  {Borrelli}(2021{\natexlab{b}})}]{DeFalco:2021uak}%
  \BibitemOpen
  \bibfield  {author} {\bibinfo {author} {\bibfnamefont {V.}~\bibnamefont
  {De~Falco}}\ and\ \bibinfo {author} {\bibfnamefont {W.}~\bibnamefont
  {Borrelli}},\ }\href {https://doi.org/10.1103/PhysRevD.103.124012} {\bibinfo
  {title} {{Timescales of the chaos onset in the general relativistic
  Poynting-Robertson effect}}} (\bibinfo {year} {2021}{\natexlab{b}}),\ \Eprint
  {https://arxiv.org/abs/2105.00965} {arXiv:2105.00965 [gr-qc]} \BibitemShut
  {NoStop}%
\bibitem [{\citenamefont {Vieira}\ and\ \citenamefont
  {Letelier}(1996)}]{Vieira:1996zf}%
  \BibitemOpen
  \bibfield  {author} {\bibinfo {author} {\bibfnamefont {W.~M.}\ \bibnamefont
  {Vieira}}\ and\ \bibinfo {author} {\bibfnamefont {P.~S.}\ \bibnamefont
  {Letelier}},\ }\bibfield  {title} {\bibinfo {title} {{Chaos around a
  Henon-Heiles inspired exact perturbation of a black hole}},\ }\href
  {https://doi.org/10.1103/PhysRevLett.76.1409} {\bibfield  {journal} {\bibinfo
   {journal} {Phys. Rev. Lett.}\ }\textbf {\bibinfo {volume} {76}},\ \bibinfo
  {pages} {1409} (\bibinfo {year} {1996})},\ \Eprint
  {https://arxiv.org/abs/gr-qc/9604037} {arXiv:gr-qc/9604037} \BibitemShut
  {NoStop}%
\bibitem [{\citenamefont {de~Moura}\ and\ \citenamefont
  {Letelier}(2000)}]{deMoura:1999wf}%
  \BibitemOpen
  \bibfield  {author} {\bibinfo {author} {\bibfnamefont {A.~P.~S.}\
  \bibnamefont {de~Moura}}\ and\ \bibinfo {author} {\bibfnamefont {P.~S.}\
  \bibnamefont {Letelier}},\ }\bibfield  {title} {\bibinfo {title} {{Chaos and
  fractals in geodesic motions around a nonrotating black hole with an external
  halo}},\ }\href {https://doi.org/10.1103/PhysRevE.61.6506} {\bibfield
  {journal} {\bibinfo  {journal} {Phys. Rev. E}\ }\textbf {\bibinfo {volume}
  {61}},\ \bibinfo {pages} {6506} (\bibinfo {year} {2000})},\ \Eprint
  {https://arxiv.org/abs/chao-dyn/9910035} {arXiv:chao-dyn/9910035}
  \BibitemShut {NoStop}%
\bibitem [{\citenamefont {{Will}}(1974)}]{Will:1974}%
  \BibitemOpen
  \bibfield  {author} {\bibinfo {author} {\bibfnamefont {C.~M.}\ \bibnamefont
  {{Will}}},\ }\bibfield  {title} {\bibinfo {title} {{Perturbation of a Slowly
  Rotating Black Hole by a Stationary Axisymmetric Ring of Matter. I.
  Equilibrium Configurations}},\ }\href {https://doi.org/10.1086/152992}
  {\bibfield  {journal} {\bibinfo  {journal} {\apj}\ }\textbf {\bibinfo
  {volume} {191}},\ \bibinfo {pages} {521} (\bibinfo {year}
  {1974})}\BibitemShut {NoStop}%
\bibitem [{\citenamefont {Lemos}\ and\ \citenamefont
  {Letelier}(1994)}]{Lemos:1993qp}%
  \BibitemOpen
  \bibfield  {author} {\bibinfo {author} {\bibfnamefont {J.~P.~S.}\
  \bibnamefont {Lemos}}\ and\ \bibinfo {author} {\bibfnamefont {P.~S.}\
  \bibnamefont {Letelier}},\ }\bibfield  {title} {\bibinfo {title} {{Exact
  general relativistic thin disks around black holes}},\ }\href
  {https://doi.org/10.1103/PhysRevD.49.5135} {\bibfield  {journal} {\bibinfo
  {journal} {Phys. Rev. D}\ }\textbf {\bibinfo {volume} {49}},\ \bibinfo
  {pages} {5135} (\bibinfo {year} {1994})}\BibitemShut {NoStop}%
\bibitem [{\citenamefont {Kotla{\v{r}}{\'\i}k}\ and\ \citenamefont
  {Kofro{\v{n}}}(2022)}]{Kotlarik:2022spo}%
  \BibitemOpen
  \bibfield  {author} {\bibinfo {author} {\bibfnamefont {P.}~\bibnamefont
  {Kotla{\v{r}}{\'\i}k}}\ and\ \bibinfo {author} {\bibfnamefont
  {D.}~\bibnamefont {Kofro{\v{n}}}},\ }\bibfield  {title} {\bibinfo {title}
  {{Black Hole Encircled by a Thin Disk: Fully Relativistic Solution*}},\
  }\href {https://doi.org/10.3847/1538-4357/ac9620} {\bibfield  {journal}
  {\bibinfo  {journal} {Astrophys. J.}\ }\textbf {\bibinfo {volume} {941}},\
  \bibinfo {pages} {25} (\bibinfo {year} {2022})},\ \Eprint
  {https://arxiv.org/abs/2211.04823} {arXiv:2211.04823 [gr-qc]} \BibitemShut
  {NoStop}%
\bibitem [{\citenamefont {Einstein}(1939)}]{Einstein:1939ms}%
  \BibitemOpen
  \bibfield  {author} {\bibinfo {author} {\bibfnamefont {A.}~\bibnamefont
  {Einstein}},\ }\bibfield  {title} {\bibinfo {title} {{On a stationary system
  with spherical symmetry consisting of many gravitating masses}},\ }\href
  {https://doi.org/10.2307/1968902} {\bibfield  {journal} {\bibinfo  {journal}
  {Annals Math.}\ }\textbf {\bibinfo {volume} {40}},\ \bibinfo {pages} {922}
  (\bibinfo {year} {1939})}\BibitemShut {NoStop}%
\bibitem [{\citenamefont
  {Hogan}(1978)}]{hoganReconstructionMinkowskianSpacetime1978}%
  \BibitemOpen
  \bibfield  {author} {\bibinfo {author} {\bibfnamefont {P.~A.}\ \bibnamefont
  {Hogan}},\ }\bibfield  {title} {\bibinfo {title} {A reconstruction in
  {{Minkowskian}} space-time of {{Einstein}}'s assembly of test particles},\
  }\href {https://doi.org/10.1007/BF00784662} {\bibfield  {journal} {\bibinfo
  {journal} {General Relativity and Gravitation}\ }\textbf {\bibinfo {volume}
  {9}},\ \bibinfo {pages} {1021} (\bibinfo {year} {1978})}\BibitemShut
  {NoStop}%
\bibitem [{\citenamefont {{Zapolsky}}(1968)}]{1968ApJ...153L.163Z}%
  \BibitemOpen
  \bibfield  {author} {\bibinfo {author} {\bibfnamefont {H.~S.}\ \bibnamefont
  {{Zapolsky}}},\ }\bibfield  {title} {\bibinfo {title} {{Can the Redshifts of
  Quasi-Stellar Objects BE Gravitational?}},\ }\href
  {https://doi.org/10.1086/180244} {\bibfield  {journal} {\bibinfo  {journal}
  {The Astrophysical Journal Letters}\ }\textbf {\bibinfo {volume} {153}},\
  \bibinfo {pages} {L163} (\bibinfo {year} {1968})}\BibitemShut {NoStop}%
\bibitem [{\citenamefont {{Florides}}(1974)}]{1974RSPSA.337..529F}%
  \BibitemOpen
  \bibfield  {author} {\bibinfo {author} {\bibfnamefont {P.~S.}\ \bibnamefont
  {{Florides}}},\ }\bibfield  {title} {\bibinfo {title} {{A New Interior
  Schwarzschild Solution}},\ }\href {https://doi.org/10.1098/rspa.1974.0065}
  {\bibfield  {journal} {\bibinfo  {journal} {Proceedings of the Royal Society
  of London Series A}\ }\textbf {\bibinfo {volume} {337}},\ \bibinfo {pages}
  {529} (\bibinfo {year} {1974})}\BibitemShut {NoStop}%
\bibitem [{\citenamefont {Comer}\ and\ \citenamefont
  {Katz}(1993)}]{Comer:1993rx}%
  \BibitemOpen
  \bibfield  {author} {\bibinfo {author} {\bibfnamefont {G.~L.}\ \bibnamefont
  {Comer}}\ and\ \bibinfo {author} {\bibfnamefont {J.}~\bibnamefont {Katz}},\
  }\bibfield  {title} {\bibinfo {title} {{Thick Einstein shells and their
  mechanical stability}},\ }\href {https://doi.org/10.1088/0264-9381/10/9/017}
  {\bibfield  {journal} {\bibinfo  {journal} {Class. Quant. Grav.}\ }\textbf
  {\bibinfo {volume} {10}},\ \bibinfo {pages} {1751} (\bibinfo {year}
  {1993})}\BibitemShut {NoStop}%
\bibitem [{\citenamefont
  {Kumar~Datta}(1970)}]{kumardattaNonstaticSphericallySymmetric1970}%
  \BibitemOpen
  \bibfield  {author} {\bibinfo {author} {\bibfnamefont {B.}~\bibnamefont
  {Kumar~Datta}},\ }\bibfield  {title} {\bibinfo {title} {Non-static
  spherically symmetric clusters of particles in general relativity: I},\
  }\href {https://doi.org/10.1007/BF00759199} {\bibfield  {journal} {\bibinfo
  {journal} {General Relativity and Gravitation}\ }\textbf {\bibinfo {volume}
  {1}},\ \bibinfo {pages} {19} (\bibinfo {year} {1970})}\BibitemShut {NoStop}%
\bibitem [{\citenamefont {Bondi}(1971)}]{bondiDattasSphericallySymmetric1971}%
  \BibitemOpen
  \bibfield  {author} {\bibinfo {author} {\bibfnamefont {H.}~\bibnamefont
  {Bondi}},\ }\bibfield  {title} {\bibinfo {title} {On datta's spherically
  symmetric systems in general relativity},\ }\href
  {https://doi.org/10.1007/BF00758151} {\bibfield  {journal} {\bibinfo
  {journal} {General Relativity and Gravitation}\ }\textbf {\bibinfo {volume}
  {2}},\ \bibinfo {pages} {321} (\bibinfo {year} {1971})}\BibitemShut {NoStop}%
\bibitem [{\citenamefont {Gair}(2001)}]{Gair:2001qu}%
  \BibitemOpen
  \bibfield  {author} {\bibinfo {author} {\bibfnamefont {J.~R.}\ \bibnamefont
  {Gair}},\ }\bibfield  {title} {\bibinfo {title} {{Spherical universes with
  anisotropic pressure}},\ }\href {https://doi.org/10.1088/0264-9381/18/22/313}
  {\bibfield  {journal} {\bibinfo  {journal} {Class. Quant. Grav.}\ }\textbf
  {\bibinfo {volume} {18}},\ \bibinfo {pages} {4897} (\bibinfo {year}
  {2001})},\ \Eprint {https://arxiv.org/abs/gr-qc/0110017}
  {arXiv:gr-qc/0110017} \BibitemShut {NoStop}%
\bibitem [{\citenamefont {Szybka}\ and\ \citenamefont
  {Rutkowski}(2020)}]{Szybka:2018hoe}%
  \BibitemOpen
  \bibfield  {author} {\bibinfo {author} {\bibfnamefont {S.~J.}\ \bibnamefont
  {Szybka}}\ and\ \bibinfo {author} {\bibfnamefont {M.}~\bibnamefont
  {Rutkowski}},\ }\bibfield  {title} {\bibinfo {title} {{Einstein clusters as
  models of inhomogeneous spacetimes}},\ }\href
  {https://doi.org/10.1140/epjc/s10052-020-7948-0} {\bibfield  {journal}
  {\bibinfo  {journal} {Eur. Phys. J. C}\ }\textbf {\bibinfo {volume} {80}},\
  \bibinfo {pages} {397} (\bibinfo {year} {2020})},\ \Eprint
  {https://arxiv.org/abs/1812.11112} {arXiv:1812.11112 [gr-qc]} \BibitemShut
  {NoStop}%
\bibitem [{\citenamefont {Mahajan}\ \emph {et~al.}(2007)\citenamefont
  {Mahajan}, \citenamefont {Harada}, \citenamefont {Joshi},\ and\ \citenamefont
  {Nakao}}]{Mahajan:2007vw}%
  \BibitemOpen
  \bibfield  {author} {\bibinfo {author} {\bibfnamefont {A.}~\bibnamefont
  {Mahajan}}, \bibinfo {author} {\bibfnamefont {T.}~\bibnamefont {Harada}},
  \bibinfo {author} {\bibfnamefont {P.~S.}\ \bibnamefont {Joshi}},\ and\
  \bibinfo {author} {\bibfnamefont {K.-i.}\ \bibnamefont {Nakao}},\ }\bibfield
  {title} {\bibinfo {title} {{Critical Collapse of Einstein Cluster}},\ }\href
  {https://doi.org/10.1143/PTP.118.865} {\bibfield  {journal} {\bibinfo
  {journal} {Prog. Theor. Phys.}\ }\textbf {\bibinfo {volume} {118}},\ \bibinfo
  {pages} {865} (\bibinfo {year} {2007})},\ \Eprint
  {https://arxiv.org/abs/0710.4315} {arXiv:0710.4315 [gr-qc]} \BibitemShut
  {NoStop}%
\bibitem [{\citenamefont {Magli}(1998)}]{Magli:1997qf}%
  \BibitemOpen
  \bibfield  {author} {\bibinfo {author} {\bibfnamefont {G.}~\bibnamefont
  {Magli}},\ }\bibfield  {title} {\bibinfo {title} {{Gravitational collapse
  with nonvanishing tangential stresses. 2. Extension to the charged case and
  general solution}},\ }\href {https://doi.org/10.1088/0264-9381/15/10/022}
  {\bibfield  {journal} {\bibinfo  {journal} {Class. Quant. Grav.}\ }\textbf
  {\bibinfo {volume} {15}},\ \bibinfo {pages} {3215} (\bibinfo {year}
  {1998})},\ \Eprint {https://arxiv.org/abs/gr-qc/9711082}
  {arXiv:gr-qc/9711082} \BibitemShut {NoStop}%
\bibitem [{\citenamefont {Acharyya}\ \emph {et~al.}(2024)\citenamefont
  {Acharyya}, \citenamefont {Banerjee},\ and\ \citenamefont
  {Kar}}]{Acharyya:2023rnq}%
  \BibitemOpen
  \bibfield  {author} {\bibinfo {author} {\bibfnamefont {R.}~\bibnamefont
  {Acharyya}}, \bibinfo {author} {\bibfnamefont {P.}~\bibnamefont {Banerjee}},\
  and\ \bibinfo {author} {\bibfnamefont {S.}~\bibnamefont {Kar}},\ }\bibfield
  {title} {\bibinfo {title} {{Modelling Einstein cluster using Einasto
  profile}},\ }\href {https://doi.org/10.1088/1475-7516/2024/04/070} {\bibfield
   {journal} {\bibinfo  {journal} {JCAP}\ }\textbf {\bibinfo {volume} {04}},\
  \bibinfo {pages} {070}},\ \Eprint {https://arxiv.org/abs/2311.18622}
  {arXiv:2311.18622 [gr-qc]} \BibitemShut {NoStop}%
\bibitem [{\citenamefont {Boehmer}\ and\ \citenamefont
  {Harko}(2007)}]{Boehmer:2007az}%
  \BibitemOpen
  \bibfield  {author} {\bibinfo {author} {\bibfnamefont {C.~G.}\ \bibnamefont
  {Boehmer}}\ and\ \bibinfo {author} {\bibfnamefont {T.}~\bibnamefont
  {Harko}},\ }\bibfield  {title} {\bibinfo {title} {{On Einstein clusters as
  galactic dark matter halos}},\ }\href
  {https://doi.org/10.1111/j.1365-2966.2007.11977.x} {\bibfield  {journal}
  {\bibinfo  {journal} {Mon. Not. Roy. Astron. Soc.}\ }\textbf {\bibinfo
  {volume} {379}},\ \bibinfo {pages} {393} (\bibinfo {year} {2007})},\ \Eprint
  {https://arxiv.org/abs/0705.1756} {arXiv:0705.1756 [gr-qc]} \BibitemShut
  {NoStop}%
\bibitem [{\citenamefont {Lake}(2006)}]{Lake:2006pp}%
  \BibitemOpen
  \bibfield  {author} {\bibinfo {author} {\bibfnamefont {K.}~\bibnamefont
  {Lake}},\ }\href@noop {} {\bibinfo {title} {{Galactic halos are Einstein
  clusters of WIMPs}}} (\bibinfo {year} {2006}),\ \Eprint
  {https://arxiv.org/abs/gr-qc/0607057} {arXiv:gr-qc/0607057} \BibitemShut
  {NoStop}%
\bibitem [{\citenamefont {Geralico}\ \emph {et~al.}(2012)\citenamefont
  {Geralico}, \citenamefont {Pompi},\ and\ \citenamefont
  {Ruffini}}]{Geralico:2012jt}%
  \BibitemOpen
  \bibfield  {author} {\bibinfo {author} {\bibfnamefont {A.}~\bibnamefont
  {Geralico}}, \bibinfo {author} {\bibfnamefont {F.}~\bibnamefont {Pompi}},\
  and\ \bibinfo {author} {\bibfnamefont {R.}~\bibnamefont {Ruffini}},\
  }\bibfield  {title} {\bibinfo {title} {{On Einstein clusters}},\ }\href
  {https://doi.org/10.1142/S2010194512006356} {\bibfield  {journal} {\bibinfo
  {journal} {Int. J. Mod. Phys. Conf. Ser.}\ }\textbf {\bibinfo {volume}
  {12}},\ \bibinfo {pages} {146} (\bibinfo {year} {2012})}\BibitemShut
  {NoStop}%
\bibitem [{\citenamefont {Fernandes}\ and\ \citenamefont
  {Cardoso}(2025{\natexlab{b}})}]{Fernandes:2025lon}%
  \BibitemOpen
  \bibfield  {author} {\bibinfo {author} {\bibfnamefont {P.~G.~S.}\
  \bibnamefont {Fernandes}}\ and\ \bibinfo {author} {\bibfnamefont
  {V.}~\bibnamefont {Cardoso}},\ }\href@noop {} {\bibinfo {title} {{Dark matter
  as a vector field: an action principle for the Einstein cluster}}} (\bibinfo
  {year} {2025}{\natexlab{b}}),\ \Eprint {https://arxiv.org/abs/2505.00563}
  {arXiv:2505.00563 [gr-qc]} \BibitemShut {NoStop}%
\bibitem [{\citenamefont {King}(1962)}]{King:1962wi}%
  \BibitemOpen
  \bibfield  {author} {\bibinfo {author} {\bibfnamefont {I.}~\bibnamefont
  {King}},\ }\bibfield  {title} {\bibinfo {title} {{The structure of star
  clusters. I. An Empirical density law}},\ }\href
  {https://doi.org/10.1086/108756} {\bibfield  {journal} {\bibinfo  {journal}
  {Astron. J.}\ }\textbf {\bibinfo {volume} {67}},\ \bibinfo {pages} {471}
  (\bibinfo {year} {1962})}\BibitemShut {NoStop}%
\bibitem [{\citenamefont {Jaffe}(1983)}]{Jaffe}%
  \BibitemOpen
  \bibfield  {author} {\bibinfo {author} {\bibfnamefont {W.}~\bibnamefont
  {Jaffe}},\ }\bibfield  {title} {\bibinfo {title} {{A simple model for the
  distribution of light in spherical galaxies}},\ }\href
  {https://doi.org/10.1093/mnras/202.4.995} {\bibfield  {journal} {\bibinfo
  {journal} {Monthly Notices of the Royal Astronomical Society}\ }\textbf
  {\bibinfo {volume} {202}},\ \bibinfo {pages} {995} (\bibinfo {year}
  {1983})},\ \Eprint
  {https://arxiv.org/abs/https://academic.oup.com/mnras/article-pdf/202/4/995/18194452/mnras202-0995.pdf}
  {https://academic.oup.com/mnras/article-pdf/202/4/995/18194452/mnras202-0995.pdf}
  \BibitemShut {NoStop}%
\bibitem [{\citenamefont {Navarro}\ \emph {et~al.}(1996)\citenamefont
  {Navarro}, \citenamefont {Frenk},\ and\ \citenamefont
  {White}}]{Navarro:1995iw}%
  \BibitemOpen
  \bibfield  {author} {\bibinfo {author} {\bibfnamefont {J.~F.}\ \bibnamefont
  {Navarro}}, \bibinfo {author} {\bibfnamefont {C.~S.}\ \bibnamefont {Frenk}},\
  and\ \bibinfo {author} {\bibfnamefont {S.~D.~M.}\ \bibnamefont {White}},\
  }\bibfield  {title} {\bibinfo {title} {{The Structure of cold dark matter
  halos}},\ }\href {https://doi.org/10.1086/177173} {\bibfield  {journal}
  {\bibinfo  {journal} {Astrophys. J.}\ }\textbf {\bibinfo {volume} {462}},\
  \bibinfo {pages} {563} (\bibinfo {year} {1996})},\ \Eprint
  {https://arxiv.org/abs/astro-ph/9508025} {arXiv:astro-ph/9508025}
  \BibitemShut {NoStop}%
\bibitem [{\citenamefont {Zhao}(1996)}]{Zhao:1995cp}%
  \BibitemOpen
  \bibfield  {author} {\bibinfo {author} {\bibfnamefont {H.}~\bibnamefont
  {Zhao}},\ }\bibfield  {title} {\bibinfo {title} {{Analytical models for
  galactic nuclei}},\ }\href {https://doi.org/10.1093/mnras/278.2.488}
  {\bibfield  {journal} {\bibinfo  {journal} {Mon. Not. Roy. Astron. Soc.}\
  }\textbf {\bibinfo {volume} {278}},\ \bibinfo {pages} {488} (\bibinfo {year}
  {1996})},\ \Eprint {https://arxiv.org/abs/astro-ph/9509122}
  {arXiv:astro-ph/9509122} \BibitemShut {NoStop}%
\bibitem [{\citenamefont {Fernandes}\ and\ \citenamefont
  {Cardoso}()}]{HighPrecisionSpinningBHs}%
  \BibitemOpen
  \bibfield  {author} {\bibinfo {author} {\bibfnamefont {P.~G.~S.}\
  \bibnamefont {Fernandes}}\ and\ \bibinfo {author} {\bibfnamefont
  {V.}~\bibnamefont {Cardoso}},\ }\href@noop {} {}\bibinfo {howpublished}
  {\url{https://github.com/pgsfernandes/SpinningGalacticBlackHoles}, see also
  \url{https://the-center-of-gravity.com/data-and-routines/}}\BibitemShut
  {NoStop}%
\bibitem [{\citenamefont {Fernandes}\ and\ \citenamefont
  {Mulryne}(2023)}]{Fernandes:2022gde}%
  \BibitemOpen
  \bibfield  {author} {\bibinfo {author} {\bibfnamefont {P.~G.~S.}\
  \bibnamefont {Fernandes}}\ and\ \bibinfo {author} {\bibfnamefont {D.~J.}\
  \bibnamefont {Mulryne}},\ }\bibfield  {title} {\bibinfo {title} {{A new
  approach and code for spinning black holes in modified gravity}},\ }\href
  {https://doi.org/10.1088/1361-6382/ace232} {\bibfield  {journal} {\bibinfo
  {journal} {Class. Quant. Grav.}\ }\textbf {\bibinfo {volume} {40}},\ \bibinfo
  {pages} {165001} (\bibinfo {year} {2023})},\ \Eprint
  {https://arxiv.org/abs/2212.07293} {arXiv:2212.07293 [gr-qc]} \BibitemShut
  {NoStop}%
\bibitem [{\citenamefont {Boyer}\ and\ \citenamefont
  {Lindquist}(1967)}]{Boyer:1966qh}%
  \BibitemOpen
  \bibfield  {author} {\bibinfo {author} {\bibfnamefont {R.~H.}\ \bibnamefont
  {Boyer}}\ and\ \bibinfo {author} {\bibfnamefont {R.~W.}\ \bibnamefont
  {Lindquist}},\ }\bibfield  {title} {\bibinfo {title} {{Maximal analytic
  extension of the Kerr metric}},\ }\href {https://doi.org/10.1063/1.1705193}
  {\bibfield  {journal} {\bibinfo  {journal} {J. Math. Phys.}\ }\textbf
  {\bibinfo {volume} {8}},\ \bibinfo {pages} {265} (\bibinfo {year}
  {1967})}\BibitemShut {NoStop}%
\bibitem [{\citenamefont {Misner}\ \emph {et~al.}(1973)\citenamefont {Misner},
  \citenamefont {Thorne},\ and\ \citenamefont {Wheeler}}]{Misner:1973prb}%
  \BibitemOpen
  \bibfield  {author} {\bibinfo {author} {\bibfnamefont {C.~W.}\ \bibnamefont
  {Misner}}, \bibinfo {author} {\bibfnamefont {K.~S.}\ \bibnamefont {Thorne}},\
  and\ \bibinfo {author} {\bibfnamefont {J.~A.}\ \bibnamefont {Wheeler}},\
  }\href@noop {} {\emph {\bibinfo {title} {{Gravitation}}}}\ (\bibinfo
  {publisher} {W. H. Freeman},\ \bibinfo {address} {San Francisco},\ \bibinfo
  {year} {1973})\BibitemShut {NoStop}%
\bibitem [{\citenamefont {Kocherlakota}\ and\ \citenamefont
  {Narayan}(2025)}]{Kocherlakota:2025cwq}%
  \BibitemOpen
  \bibfield  {author} {\bibinfo {author} {\bibfnamefont {P.}~\bibnamefont
  {Kocherlakota}}\ and\ \bibinfo {author} {\bibfnamefont {R.}~\bibnamefont
  {Narayan}},\ }\href@noop {} {\bibinfo {title} {{Doubly Separable Spacetimes
  and Symmetry Constraints on their Self-Gravitating Matter Content}}}
  (\bibinfo {year} {2025}),\ \Eprint {https://arxiv.org/abs/2507.18706}
  {arXiv:2507.18706 [gr-qc]} \BibitemShut {NoStop}%
\bibitem [{\citenamefont {Lynch}\ \emph {et~al.}(2024)\citenamefont {Lynch},
  \citenamefont {Witzany}, \citenamefont {van~de Meent},\ and\ \citenamefont
  {Warburton}}]{Lynch:2024ohd}%
  \BibitemOpen
  \bibfield  {author} {\bibinfo {author} {\bibfnamefont {P.}~\bibnamefont
  {Lynch}}, \bibinfo {author} {\bibfnamefont {V.}~\bibnamefont {Witzany}},
  \bibinfo {author} {\bibfnamefont {M.}~\bibnamefont {van~de Meent}},\ and\
  \bibinfo {author} {\bibfnamefont {N.}~\bibnamefont {Warburton}},\ }\bibfield
  {title} {\bibinfo {title} {{Fast inspirals and the treatment of orbital
  resonances}},\ }\href {https://doi.org/10.1088/1361-6382/ad7dc9} {\bibfield
  {journal} {\bibinfo  {journal} {Class. Quant. Grav.}\ }\textbf {\bibinfo
  {volume} {41}},\ \bibinfo {pages} {225002} (\bibinfo {year} {2024})},\
  \Eprint {https://arxiv.org/abs/2405.21072} {arXiv:2405.21072 [gr-qc]}
  \BibitemShut {NoStop}%
\bibitem [{\citenamefont {Levati}\ \emph {et~al.}(2025)\citenamefont {Levati},
  \citenamefont {C{\'a}rdenas-Avenda{\~n}o}, \citenamefont {Destounis},\ and\
  \citenamefont {Pani}}]{Levati:2025ybi}%
  \BibitemOpen
  \bibfield  {author} {\bibinfo {author} {\bibfnamefont {E.}~\bibnamefont
  {Levati}}, \bibinfo {author} {\bibfnamefont {A.}~\bibnamefont
  {C{\'a}rdenas-Avenda{\~n}o}}, \bibinfo {author} {\bibfnamefont
  {K.}~\bibnamefont {Destounis}},\ and\ \bibinfo {author} {\bibfnamefont
  {P.}~\bibnamefont {Pani}},\ }\bibfield  {title} {\bibinfo {title}
  {{Cumulative effect of orbital resonances in extreme-mass-ratio inspirals}},\
  }\href {https://doi.org/10.1103/PhysRevD.111.104006} {\bibfield  {journal}
  {\bibinfo  {journal} {Phys. Rev. D}\ }\textbf {\bibinfo {volume} {111}},\
  \bibinfo {pages} {104006} (\bibinfo {year} {2025})},\ \Eprint
  {https://arxiv.org/abs/2502.20457} {arXiv:2502.20457 [gr-qc]} \BibitemShut
  {NoStop}%
\bibitem [{\citenamefont {Lichtenberg}\ and\ \citenamefont
  {Lieberman}(2013)}]{lichtenberg2013regular}%
  \BibitemOpen
  \bibfield  {author} {\bibinfo {author} {\bibfnamefont {A.}~\bibnamefont
  {Lichtenberg}}\ and\ \bibinfo {author} {\bibfnamefont {M.}~\bibnamefont
  {Lieberman}},\ }\href {https://books.google.pt/books?id=pR3aBwAAQBAJ} {\emph
  {\bibinfo {title} {Regular and Chaotic Dynamics}}},\ Applied Mathematical
  Sciences\ (\bibinfo  {publisher} {Springer New York},\ \bibinfo {year}
  {2013})\BibitemShut {NoStop}%
\bibitem [{\citenamefont {Möser}(1962)}]{Moser:430015}%
  \BibitemOpen
  \bibfield  {author} {\bibinfo {author} {\bibfnamefont {J.}~\bibnamefont
  {Möser}},\ }\bibfield  {title} {\bibinfo {title} {{On invariant curves of
  area-preserving mappings of an annulus}},\ }\href
  {https://cds.cern.ch/record/430015} {\bibfield  {journal} {\bibinfo
  {journal} {Nachr. Akad. Wiss. Göttingen, II}\ ,\ \bibinfo {pages} {1}}
  (\bibinfo {year} {1962})}\BibitemShut {NoStop}%
\bibitem [{\citenamefont {Arnol'd}(1963)}]{Arnold_1963}%
  \BibitemOpen
  \bibfield  {author} {\bibinfo {author} {\bibfnamefont {V.~I.}\ \bibnamefont
  {Arnol'd}},\ }\bibfield  {title} {\bibinfo {title} {Proof of a theorem of
  a. n. kolmogorov on the invariance of quasi-periodic motions under small
  perturbations of the hamiltonian},\ }\href
  {https://doi.org/10.1070/RM1963v018n05ABEH004130} {\bibfield  {journal}
  {\bibinfo  {journal} {Russian Mathematical Surveys}\ }\textbf {\bibinfo
  {volume} {18}},\ \bibinfo {pages} {9} (\bibinfo {year} {1963})}\BibitemShut
  {NoStop}%
\bibitem [{\citenamefont {Birkhoff}(1913)}]{Birkhoff:1913}%
  \BibitemOpen
  \bibfield  {author} {\bibinfo {author} {\bibfnamefont {G.~D.}\ \bibnamefont
  {Birkhoff}},\ }\bibfield  {title} {\bibinfo {title} {Proof of poincaré's
  geometric theorem},\ }\href {http://www.jstor.org/stable/1988766} {\bibfield
  {journal} {\bibinfo  {journal} {Transactions of the American Mathematical
  Society}\ }\textbf {\bibinfo {volume} {14}},\ \bibinfo {pages} {14} (\bibinfo
  {year} {1913})}\BibitemShut {NoStop}%
\bibitem [{\citenamefont {Barack}(2009)}]{Barack:2009ux}%
  \BibitemOpen
  \bibfield  {author} {\bibinfo {author} {\bibfnamefont {L.}~\bibnamefont
  {Barack}},\ }\bibfield  {title} {\bibinfo {title} {{Gravitational self force
  in extreme mass-ratio inspirals}},\ }\href
  {https://doi.org/10.1088/0264-9381/26/21/213001} {\bibfield  {journal}
  {\bibinfo  {journal} {Class. Quant. Grav.}\ }\textbf {\bibinfo {volume}
  {26}},\ \bibinfo {pages} {213001} (\bibinfo {year} {2009})},\ \Eprint
  {https://arxiv.org/abs/0908.1664} {arXiv:0908.1664 [gr-qc]} \BibitemShut
  {NoStop}%
\bibitem [{\citenamefont
  {Lukes-Gerakopoulos}(2014)}]{Lukes-Gerakopoulos:2013qva}%
  \BibitemOpen
  \bibfield  {author} {\bibinfo {author} {\bibfnamefont {G.}~\bibnamefont
  {Lukes-Gerakopoulos}},\ }\bibfield  {title} {\bibinfo {title} {{Adjusting
  chaotic indicators to curved spacetimes}},\ }\href
  {https://doi.org/10.1103/PhysRevD.89.043002} {\bibfield  {journal} {\bibinfo
  {journal} {Phys. Rev. D}\ }\textbf {\bibinfo {volume} {89}},\ \bibinfo
  {pages} {043002} (\bibinfo {year} {2014})},\ \Eprint
  {https://arxiv.org/abs/1311.6281} {arXiv:1311.6281 [gr-qc]} \BibitemShut
  {NoStop}%
\bibitem [{\citenamefont {Guti{\'e}rrez-Ruiz}\ \emph
  {et~al.}(2021)\citenamefont {Guti{\'e}rrez-Ruiz}, \citenamefont
  {C{\'a}rdenas-Avenda{\~n}o}, \citenamefont {Yunes},\ and\ \citenamefont
  {Pach{\'o}n}}]{Gutierrez-Ruiz:2018tre}%
  \BibitemOpen
  \bibfield  {author} {\bibinfo {author} {\bibfnamefont {A.~F.}\ \bibnamefont
  {Guti{\'e}rrez-Ruiz}}, \bibinfo {author} {\bibfnamefont {A.}~\bibnamefont
  {C{\'a}rdenas-Avenda{\~n}o}}, \bibinfo {author} {\bibfnamefont
  {N.}~\bibnamefont {Yunes}},\ and\ \bibinfo {author} {\bibfnamefont {L.~A.}\
  \bibnamefont {Pach{\'o}n}},\ }\bibfield  {title} {\bibinfo {title} {{Stealth
  chaos due to frame-dragging}},\ }\href
  {https://doi.org/10.1088/1361-6382/abff99} {\bibfield  {journal} {\bibinfo
  {journal} {Class. Quant. Grav.}\ }\textbf {\bibinfo {volume} {38}},\ \bibinfo
  {pages} {145013} (\bibinfo {year} {2021})},\ \Eprint
  {https://arxiv.org/abs/1806.06476} {arXiv:1806.06476 [gr-qc]} \BibitemShut
  {NoStop}%
\bibitem [{\citenamefont {Pan}\ \emph {et~al.}(2022)\citenamefont {Pan},
  \citenamefont {Lyu},\ and\ \citenamefont {Yang}}]{Pan:2021lyw}%
  \BibitemOpen
  \bibfield  {author} {\bibinfo {author} {\bibfnamefont {Z.}~\bibnamefont
  {Pan}}, \bibinfo {author} {\bibfnamefont {Z.}~\bibnamefont {Lyu}},\ and\
  \bibinfo {author} {\bibfnamefont {H.}~\bibnamefont {Yang}},\ }\bibfield
  {title} {\bibinfo {title} {{Mass-gap extreme mass ratio inspirals}},\ }\href
  {https://doi.org/10.1103/PhysRevD.105.083005} {\bibfield  {journal} {\bibinfo
   {journal} {Phys. Rev. D}\ }\textbf {\bibinfo {volume} {105}},\ \bibinfo
  {pages} {083005} (\bibinfo {year} {2022})},\ \Eprint
  {https://arxiv.org/abs/2112.10237} {arXiv:2112.10237 [astro-ph.HE]}
  \BibitemShut {NoStop}%
\bibitem [{\citenamefont {Destounis}\ and\ \citenamefont
  {Kokkotas}(2023)}]{Destounis:2023cim}%
  \BibitemOpen
  \bibfield  {author} {\bibinfo {author} {\bibfnamefont {K.}~\bibnamefont
  {Destounis}}\ and\ \bibinfo {author} {\bibfnamefont {K.~D.}\ \bibnamefont
  {Kokkotas}},\ }\bibfield  {title} {\bibinfo {title} {{Slowly-rotating compact
  objects: the nonintegrability of Hartle{\textendash}Thorne particle
  geodesics}},\ }\href {https://doi.org/10.1007/s10714-023-03170-z} {\bibfield
  {journal} {\bibinfo  {journal} {Gen. Rel. Grav.}\ }\textbf {\bibinfo {volume}
  {55}},\ \bibinfo {pages} {123} (\bibinfo {year} {2023})},\ \Eprint
  {https://arxiv.org/abs/2305.18522} {arXiv:2305.18522 [gr-qc]} \BibitemShut
  {NoStop}%
\bibitem [{\citenamefont {Cornish}\ and\ \citenamefont
  {Littenberg}(2015)}]{Cornish:2014kda}%
  \BibitemOpen
  \bibfield  {author} {\bibinfo {author} {\bibfnamefont {N.~J.}\ \bibnamefont
  {Cornish}}\ and\ \bibinfo {author} {\bibfnamefont {T.~B.}\ \bibnamefont
  {Littenberg}},\ }\bibfield  {title} {\bibinfo {title} {{BayesWave: Bayesian
  Inference for Gravitational Wave Bursts and Instrument Glitches}},\ }\href
  {https://doi.org/10.1088/0264-9381/32/13/135012} {\bibfield  {journal}
  {\bibinfo  {journal} {Class. Quant. Grav.}\ }\textbf {\bibinfo {volume}
  {32}},\ \bibinfo {pages} {135012} (\bibinfo {year} {2015})},\ \Eprint
  {https://arxiv.org/abs/1410.3835} {arXiv:1410.3835 [gr-qc]} \BibitemShut
  {NoStop}%
\bibitem [{\citenamefont {Coughlin}\ \emph {et~al.}(2019)\citenamefont
  {Coughlin} \emph {et~al.}}]{Coughlin:2019ref}%
  \BibitemOpen
  \bibfield  {author} {\bibinfo {author} {\bibfnamefont {S.~B.}\ \bibnamefont
  {Coughlin}} \emph {et~al.},\ }\bibfield  {title} {\bibinfo {title}
  {{Classifying the unknown: discovering novel gravitational-wave detector
  glitches using similarity learning}},\ }\href
  {https://doi.org/10.1103/PhysRevD.99.082002} {\bibfield  {journal} {\bibinfo
  {journal} {Phys. Rev. D}\ }\textbf {\bibinfo {volume} {99}},\ \bibinfo
  {pages} {082002} (\bibinfo {year} {2019})},\ \Eprint
  {https://arxiv.org/abs/1903.04058} {arXiv:1903.04058 [astro-ph.IM]}
  \BibitemShut {NoStop}%
\bibitem [{\citenamefont {Cabero}\ \emph {et~al.}(2019)\citenamefont {Cabero}
  \emph {et~al.}}]{Cabero:2019orq}%
  \BibitemOpen
  \bibfield  {author} {\bibinfo {author} {\bibfnamefont {M.}~\bibnamefont
  {Cabero}} \emph {et~al.},\ }\bibfield  {title} {\bibinfo {title} {{Blip
  glitches in Advanced LIGO data}},\ }\href
  {https://doi.org/10.1088/1361-6382/ab2e14} {\bibfield  {journal} {\bibinfo
  {journal} {Class. Quant. Grav.}\ }\textbf {\bibinfo {volume} {36}},\ \bibinfo
  {pages} {15} (\bibinfo {year} {2019})},\ \Eprint
  {https://arxiv.org/abs/1901.05093} {arXiv:1901.05093 [physics.ins-det]}
  \BibitemShut {NoStop}%
\bibitem [{\citenamefont {Edwards}\ \emph {et~al.}(2020)\citenamefont
  {Edwards}, \citenamefont {Maturana-Russel}, \citenamefont {Meyer},
  \citenamefont {Gair}, \citenamefont {Korsakova},\ and\ \citenamefont
  {Christensen}}]{Edwards:2020tlp}%
  \BibitemOpen
  \bibfield  {author} {\bibinfo {author} {\bibfnamefont {M.~C.}\ \bibnamefont
  {Edwards}}, \bibinfo {author} {\bibfnamefont {P.}~\bibnamefont
  {Maturana-Russel}}, \bibinfo {author} {\bibfnamefont {R.}~\bibnamefont
  {Meyer}}, \bibinfo {author} {\bibfnamefont {J.}~\bibnamefont {Gair}},
  \bibinfo {author} {\bibfnamefont {N.}~\bibnamefont {Korsakova}},\ and\
  \bibinfo {author} {\bibfnamefont {N.}~\bibnamefont {Christensen}},\
  }\bibfield  {title} {\bibinfo {title} {{Identifying and Addressing
  Nonstationary LISA Noise}},\ }\href
  {https://doi.org/10.1103/PhysRevD.102.084062} {\bibfield  {journal} {\bibinfo
   {journal} {Phys. Rev. D}\ }\textbf {\bibinfo {volume} {102}},\ \bibinfo
  {pages} {084062} (\bibinfo {year} {2020})},\ \Eprint
  {https://arxiv.org/abs/2004.07515} {arXiv:2004.07515 [gr-qc]} \BibitemShut
  {NoStop}%
\bibitem [{\citenamefont {Cornish}\ \emph {et~al.}(2021)\citenamefont
  {Cornish}, \citenamefont {Littenberg}, \citenamefont {B{\'e}csy},
  \citenamefont {Chatziioannou}, \citenamefont {Clark}, \citenamefont
  {Ghonge},\ and\ \citenamefont {Millhouse}}]{Cornish:2020dwh}%
  \BibitemOpen
  \bibfield  {author} {\bibinfo {author} {\bibfnamefont {N.~J.}\ \bibnamefont
  {Cornish}}, \bibinfo {author} {\bibfnamefont {T.~B.}\ \bibnamefont
  {Littenberg}}, \bibinfo {author} {\bibfnamefont {B.}~\bibnamefont
  {B{\'e}csy}}, \bibinfo {author} {\bibfnamefont {K.}~\bibnamefont
  {Chatziioannou}}, \bibinfo {author} {\bibfnamefont {J.~A.}\ \bibnamefont
  {Clark}}, \bibinfo {author} {\bibfnamefont {S.}~\bibnamefont {Ghonge}},\ and\
  \bibinfo {author} {\bibfnamefont {M.}~\bibnamefont {Millhouse}},\ }\bibfield
  {title} {\bibinfo {title} {{BayesWave analysis pipeline in the era of
  gravitational wave observations}},\ }\href
  {https://doi.org/10.1103/PhysRevD.103.044006} {\bibfield  {journal} {\bibinfo
   {journal} {Phys. Rev. D}\ }\textbf {\bibinfo {volume} {103}},\ \bibinfo
  {pages} {044006} (\bibinfo {year} {2021})},\ \Eprint
  {https://arxiv.org/abs/2011.09494} {arXiv:2011.09494 [gr-qc]} \BibitemShut
  {NoStop}%
\bibitem [{\citenamefont {Chatziioannou}\ \emph {et~al.}(2021)\citenamefont
  {Chatziioannou}, \citenamefont {Cornish}, \citenamefont {Wijngaarden},\ and\
  \citenamefont {Littenberg}}]{Chatziioannou:2021ezd}%
  \BibitemOpen
  \bibfield  {author} {\bibinfo {author} {\bibfnamefont {K.}~\bibnamefont
  {Chatziioannou}}, \bibinfo {author} {\bibfnamefont {N.}~\bibnamefont
  {Cornish}}, \bibinfo {author} {\bibfnamefont {M.}~\bibnamefont
  {Wijngaarden}},\ and\ \bibinfo {author} {\bibfnamefont {T.~B.}\ \bibnamefont
  {Littenberg}},\ }\bibfield  {title} {\bibinfo {title} {{Modeling compact
  binary signals and instrumental glitches in gravitational wave data}},\
  }\href {https://doi.org/10.1103/PhysRevD.103.044013} {\bibfield  {journal}
  {\bibinfo  {journal} {Phys. Rev. D}\ }\textbf {\bibinfo {volume} {103}},\
  \bibinfo {pages} {044013} (\bibinfo {year} {2021})},\ \Eprint
  {https://arxiv.org/abs/2101.01200} {arXiv:2101.01200 [gr-qc]} \BibitemShut
  {NoStop}%
\bibitem [{\citenamefont {Muratore}\ \emph {et~al.}(2025)\citenamefont
  {Muratore}, \citenamefont {Gair}, \citenamefont {Hartwig}, \citenamefont
  {Katz},\ and\ \citenamefont {Toubiana}}]{Muratore:2025knh}%
  \BibitemOpen
  \bibfield  {author} {\bibinfo {author} {\bibfnamefont {M.}~\bibnamefont
  {Muratore}}, \bibinfo {author} {\bibfnamefont {J.}~\bibnamefont {Gair}},
  \bibinfo {author} {\bibfnamefont {O.}~\bibnamefont {Hartwig}}, \bibinfo
  {author} {\bibfnamefont {M.~L.}\ \bibnamefont {Katz}},\ and\ \bibinfo
  {author} {\bibfnamefont {A.}~\bibnamefont {Toubiana}},\ }\href@noop {}
  {\bibinfo {title} {{A pipeline for searching and fitting instrumental
  glitches in LISA data}}} (\bibinfo {year} {2025}),\ \Eprint
  {https://arxiv.org/abs/2505.19870} {arXiv:2505.19870 [gr-qc]} \BibitemShut
  {NoStop}%
\bibitem [{\citenamefont {Vicente}\ and\ \citenamefont
  {Cardoso}(2022)}]{Vicente:2022ivh}%
  \BibitemOpen
  \bibfield  {author} {\bibinfo {author} {\bibfnamefont {R.}~\bibnamefont
  {Vicente}}\ and\ \bibinfo {author} {\bibfnamefont {V.}~\bibnamefont
  {Cardoso}},\ }\bibfield  {title} {\bibinfo {title} {{Dynamical friction of
  black holes in ultralight dark matter}},\ }\href
  {https://doi.org/10.1103/PhysRevD.105.083008} {\bibfield  {journal} {\bibinfo
   {journal} {Phys. Rev. D}\ }\textbf {\bibinfo {volume} {105}},\ \bibinfo
  {pages} {083008} (\bibinfo {year} {2022})},\ \Eprint
  {https://arxiv.org/abs/2201.08854} {arXiv:2201.08854 [gr-qc]} \BibitemShut
  {NoStop}%
\bibitem [{\citenamefont {Flanagan}\ \emph {et~al.}(2014)\citenamefont
  {Flanagan}, \citenamefont {Hughes},\ and\ \citenamefont
  {Ruangsri}}]{Flanagan:2012kg}%
  \BibitemOpen
  \bibfield  {author} {\bibinfo {author} {\bibfnamefont {E.~E.}\ \bibnamefont
  {Flanagan}}, \bibinfo {author} {\bibfnamefont {S.~A.}\ \bibnamefont
  {Hughes}},\ and\ \bibinfo {author} {\bibfnamefont {U.}~\bibnamefont
  {Ruangsri}},\ }\bibfield  {title} {\bibinfo {title} {{Resonantly enhanced and
  diminished strong-field gravitational-wave fluxes}},\ }\href
  {https://doi.org/10.1103/PhysRevD.89.084028} {\bibfield  {journal} {\bibinfo
  {journal} {Phys. Rev. D}\ }\textbf {\bibinfo {volume} {89}},\ \bibinfo
  {pages} {084028} (\bibinfo {year} {2014})},\ \Eprint
  {https://arxiv.org/abs/1208.3906} {arXiv:1208.3906 [gr-qc]} \BibitemShut
  {NoStop}%
\bibitem [{\citenamefont {Pan}\ \emph {et~al.}(2023)\citenamefont {Pan},
  \citenamefont {Yang}, \citenamefont {Bernard},\ and\ \citenamefont
  {Bonga}}]{Pan:2023wau}%
  \BibitemOpen
  \bibfield  {author} {\bibinfo {author} {\bibfnamefont {Z.}~\bibnamefont
  {Pan}}, \bibinfo {author} {\bibfnamefont {H.}~\bibnamefont {Yang}}, \bibinfo
  {author} {\bibfnamefont {L.}~\bibnamefont {Bernard}},\ and\ \bibinfo {author}
  {\bibfnamefont {B.}~\bibnamefont {Bonga}},\ }\bibfield  {title} {\bibinfo
  {title} {{Resonant dynamics of extreme mass-ratio inspirals in a perturbed
  Kerr spacetime}},\ }\href {https://doi.org/10.1103/PhysRevD.108.104026}
  {\bibfield  {journal} {\bibinfo  {journal} {Phys. Rev. D}\ }\textbf {\bibinfo
  {volume} {108}},\ \bibinfo {pages} {104026} (\bibinfo {year} {2023})},\
  \Eprint {https://arxiv.org/abs/2306.06576} {arXiv:2306.06576 [gr-qc]}
  \BibitemShut {NoStop}%
\bibitem [{\citenamefont {Cardoso}\ \emph {et~al.}(2021)\citenamefont
  {Cardoso}, \citenamefont {Duque},\ and\ \citenamefont
  {Khanna}}]{Cardoso:2021vjq}%
  \BibitemOpen
  \bibfield  {author} {\bibinfo {author} {\bibfnamefont {V.}~\bibnamefont
  {Cardoso}}, \bibinfo {author} {\bibfnamefont {F.}~\bibnamefont {Duque}},\
  and\ \bibinfo {author} {\bibfnamefont {G.}~\bibnamefont {Khanna}},\
  }\bibfield  {title} {\bibinfo {title} {{Gravitational tuning forks and
  hierarchical triple systems}},\ }\href
  {https://doi.org/10.1103/PhysRevD.103.L081501} {\bibfield  {journal}
  {\bibinfo  {journal} {Phys. Rev. D}\ }\textbf {\bibinfo {volume} {103}},\
  \bibinfo {pages} {L081501} (\bibinfo {year} {2021})},\ \Eprint
  {https://arxiv.org/abs/2101.01186} {arXiv:2101.01186 [gr-qc]} \BibitemShut
  {NoStop}%
\bibitem [{\citenamefont {Kuntz}(2022)}]{Kuntz:2021hhm}%
  \BibitemOpen
  \bibfield  {author} {\bibinfo {author} {\bibfnamefont {A.}~\bibnamefont
  {Kuntz}},\ }\bibfield  {title} {\bibinfo {title} {{Precession resonances in
  hierarchical triple systems}},\ }\href
  {https://doi.org/10.1103/PhysRevD.105.024017} {\bibfield  {journal} {\bibinfo
   {journal} {Phys. Rev. D}\ }\textbf {\bibinfo {volume} {105}},\ \bibinfo
  {pages} {024017} (\bibinfo {year} {2022})},\ \Eprint
  {https://arxiv.org/abs/2112.05167} {arXiv:2112.05167 [gr-qc]} \BibitemShut
  {NoStop}%
\bibitem [{\citenamefont {Kuntz}\ \emph {et~al.}(2021)\citenamefont {Kuntz},
  \citenamefont {Serra},\ and\ \citenamefont {Trincherini}}]{Kuntz:2021ohi}%
  \BibitemOpen
  \bibfield  {author} {\bibinfo {author} {\bibfnamefont {A.}~\bibnamefont
  {Kuntz}}, \bibinfo {author} {\bibfnamefont {F.}~\bibnamefont {Serra}},\ and\
  \bibinfo {author} {\bibfnamefont {E.}~\bibnamefont {Trincherini}},\
  }\bibfield  {title} {\bibinfo {title} {{Effective two-body approach to the
  hierarchical three-body problem}},\ }\href
  {https://doi.org/10.1103/PhysRevD.104.024016} {\bibfield  {journal} {\bibinfo
   {journal} {Phys. Rev. D}\ }\textbf {\bibinfo {volume} {104}},\ \bibinfo
  {pages} {024016} (\bibinfo {year} {2021})},\ \Eprint
  {https://arxiv.org/abs/2104.13387} {arXiv:2104.13387 [hep-th]} \BibitemShut
  {NoStop}%
\bibitem [{\citenamefont {Kuntz}\ \emph {et~al.}(2023)\citenamefont {Kuntz},
  \citenamefont {Serra},\ and\ \citenamefont {Trincherini}}]{Kuntz:2022onu}%
  \BibitemOpen
  \bibfield  {author} {\bibinfo {author} {\bibfnamefont {A.}~\bibnamefont
  {Kuntz}}, \bibinfo {author} {\bibfnamefont {F.}~\bibnamefont {Serra}},\ and\
  \bibinfo {author} {\bibfnamefont {E.}~\bibnamefont {Trincherini}},\
  }\bibfield  {title} {\bibinfo {title} {{Effective two-body approach to the
  hierarchical three-body problem: Quadrupole to 1PN}},\ }\href
  {https://doi.org/10.1103/PhysRevD.107.044011} {\bibfield  {journal} {\bibinfo
   {journal} {Phys. Rev. D}\ }\textbf {\bibinfo {volume} {107}},\ \bibinfo
  {pages} {044011} (\bibinfo {year} {2023})},\ \Eprint
  {https://arxiv.org/abs/2210.13493} {arXiv:2210.13493 [gr-qc]} \BibitemShut
  {NoStop}%
\bibitem [{\citenamefont {Camilloni}\ \emph {et~al.}(2024)\citenamefont
  {Camilloni}, \citenamefont {Harmark}, \citenamefont {Grignani}, \citenamefont
  {Orselli},\ and\ \citenamefont {Pica}}]{Camilloni:2023xvf}%
  \BibitemOpen
  \bibfield  {author} {\bibinfo {author} {\bibfnamefont {F.}~\bibnamefont
  {Camilloni}}, \bibinfo {author} {\bibfnamefont {T.}~\bibnamefont {Harmark}},
  \bibinfo {author} {\bibfnamefont {G.}~\bibnamefont {Grignani}}, \bibinfo
  {author} {\bibfnamefont {M.}~\bibnamefont {Orselli}},\ and\ \bibinfo {author}
  {\bibfnamefont {D.}~\bibnamefont {Pica}},\ }\bibfield  {title} {\bibinfo
  {title} {{Binary mergers in strong gravity background of Kerr black hole}},\
  }\href {https://doi.org/10.1093/mnras/stae1093} {\bibfield  {journal}
  {\bibinfo  {journal} {Mon. Not. Roy. Astron. Soc.}\ }\textbf {\bibinfo
  {volume} {531}},\ \bibinfo {pages} {1884} (\bibinfo {year} {2024})},\ \Eprint
  {https://arxiv.org/abs/2310.06894} {arXiv:2310.06894 [gr-qc]} \BibitemShut
  {NoStop}%
\bibitem [{\citenamefont {Camilloni}\ \emph {et~al.}(2023)\citenamefont
  {Camilloni}, \citenamefont {Grignani}, \citenamefont {Harmark}, \citenamefont
  {Oliveri}, \citenamefont {Orselli},\ and\ \citenamefont
  {Pica}}]{Camilloni:2023rra}%
  \BibitemOpen
  \bibfield  {author} {\bibinfo {author} {\bibfnamefont {F.}~\bibnamefont
  {Camilloni}}, \bibinfo {author} {\bibfnamefont {G.}~\bibnamefont {Grignani}},
  \bibinfo {author} {\bibfnamefont {T.}~\bibnamefont {Harmark}}, \bibinfo
  {author} {\bibfnamefont {R.}~\bibnamefont {Oliveri}}, \bibinfo {author}
  {\bibfnamefont {M.}~\bibnamefont {Orselli}},\ and\ \bibinfo {author}
  {\bibfnamefont {D.}~\bibnamefont {Pica}},\ }\bibfield  {title} {\bibinfo
  {title} {{Tidal deformations of a binary system induced by an external Kerr
  black hole}},\ }\href {https://doi.org/10.1103/PhysRevD.107.084011}
  {\bibfield  {journal} {\bibinfo  {journal} {Phys. Rev. D}\ }\textbf {\bibinfo
  {volume} {107}},\ \bibinfo {pages} {084011} (\bibinfo {year} {2023})},\
  \Eprint {https://arxiv.org/abs/2301.04879} {arXiv:2301.04879 [gr-qc]}
  \BibitemShut {NoStop}%
\bibitem [{\citenamefont {Grilli}\ \emph {et~al.}(2025)\citenamefont {Grilli},
  \citenamefont {Orselli}, \citenamefont {Pere{\~n}iguez},\ and\ \citenamefont
  {Pica}}]{Grilli:2024fds}%
  \BibitemOpen
  \bibfield  {author} {\bibinfo {author} {\bibfnamefont {E.}~\bibnamefont
  {Grilli}}, \bibinfo {author} {\bibfnamefont {M.}~\bibnamefont {Orselli}},
  \bibinfo {author} {\bibfnamefont {D.}~\bibnamefont {Pere{\~n}iguez}},\ and\
  \bibinfo {author} {\bibfnamefont {D.}~\bibnamefont {Pica}},\ }\bibfield
  {title} {\bibinfo {title} {{Charged binaries in gravitational tides}},\
  }\href {https://doi.org/10.1088/1475-7516/2025/02/028} {\bibfield  {journal}
  {\bibinfo  {journal} {JCAP}\ }\textbf {\bibinfo {volume} {02}},\ \bibinfo
  {pages} {028}},\ \Eprint {https://arxiv.org/abs/2411.08089} {arXiv:2411.08089
  [gr-qc]} \BibitemShut {NoStop}%
\bibitem [{\citenamefont {Cocco}\ \emph {et~al.}(2025)\citenamefont {Cocco},
  \citenamefont {Grignani}, \citenamefont {Harmark}, \citenamefont {Orselli},\
  and\ \citenamefont {Pica}}]{Cocco:2025adu}%
  \BibitemOpen
  \bibfield  {author} {\bibinfo {author} {\bibfnamefont {M.}~\bibnamefont
  {Cocco}}, \bibinfo {author} {\bibfnamefont {G.}~\bibnamefont {Grignani}},
  \bibinfo {author} {\bibfnamefont {T.}~\bibnamefont {Harmark}}, \bibinfo
  {author} {\bibfnamefont {M.}~\bibnamefont {Orselli}},\ and\ \bibinfo {author}
  {\bibfnamefont {D.}~\bibnamefont {Pica}},\ }\bibfield  {title} {\bibinfo
  {title} {{Strong-gravity precession resonances for binary systems orbiting a
  Schwarzschild black hole}},\ }\href {https://doi.org/10.1103/l542-s32g}
  {\bibfield  {journal} {\bibinfo  {journal} {Phys. Rev. D}\ }\textbf {\bibinfo
  {volume} {112}},\ \bibinfo {pages} {044010} (\bibinfo {year} {2025})},\
  \Eprint {https://arxiv.org/abs/2505.15901} {arXiv:2505.15901 [gr-qc]}
  \BibitemShut {NoStop}%
\end{thebibliography}%

\end{document}